\documentclass[a4paper,oneside,11pt,notitlepage]{report}
\usepackage{mathrsfs}
\usepackage{amssymb}
\usepackage{amsmath}
\usepackage[latin1]{inputenc}
\usepackage[dvips]{graphicx}
\usepackage{feynmp}
\usepackage{cite}
\usepackage{mcite}
\newcommand{\fslash}[1]{\hbox{$#1$}\!\!\!\!/\;}

\newcommand{\half}{\frac{1}{2}}
\newcommand{\dF}{d_\mathrm{F}}
\newcommand{\dA}{d_\mathrm{A}}
\newcommand{\CF}{C_\mathrm{F}}
\newcommand{\CA}{C_\mathrm{A}}
\newcommand{\TF}{T_\mathrm{F}}

\newcommand{\mD}{m_\mathrm{D}}
\newcommand{\nF}{\frac{n_\mathrm{F}}{2}}
\newcommand{\NF}{n_\mathrm{F}}
\newcommand{\nS}{n_\mathrm{S}}
\newcommand{\Nc}{N_\mathrm{c}}

\newcommand{\pE}{p_\mathrm{E}}
\newcommand{\pM}{p_\mathrm{M}}

\newcommand{\dd}{\mathrm{d}}

\begin{document}

\thispagestyle{empty}

\begin{titlepage}
\begin{flushright}
HU-P-D130\\
hep-ph/0609226
\end{flushright}
\begin{centering}
\vfill

{\Large{\bf Thermodynamics of electroweak matter}}

\vspace{0.8cm}

A.~Gynther\footnote{Present address: Department of Physics \& Astronomy, Brandon University, Brandon, MB R7A 6A9, Canada. E-mail: gynthera@brandonu.ca.}

\vspace{0.8cm}

\vspace{0.3cm}

\emph{
Theoretical Physics Division,
Department of Physical Sciences and Helsinki Institute of Physics,\\
P.O.Box~64, FIN-00014 University of Helsinki, Finland\\}

\vspace*{0.8cm}

\end{centering}

\begin{abstract}
This paper is a slightly modified version of the introductory part of a PhD thesis, also containing the articles hep-ph/0303019, hep-ph/0510375 and hep-ph/0512177. We provide a short history of the research of electroweak thermodynamics and a brief introduction to the theory as well as to the necessary theoretical tools needed to work at finite temperatures. We then review computations regarding the pressure of electroweak matter at high temperatures (the full expression of the perturbative expansion of the pressure is given in the appendix) and the electroweak phase diagram at finite chemical potentials. Finally, we compare electroweak and QCD thermodynamics. 
\end{abstract}

\vfill
\noindent

\vspace*{1cm}

\vfill

\end{titlepage}

\tableofcontents

\chapter{Introduction}
\setcounter{page}{1}
\pagenumbering{arabic}

Understanding the properties of matter is one of the main goals of modern physics. Under normal conditions, matter is composed of atoms, molecules and free electrons and interactions between them can be described by (quantum) electrodynamics. However, when the temperature and density of matter are increased, these basic building blocks of matter as we know them begin to break apart and matter eventually becomes a collection of elementary particles. Then we need to resort to theories of particle physics to describe its properties.

The elementary particles as well as the strong, weak and electromagnetic interactions between them are to high accuracy described by the standard model of particle physics. It contains a rich collection of interesting features at extreme conditions, related to, for example, different phase transitions. At present, only the properties of strongly interacting matter, QCD plasma and hadron gas, are within the reach of experimental studies. Knowledge about the behavior of electroweak matter is, on the other hand, purely theoretical at the moment. However, it is important that we understand properties of weakly interacting matter as well, since it is suggested by modern cosmology that in the very early universe temperature was so high that such matter existed.

In this thesis we will study some aspects of electroweak matter. The thesis consists of this introductory part and of three research publications \cite{Gynther:2003za,Gynther:2005dj,Gynther:2005av} . The first paper \cite{Gynther:2003za} considers the electroweak phase diagram. In it, high temperature dimensional reduction of the standard model was formulated in the presence of non-zero chemical potentials for baryon and lepton numbers and the computation was then applied to calculate the phase diagram. Special attention was given to the location of the endpoint of the first order phase transition line. The papers \cite{Gynther:2005dj,Gynther:2005av} are concerned with calculating the pressure of electroweak matter at finite temperature and zero chemical potentials. In Ref.~\cite{Gynther:2005dj} the pressure was calculated at high temperatures to three loops, or to order $g^5$ in the coupling constants and its properties, such as scale dependence, were analyzed. Comparison with QCD pressure was also performed. In Ref.~\cite{Gynther:2005av} the previous calculation was extended to lower temperatures (temperatures around the electroweak crossover) by reorganizing the effective field theories used to calculate the pressure.

This introductory part is organized as follows. We will first review some history of electroweak thermodynamics and then in chapter 2 present the basic structure of the electroweak theory. In chapter 3 we consider field theories at finite temperatures and chemical potentials in general. The actual calculations concerning the pressure of the standard model (chapter 4) and electroweak phase diagram (chapter 5) are discussed next. In the final chapter we consider the similarities and differences between the QCD and electroweak thermodynamics relevant to the thesis. Expansion of the pressure of the standard model to three loop order is given in detail in the appendix A.  

\section{History}

Soon after the electroweak model was constructed, interest about its consequences when applied to thermal systems arose. Based on the close analogy between the bosonic sector of the electroweak model and the Ginzburg-Landau theory of superconductivity, Kirzhnits and Linde proposed that the symmetry that is spontaneously broken at low temperatures would be restored at high temperatures \cite{Kirzhnits:1972iw,Kirzhnits:1972ut}. The reason is that, unlike at low temperatures where the equilibrium state of the system is such that energy is minimized, at high temperatures thermal equilibrium is achieved when entropy of the system is maximized,\footnote{We are considering canonical ensembles.} \emph{i.e.}, when the symmetry of the theory is restored (no order in the system). They estimated that the restoration of symmetry would happen in a second order phase transition at temperatures of the order of $T \sim G_\mu^{-1/2} \sim 10^2$~GeV, where $G_\mu$ is the Fermi coupling constant. A more systematic approach using finite temperature effective potentials was then developed by Dolan and Jackiw \cite{Dolan:1973qd} and by Weinberg \cite{Weinberg:1974hy}. Those studies confirmed the validity of earlier, more heuristic, arguments and a more quantitative estimate for the critical temperature was derived for a number of different theories. Electroweak phase transition thus gained a status as a basic ingredient in cosmology. The possibility that the transition would be of first order was shortly after taken into account by Kirzhnits and Linde \cite{Kirzhnits:1976ts} who calculated leading order corrections to the previously evaluated effective potentials. Similar phase transitions in grand unified theories lead to the development of inflationary models in cosmology \cite{Guth:1980zm}.

Need for a more detailed understanding of the electroweak symmetry restoration became relevant when it was realized that the electroweak phase transition might have a crucial role in understanding the baryon number asymmetry in the universe \cite{Dimopoulos:1978kv,Kuzmin:1985mm}. Electroweak vacuum has a non-trivial topological structure and transitions between topologically distinct vacua violate conservation of baryon number which could lead to erasure of any baryon number asymmetry in the universe. On the other hand, the same transitions could also provide a mechanism for producing the baryon number asymmetry, depending on the properties of the electroweak phase transition \cite{Shaposhnikov:1986jp,*Shaposhnikov:1987tw,*Shaposhnikov:1987pf,*Bochkarev:1989kp}. If the transition is strongly first order, a baryon number asymmetry could be generated during the transition. The prospect of understanding the roots of the baryon number asymmetry in the universe lead to a renewed interest in quantitative description of the electroweak phase transition.

Building on the work laid out by Dolan, Jackiw and Weinberg, the electroweak phase transition was first studied in detail with one-loop effective potential calculations \cite{Anderson:1991zb,Carrington:1991hz,Dine:1992wr}. At high temperatures and assuming that the Higgs boson is sufficiently light, the strictly one-loop expansion of the effective potential is given by \cite{Anderson:1991zb}
\begin{equation}
\mathcal{V}(\varphi,T) = D(T^2-T_0^2)\varphi^2 - ET\varphi^3 + \frac{\lambda_T}{4}\varphi^4,
\end{equation}
where $\varphi$ is the expectation value of the Higgs field and the different coefficients can be computed to be
\begin{eqnarray}
D & = & \frac{1}{8\varphi_0^2}\left(2m_W^2 + m_Z^2 + 2m_t^2\right), \nonumber \\
E & = & \frac{1}{4\pi\varphi_0^3}\left(2m_W^3 + m_Z^3\right), \nonumber \\
T_0^2 & = & \frac{1}{4D}\left(m_H^2-8B\varphi_0^2\right), \nonumber \\
\lambda_T & = & \lambda - \frac{3}{16\pi^2\varphi_0^4}\left(2m_W^4\ln\frac{m_W^2}{c_BT^2} + m_Z^4\ln\frac{m_Z^2}{c_BT^2} - 4m_t^4\ln\frac{m_t^2}{c_FT^2}\right), \nonumber \\   
B & = & \frac{3}{64\pi^2\varphi_0^4}\left(2m_W^4+m_Z^4 - 4m_t^4\right), \nonumber \\
\ln\;c_B & = & \frac{3}{2}+2\ln 4\pi - 2\gamma, \quad \ln\;c_F \;=\;\frac{3}{2}+2\ln \pi - 2\gamma.
\end{eqnarray} 
Here all the masses are measured at zero temperature and $\varphi_0 = 246$~GeV is the value of the scalar condensate at $T=0$.

\begin{figure}[tb]
\begin{center}
\includegraphics[width=0.87\textwidth]{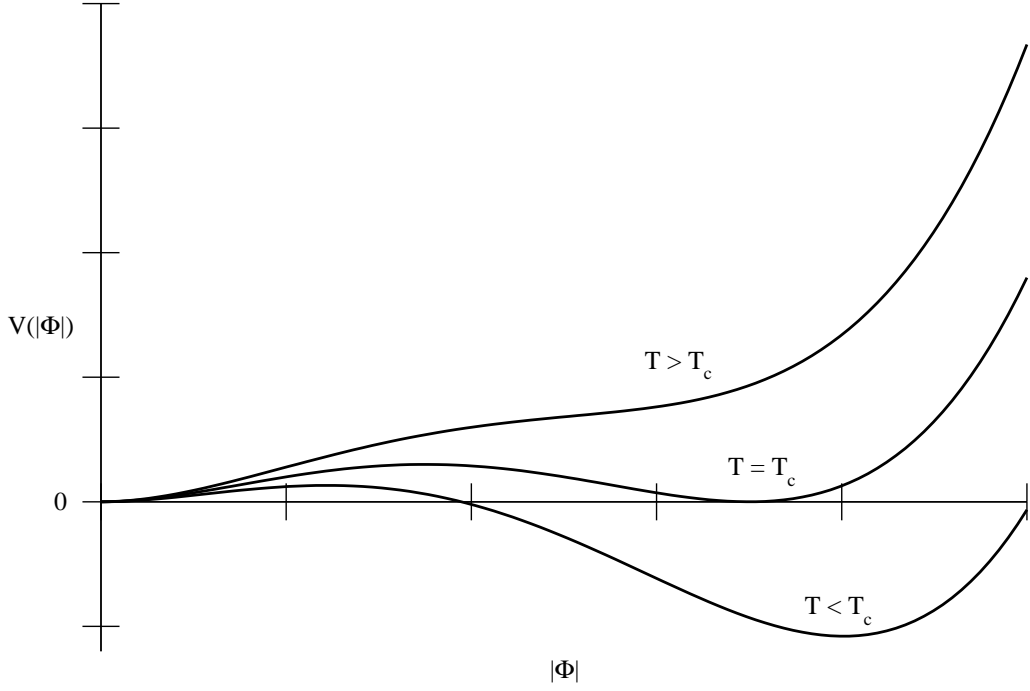}
\caption{A schematic plot of the evolution of the scalar potential as temperature is decreased below the critical temperature.}
\label{fig:finite_temperature_potential}
\end{center}
\end{figure}

At high temperatures the minimum of this potential is achieved when the expectation value of the scalar field vanishes, $\varphi = 0$, and thus the symmetry is exact. However, as the temperature is decreased, another local minimum appears, which becomes the global minimum at some critical temperature $T_c$ and it becomes favorable for the system to reside there (see Fig.~\ref{fig:finite_temperature_potential}). Thus, the symmetry of the theory gets spontaneously broken. Using the expression for the effective potential above, it is possible to calculate many of the essential features of this phase transition. For example, the critical temperature is easily evaluated to satisfy
\begin{eqnarray}
T_c & = & \frac{T_0}{\sqrt{1-\frac{E^2}{\lambda_{T_c}D}}}.
\end{eqnarray}
Another important quantity is the value of the expectation value of the scalar condensate in the broken symmetry phase at the moment of the phase transition, $\varphi_c = 2ET_c/\lambda_{T_c}$. Baryogenesis can be explained within electroweak physics only if the ratio $\varphi_c/T_c$ is large enough. The baryon number violating processes must be cut-off after the transition or any baryon number asymmetry that was generated in the transition will be washed out. Sufficient criterion for this is that  $\varphi_c/T_c \gtrsim 1$ is fulfilled \cite{Shaposhnikov:1986jp,*Shaposhnikov:1987tw,*Shaposhnikov:1987pf,*Bochkarev:1989kp}.

Although the one-loop approximation can be used to calculate the characteristics of the phase transition, it is not guaranteed to be a reliable method. Indeed, perturbative calculations in gauge field theories are known to suffer from infrared problems at high temperatures \cite{Linde:1980ts,Gross:1981br}. Collective phenomena such as screening of electric fields (Debye screening), not contained in the strict loop expansion, take place. The Debye screening can be taken into account by introducing a thermal mass for the static Coulomb fields, $m_D \sim gT$. This corresponds to resumming an infinite class of ring diagrams to the effective potential and thus yields an improved one-loop expansion of the effective potential \cite{Dine:1992wr,Carrington:1991hz}. The essential difference to the strict one-loop expansion of the potential is that the value of the coefficient $E$ of the $\sim \varphi^3$ term, responsible for the first order nature of the transition, will be reduced by a factor of $2/3$ and thus the transition becomes weaker. Especially, the result suggests that the transition will not be strong enough to explain the baryon number asymmetry if the Higgs mass is greater than $m_H \gtrsim 45$~GeV \cite{Dine:1992wr}.

The analysis has been further improved by calculating the two-loop corrections to the effective potential, to order $g^4,\lambda$ by Arnold and Espinosa \cite{Arnold:1992rz} and to order $g^4,\lambda^2$ by Fodor and Hebecker \cite{Fodor:1994bs}. Although the two-loop calculation provides (at least formally) a more precise description of the phase transition, the qualitative features remain the same as in the one-loop calculation, predicting a first order phase transition.

With a quantitative understanding provided by the effective potentials, it is possible to study the dynamics of the electroweak phase transition in more detail \cite{Anderson:1991zb,Enqvist:1991xw,Dine:1992wr,Carrington:1993ng,Ignatius:1993qn}. A first order phase transition proceeds by nucleation of true equilibrium state bubbles into the system which subsequently grow until the whole universe resides in the true equilibrium state. Such a nucleation process leads to local departure from thermal equilibrium and can leave observable traces in the universe.

\begin{figure}[tb]
\begin{center}
\includegraphics[width=0.67\textwidth]{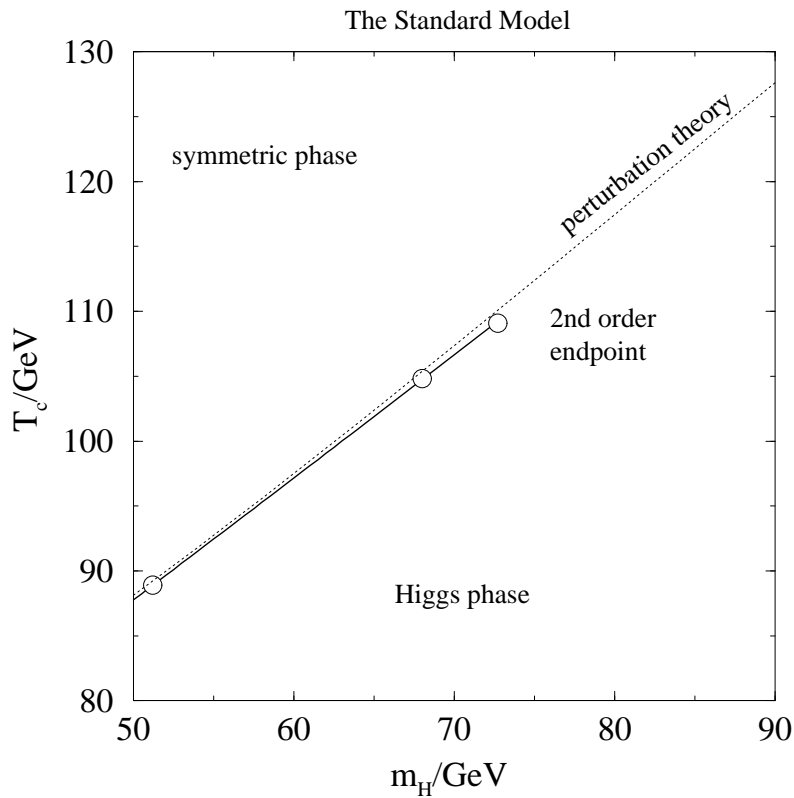}
\caption{The phase diagram of the electroweak theory.}
\label{fig:phase_diagram}
\end{center}
\end{figure}

The studies described above, based on calculating the properties of the phase transition with perturbatively computed effective potentials, though convenient, are ultimately unreliable due to infrared divergences that plague perturbative gauge field theories \cite{Linde:1980ts,Gross:1981br}. Even though the infrared problems are not manifest when calculating in the broken symmetry phase (since particles obtain masses via Higgs mechanism which serves to regulate the infrared divergences), to obtain information about the phase transition one must compare the calculations in the broken symmetry phase to those in the symmetric phase where the infrared problems reappear. To obtain reliable results, one must therefore resort to non-perturbative methods.

The non-perturbative analysis is convenient to perform by identifying the modes that are responsible for the infrared divergences in perturbation theory. These modes can be isolated to a series of effective theories \cite{Ginsparg:1980ef,Appelquist:1981vg,Jakovac:1994xg,Jakovac:1994mq,Jakovac:1996kv,Kajantie:1996dw} and we can combine the use of perturbative calculations (applied to infrared safe modes) and numerical computations (applied to modes that are infrared divergent in perturbation theory). Carrying out such calculations, the phase diagram of the electroweak theory was solved by a number of authors \cite{Kajantie:1995kf,Kajantie:1996mn,Kajantie:1996qd,Karsch:1996yh,Gurtler:1997hr}. What is observed (see the phase diagram in Fig.~\ref{fig:phase_diagram}, given in \cite{Laine:1998jb}) is that, although for small Higgs masses there is a first order phase transition in the electroweak theory, as the Higgs mass becomes larger, the phase transition weakens. The first order phase transition line has a 2nd order endpoint at $m_H \approx 72$~GeV and for larger Higgs masses no phase transition is observed, just a smooth crossover. Furthermore, the endpoint has been observed to be of $3$d Ising universality class \cite{Rummukainen:1998as}. Since there is no first order phase transition in the (minimal) electroweak theory for physical values of the Higgs mass, electroweak baryogenesis (within minimal standard model) has been ruled out.

In addition to considering just the effects of high temperatures, properties of electroweak matter have also been studied when density of matter is high. It was noted by Linde \cite{Linde:1976kh}, using again analogy to superconductivity, that increasing fermion number density would lead to an increase in the symmetry breaking, \emph{i.e.}, the expectation value of the scalar condensate would increase (conversely, non-zero fermion current density tends to restore the symmetry). These considerations were brought to a firmer ground by computing the one-loop effective potential with finite chemical potentials associated with conserved fermionic numbers taken into account \cite{PerezRojas:1987ni,Kalashnikov:1987te,Ferrer:1987jc,Ferrer:1987ag,Kapusta:1990qc}. Similarities between spontaneous symmetry breaking and Bose-Einstein condensation by considering finite bosonic chemical potentials (related to conserved gauge charges) have also been studied \cite{Kapusta:1981aa,Sannino:2003mt}.

Complete thermodynamic description of the standard model is given when, in addition to temperature and chemical potentials for the conserved particle numbers, also the effects of an external U(1) magnetic field are taken into account. It is much speculated that strong magnetic fields might have been present in the very early universe, seeding the galactic and inter-galactic magnetic fields that we observe today \cite{Enqvist:1998fw}. The effect that such magnetic fields could have on the electroweak phase transition has been studied \cite{Kajantie:1998rz} and it was observed that they can strengthen the transition somewhat and lower the transition temperature, but not enough to save electroweak baryogenesis. It was also noted that no exotic phase, analogous to the Abrikosov vortex lattice of type-II superconductors and implied by perturbative calculations \cite{Ambjorn:1988tm,*Ambjorn:1988gb,*Ambjorn:1989bd,*Ambjorn:1989sz}, could be found. 

\chapter{Basic structure of the electroweak theory}
\label{sec:ewatft}

The standard model of particle physics is a quantum gauge field theory based on local $\textrm{SU}(3)_C\times\textrm{SU}(2)_L\times\textrm{U}(1)_Y$ symmetry. It describes dynamics between leptons ($\textrm{SU}(3)_C$ singlets) and quarks (belonging to the fundamental representation of $\textrm{SU}(3)_C$). These are divided into $3$ families, each composed of $2$ leptons and $2$ quarks. The families are identical to each other in every other respect apart from the Yukawa couplings giving masses to the particles. This theory has been confirmed experimentally with an extraordinary precision and it forms the cornerstone of modern particle physics.

In this thesis we will concentrate on the electroweak sector of the standard model. The electroweak theory was first formulated by Weinberg, Glashow and Salam \cite{Weinberg:1967tq,Glashow:1961tr,Salam:1968rm} and it replaced Fermi's theory\footnote{It was already known before the electroweak theory was formulated that the Fermi coupling could be understood as an exchange of massive, charged vector bosons between fermions.} of $\beta$-decay as a description of weak interactions. It predicted neutral current processes, unknown at the time, and the discovery of those \cite{Hasert:1973ff} and of the weak gauge bosons \cite{Arnison:1983rp,*Banner:1983jy,*Arnison:1983mk,*Bagnaia:1983zx} established the status of the theory.

In this chapter we will briefly review the basic setting of the electroweak theory and its properties relevant to the thesis. The theory is defined by the Lagrangian
\begin{eqnarray}
\mathcal{L} & = & \mathcal{L}_\mathrm{bos.} + \mathcal{L}_\mathrm{ferm.} + \mathcal{L}_\mathrm{QCD} + \mathcal{L}_\mathrm{top\;mass},
\end{eqnarray}
where the different parts are defined in the following. The gauge fixing and ghost terms in the Lagrangian are not explicitly written down.

\section{Higgs mechanism}

A distinct feature of electroweak interactions is that the $W^\pm$ and $Z^0$ bosons mediating the interactions are massive. Since gauge invariance protects gauge bosons from acquiring masses, it would then seem impossible to describe weak interactions in terms of a gauge field theory. However, even though the theory has a symmetry, it is not necessary that the ground state of the theory has the same symmetry, that is, the symmetry may be spontaneously broken. This is a sufficient requirement for producing masses for gauge bosons. In the standard model, this is accomplished by introducing a scalar field, called the Higgs scalar, into the theory. The scalar belongs to the fundamental representation of the SU($2$) symmetry group and acquires a non-zero vacuum expectation value that serves to spontaneously break the symmetry. 

\begin{figure}[tb]
\begin{center}
\includegraphics[width=0.87\textwidth]{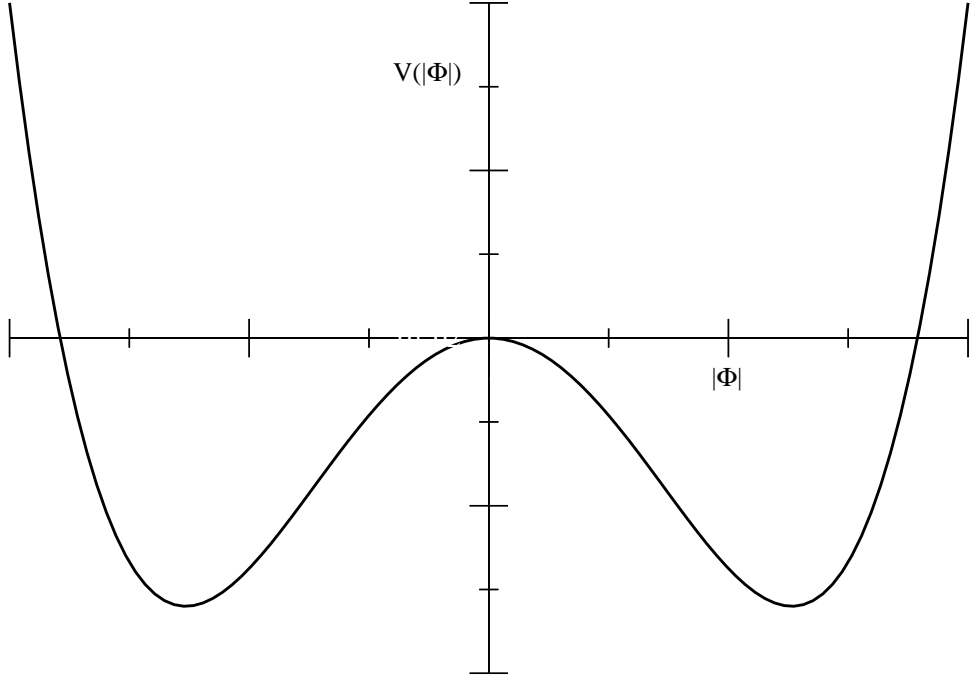}
\caption{A schematic plot of the Higgs potential.}
\label{fig:vacuumpotential}
\end{center}
\end{figure}

Consider the bosonic sector of the electroweak theory. The Lagrangian describing the interactions is
\begin{equation}
\mathcal{L}_\mathrm{bos.} = -\frac{1}{4}G_{\mu\nu}^a G^{\mu\nu,a} - \frac{1}{4}F_{\mu\nu} F^{\mu\nu} + D_\mu \Phi^\dagger D^\mu \Phi + \nu^2\Phi^\dagger\Phi - \lambda\left(\Phi^\dagger\Phi\right)^2,
\end{equation}
where $\Phi$ is the scalar doublet in the fundamental representation of SU($2$), $D_\mu \Phi = \partial_\mu \Phi - ig A_\mu^a \tau^a \Phi / 2  - Yig' B_\mu \Phi$ with $\tau^a$ being the Pauli spin matrices and $Y=1/2$ is the hypercharge of the scalar doublet. The field strength tensors are given by $G_{\mu\nu}^a = \partial_\mu A_\nu^a - \partial_\nu A_\mu^a + g\epsilon^{abc}A_\mu^b A_\nu^c$ and $F_{\mu\nu} = \partial_\mu B_\nu - \partial_\nu B_\mu$ where $A_\mu^a$ and $B_\mu$ are the gauge bosons of weak and hypercharge interactions, respectively, and $\epsilon^{abc}$ are the generators of the adjoint representation of SU($2$). The symmetric vacuum $\Phi = 0$ is unstable due to the sign of the mass term for the scalar doublet (see Fig.~\ref{fig:vacuumpotential}). At tree level, a stable solution to the equations of motion satisfies instead $\Phi_0^\dagger\Phi_0 = \nu^2/(2\lambda)$ and is commonly chosen to be
\begin{equation}
\Phi_0 = \frac{1}{\sqrt{2}}\left(\begin{array}{c} 0 \\ v \end{array}\right),
\end{equation}
where $v^2 = \nu^2/\lambda$. We should consider fluctuations around this state. The vacuum is no longer invariant under the full gauge group, thus spontaneously breaking the symmetry. A residual U($1$) symmetry, generated by $Q = \tau^3/2 + Y$, remains exact. The mass eigenstates and the masses can be readily worked out
\begin{equation}
\begin{array}{rclrcl}
W_\mu^\pm & = & {\displaystyle \frac{1}{\sqrt{2}}\left(A_\mu^1 \mp iA_\mu^2\right)}, & m_W^2 & = & {\displaystyle \frac{1}{4}g^2 v^2}, \\
Z_\mu^0 & = & {\displaystyle \frac{1}{\sqrt{g^2+g'^2}}\left(gA_\mu^3 - g'B_\mu\right)}, & m_Z^2 & = & {\displaystyle \frac{1}{4}\left(g^2+g'^2\right)v^2}, \\
A_\mu & = & {\displaystyle \frac{1}{\sqrt{g^2+g'^2}}\left(g'A_\mu^3 + gB_\mu\right)}, & m_\gamma^2 & = & 0.
\end{array}
\end{equation}
The photon $A_\mu$, corresponding to the residual U($1$) symmetry, remains massless as it should. The electric coupling $e$ is related to the coupling constants $g$ and $g'$ by $e = gg'/\sqrt{g^2+g'^2}$.

Although it is experimentally verified that the electroweak gauge bosons are massive, full confirmation of the Higgs mechanism is still lacking since the Higgs boson has not been found. Today, the direct experimental lower limit for the Higgs mass is $m_H \gtrsim 114$~GeV \cite{Eidelman:2004wy}. In spite of this, we will assume in this thesis that the Higgs mechanism of the minimal standard model is valid.

\section{Fermionic sector of the theory}

Unique feature of electroweak interactions is that the different chiral projections of fermions transform differently under the gauge group. The left handed components form doublets under SU($2$) transformations while the right handed components are SU($2$) singlets. This leads to parity violation in electroweak processes. The fermions are classified into families as
\begin{equation}
\begin{array}{lclclcl}
\mathrm{leptons} & \left(\begin{array}{c} \nu_e \\ e \end{array}\right)_L, & e_R\; ; & \left(\begin{array}{c} \nu_\mu \\ \mu \end{array}\right)_L, & \mu_R\; ; & \left(\begin{array}{c} \nu_\tau \\ \tau \end{array}\right)_L, & \tau_R, \vspace{0.3cm} \\
\mathrm{quarks} & \left(\begin{array}{c} u \\ d \end{array}\right)_L, & u_R, d_R\; ; & \left(\begin{array}{c} c \\ s \end{array}\right)_L, & c_R, s_R\; ; & \left(\begin{array}{c} t \\ b \end{array}\right)_L, & t_R, b_R,
\end{array} \nonumber
\end{equation} 
where the subscripts $L/R$ refer to left/right handed components, defined by 
\begin{equation}
\begin{array}{rclrcl}
\psi_L & = & {\displaystyle \frac{1}{2}\left(1-\gamma_5\right)\psi}, & \psi_R & = & {\displaystyle \frac{1}{2}\left(1+\gamma_5\right)\psi}.
\end{array}
\end{equation}
Note that there are no right handed neutrinos in the minimal standard model. Denoting by $l_L$ and $e_R$ the left handed doublets and right handed singlets of leptons, respectively, and by $q_L$, $u_R$ and $d_R$ the left handed quark doublets and right handed $u$ and $d$ -type quark singlets, respectively, the coupling of fermions to the gauge fields is governed by
\begin{eqnarray}
\mathcal{L}_\mathrm{ferm.} & = & \sum_\psi\bar{\psi}iD\!\!\!\!/\;\psi, \quad \psi \; \in \; \{l_L,e_R,q_L,u_R,d_R\}, \nonumber \\
\quad D_\mu & = & \partial_\mu - IigA_\mu^a\tau^a - Yig'B_\mu,
\end{eqnarray}
where $I$ and $Y$ are the weak isospin and the hypercharge of the relevant fermion, respectively. The values of those are given in table \ref{tab:charges}.

\begin{table}[tb]
\begin{center}
\begin{tabular}{|c|l|l|l|}
\hline
 & $I$ & $Y$ & $Q = T^3 + Y$ \\ \hline
$l_L$ & $1/2$ & $-1/2$ & $0,-1$ \\ \hline
$e_R$ & $0$ &$ -1$ & $-1$ \\ \hline
$q_L$ & $1/2$ & $1/6$ & $2/3,-1/3$ \\ \hline
$u_R$ & $0$ & $2/3$ & $2/3$ \\ \hline
$d_R$ & $0$ & $-1/3$ & $-1/3$  \\ \hline
\end{tabular}
\caption{Values of the various charges for fermions. Here $T^3$ stands for the third component of weak isospin. For doublets, the electric charge is given for each member of the doublet.}
\label{tab:charges}
\end{center}
\end{table}

Due to the chiral transformation properties of fermions, introducing masses for them using standard mass terms $m\bar{\psi}\psi$ is not possible since such terms would violate gauge invariance. Instead, masses can be generated for them by coupling them to the Higgs field through Yukawa couplings. Since all the fermions except for the top quark are so light that their masses can be neglected for our purposes, we only need to take into account the top Yukawa coupling
\begin{eqnarray}
\mathcal{L}_\mathrm{top\;mass} & = & ig_Y\overline{q}_L\tau^2\Phi^\ast t_R + \mathrm{h.c.}
\end{eqnarray}
In the symmetry breaking the mass $m_t = g_Y v/\sqrt{2}$ is generated for the top quark. 

Since the theory contains quarks, we have to take the strong interactions between them into account. These are governed by QCD,
\begin{eqnarray}
\mathcal{L}_\mathrm{QCD} & = & -\frac{1}{4}W^a_{\mu\nu}W^{\mu\nu,a} - ig_s\sum_\mathrm{quarks} \bar{q}i\gamma^\mu C_\mu^a T^a q,
\end{eqnarray}
where $q$ are the quarks, $W^a_{\mu\nu} = \partial_\mu C^a_\nu - \partial_\nu C^a_\mu + g_s f^{abc}C_\mu^b C_\nu^c$ with $C_\mu^a$ being the gluons, $T^a$ and $f^{abc}$ the generators of the fundamental and adjoint representations of $\mathrm{SU}(3)_C$, respectively and $g_s$ is the strong coupling constant. Note that the free propagation of quarks, $\bar{q}i\partial\!\!\!/q$, is already taken into account in $\mathcal{L}_\mathrm{ferm.}$.

\subsection{Conservation of fermion numbers}

At classical level, the baryon number and the three lepton numbers are conserved independently.\footnote{The baryon numbers for each family are not separately conserved since for quarks, the weak eigenstates are not equal to the mass eigenstates. If we considered a model where neutrinos are massive, then similar mixing would be possible also in the leptonic sector and thus there would just be one conserved lepton number.} Instead, in the quantum theory, due to the chiral couplings of fermions to the weak gauge bosons, there is an anomaly \cite{Adler:1969gk,Bell:1969ts,Bardeen:1969md} and the conservation of baryon and lepton numbers is broken,\footnote{Even though the baryon and lepton number currents are not conserved, the charge currents are and hence there is no anomalous breaking of gauge invariance, which would spoil renormalizability of the theory.}
\begin{eqnarray}
\partial_\mu j^\mu_i & = & \frac{g^2}{32\pi^2}\epsilon_{\mu\nu\rho\sigma}\;\mathrm{Tr}\; G^{\mu\nu}G^{\rho\sigma}, \label{eq:non_consv}
\end{eqnarray}
where the $j^\mu_i$ correspond to the baryon and the lepton number currents. The baryon and lepton number violations are, however, related so that $(\Delta B)/3 = \Delta L_e = \Delta L_\mu = \Delta L_\tau$. Consequently, the three linear combinations
\begin{eqnarray}
X_i = \frac{1}{3}B - L_i,\quad i=e,\mu,\tau
\end{eqnarray}
are exactly conserved, while the remaining combination, $B + L$, is not. Here $B$ stands for the baryon number, $L_i$ for the lepton numbers in each family and $L = \sum_i L_i$. From now on, we will refer to the non-conserved $B+L$ number as ``baryon number'' and to the conserved $B/3 - L_i$ numbers as ``lepton numbers''.

From equation (\ref{eq:non_consv}) we see that in order for $B+L$ to change, the gauge fields must evolve so that (for a review on processes governing the baryon/lepton number non-conservation, see \cite{Rubakov:1996vz})
\begin{eqnarray}
\Delta (B+L) & = & \frac{g^2}{32\pi^2}\int_{t_0}^{t_1}\mathrm{d}t\int\mathrm{d}^3\mathbf{x}\,\epsilon_{\mu\nu\rho\sigma}\;\mathrm{Tr}\; G^{\mu\nu}G^{\rho\sigma} \neq 0.
\end{eqnarray}
In general, this requires very strong fields and thus such processes are not present under normal conditions. The non-conservation of $B+L$ is, however, relevant when one considers the vacuum structure of the theory. The vacuum has a non-trivial topological structure, there being vacuum configurations of gauge fields (that is, pure gauge configurations) that cannot be continuously deformed into each other while keeping the system in vacuum. These configurations are classified into a discrete set, characterized by an integer
\begin{eqnarray}
n(A_i) & = & \frac{g^2}{24\pi^2}\int\mathrm{d}^3\mathbf{x}\,\epsilon^{ijk}\mathrm{Tr}\left(A_iA_jA_k\right),
\end{eqnarray}
where the temporal gauge $A_0 = 0$ is assumed. It is now straightforward to note that in transitions between topologically distinct vacua, $B+L$ changes as $\Delta (B+L) = \Delta n$.

Under normal conditions, the different vacuum configurations are separated by a barrier. Processes taking the system from one vacuum to another are tunneling events and are described by instantons. However, these processes are highly improbable. At high temperatures the situation is different since thermal fluctuations can carry the system from vacuum to vacuum. Moreover, the energy of the saddle point configuration of the barrier (called sphaleron) is proportional to the $W$ mass and thus vanishes when the symmetry of the theory is restored. The baryon number violating processes are then unsuppressed. This can have severe consequences on the baryon number asymmetry in the universe and was one of the main motivations to study the electroweak phase diagram to high accuracy, as discussed in the introduction.

\section{Renormalization and notation}
\label{sec:notation}

In this thesis we will use dimensional regularization to regulate all the divergent integrals, \emph{i.e} all the integrals are evaluated in $d=4-2\epsilon$ dimensions. Furthermore, we will employ $\overline{\mathrm{MS}}$ renormalization scheme which amounts to writing the momentum integrals as
\begin{eqnarray}
\int\frac{\mathrm{d}^dp}{(2\pi)^d} & = & \mu^{-2\epsilon}\left[\Lambda^{2\epsilon}\left(\frac{\mathrm{e}^\gamma}{4\pi}\right)^\epsilon\int\frac{\mathrm{d}^dp}{(2\pi)^d}\right],
\end{eqnarray}
where $\Lambda = \mu\left(\mathrm{e}^\gamma/4\pi\right)^{-1/2}$. All the couplings and observables all implicitly scaled to their $4$d dimension by $\mu$, so for example for pressure $p(T) = \mu^{2\epsilon}\hat{p}(T)$ where $[\,\hat{p}(T)\,] = 4-2\epsilon$. 

The parameters of the theory will run with the renormalization scale. This is governed by
\begin{eqnarray}
\nu^2(\Lambda) & = & \nu^2(\mu) + \frac{1}{8\pi^2}\left(-\frac{9}{4}g^2 - \frac{3}{4}g'^2 + 3g_Y^2 + 6\lambda\right)\nu^2\ln\frac{\Lambda}{\mu}, \\
\lambda(\Lambda) & = & \lambda(\mu) + \frac{1}{8\pi^2}\left(\frac{9}{16}g^4 + \frac{3}{16}g'^4 + \frac{3}{8}g^2g'^2 - \frac{9}{2}g^2\lambda  \right.\nonumber \\
& & \hspace{2.5cm} \left.- \frac{3}{2}g'^2\lambda + 12\lambda^2 - 3g_Y^4 + 6g_Y^2\lambda\right)\ln\frac{\Lambda}{\mu}, \\
g_Y^2(\Lambda) & = & g_Y^2(\mu) + \frac{1}{8\pi^2}\left(\frac{9}{2}g_Y^2 - 8g_s^2 - \frac{9}{4}g^2 - \frac{17}{12}g'^2\right)g_Y^2\ln\frac{\Lambda}{\mu}, \\
g^2(\Lambda) & = & g^2(\mu) - \frac{19}{48\pi^2}g^4\ln\frac{\Lambda}{\mu}, \\
g'^2(\Lambda) & = & g'^2(\mu) + \frac{41}{48\pi^2}g'^4\ln\frac{\Lambda}{\mu}, \\
g_s^2(\Lambda) & = & g_s^2(\mu) - \frac{7}{8\pi^2}g_s^4\ln\frac{\Lambda}{\mu}.
\end{eqnarray}
Values of the parameters are fixed so that at $\mu = m_Z$ we have
\begin{equation}
\begin{array}{rclrcl}
\nu^2(m_Z) & = & {\displaystyle \frac{1}{2}m_H^2}, & \lambda(m_Z) & = & {\displaystyle \frac{1}{\sqrt{2}}G_\mu m_H^2}, \\
g_Y^2(m_Z) & = & 2\sqrt{2}G_\mu m_t^2, & g^2(m_Z) & = & 4\sqrt{2}G_\mu m_W^2, \\
g'^2(m_Z) & = & 4\sqrt{2}G_\mu\left(m_Z^2-m_W^2\right), & \alpha_s(m_Z) & = & {\displaystyle \frac{g_s^2(m_Z)}{4\pi}} = 0.1187,
\end{array}
\end{equation}
where $m_H$ is the unknown mass of the Higgs boson, $m_W = 80.43$~GeV, $m_Z = 91.19$~GeV and $m_t = 174.3$~GeV are the masses of the $W$ and $Z$ bosons and the top quark, and $G_\mu = 1.664\cdot~\!\!10^{-5}$~$\mathrm{GeV}^{-2}$ is the Fermi coupling constant \cite{Eidelman:2004wy}. The matching of the values of the parameters to the values of the physical observables is given here at tree level for simplicity. However, when precise numerical results are wanted (when determining the critical temperature and the location of the endpoint of the first order phase transition line in chapter 5), the one loop relations given in \cite{Kajantie:1996dw} are used.  We employ a power counting rule $\lambda \sim g'^2 \sim g_s^2 \sim g_Y^2 \sim g^2$ (unless otherwise stated) and assume the temperature to always be so high that the relation $\nu^2 \lesssim g^2T^2$ applies. This power counting convention is then used to bring all the order of magnitude estimates to simple powers of $g^2$.

As will be discussed in the following chapter, we will use the imaginary time formalism to study finite temperature field theory. This in effect means that the spacetime will be Euclidean. We define the gamma matrices there so that
\begin{equation}
\begin{array}{rclrcl}
\{\gamma_\mu,\gamma_\nu\} & = & 2\delta_{\mu\nu}, & \{\gamma_\mu,\gamma_5\} & = & 0
\end{array}
\end{equation}
and $\mathrm{Tr}\;\gamma_5\gamma_\mu\gamma_\nu\gamma_\rho\gamma_\sigma \propto \epsilon_{\mu\nu\rho\sigma} + \mathcal{O}(\epsilon)$, where $\delta_{\mu\nu}$ is the Kronecker delta symbol and $\epsilon_{\mu\nu\rho\sigma}$ is the totally antisymmetric tensor with $\epsilon_{0123} = 1$.

\chapter{Field theories at finite temperature}
\label{cha:ft}

In this chapter we will briefly review the basics of finite temperature field theory. For a more detailed analysis, see for example \cite{Kapusta:1989tk}. We use here a notation such that $\varphi$ refers to the collection of all the fields of a theory, but sometimes it is essential to make a distinction between bosons and fermions and we then use a notation such that $\phi$ refers to all the bosons and $\psi$ to all the fermions.

In the grand canonical ensemble, the equilibrium properties of any system at finite temperature are given by the density matrix,
\begin{equation}
\rho = \frac{1}{Z}\;\textrm{exp}\left[-\beta\left(H-\sum_i\mu_i N_i\right)\right],
\end{equation}
where $H$ is the Hamiltonian describing the dynamics of the system, $N_i$ are all the conserved charges with $\mu_i$ being the corresponding chemical potentials and $\beta = 1/T$. In order to lighten the notation, we will from now on write $\sum_i\mu_i N_i \equiv \mu N$. Expectation values of physical variables are given as traces over the density matrix, $\langle \mathcal{O}\rangle = \textrm{Tr}\;\rho\mathcal{O}$. 

In practice, however, it is more convenient to operate with the partition function $Z(T,\mu_i,V)$:
\begin{eqnarray}
\textrm{Tr}\;\rho = 1 \quad \rightarrow \quad Z(T,\mu_i,V) & = & \textrm{Tr}\;\textrm{exp}\left[-\beta\left(H-\mu N\right)\right] \nonumber \\
& = & \sum_{\varphi_n}\langle \varphi_n|\textrm{exp}\left[-\beta\left(H-\mu N\right)\right]|\varphi_n\rangle.
\end{eqnarray}
Here $\varphi_n$ are the eigenstates of the operator $H-\mu N$. Physical variables such as pressure $p$, energy $E$, entropy $S$ and number of particles $N_i$ are given by the standard thermodynamic relations. Defining thermodynamic potential (free energy) $F$ by $F(T,\mu_i,V) = -T\ln Z(T,\mu_i,V)$, we have
\begin{eqnarray}
p & = & -\frac{\partial F}{\partial V}, \quad S \;\; = \;\; -\frac{\partial F}{\partial T}, \nonumber \\
N_i & = & -\frac{\partial F}{\partial \mu_i}, \quad E \;\; = \;\; F + TS + \sum_i \mu_i N_i. \label{eq:basic_td_eqs}
\end{eqnarray} 

The partition function is most conveniently evaluated as a path integral. The matrix element $\langle\varphi_n|\textrm{exp}[-\beta(H-\mu N)]|\varphi_n\rangle$ can be regarded as a transition amplitude in imaginary time $\tau = it$, carrying states from $\tau = 0$ to $\tau = \beta$, when the ``time evolution'' is governed by the operator $H-\mu N$. Such a matrix element can be written as a path integral
\begin{eqnarray}
& & \langle\varphi_2|\textrm{exp}\left[-\beta\left(H-\mu N\right)\right]|\varphi_1\rangle \\ & & \nonumber \\
& = & \int_{\varphi(\mathbf{x},0)=\varphi_1}^{\varphi(\mathbf{x},\beta)=\varphi_2}\mathcal{D}\varphi\mathcal{D}\pi\;\textrm{exp}\left[\int_0^\beta\textrm{d}\tau\int\textrm{d}^3\mathbf{x}\left(i\pi(\mathbf{x},\tau)\dot{\varphi}(\mathbf{x},\tau) - \mathcal{H}(\pi,\varphi) + \mu\mathcal{N}(\pi,\varphi)\right)\right], \nonumber
\end{eqnarray}
where $\pi(\mathbf{x},\tau)$ are the canonically conjugate fields of $\varphi(\mathbf{x},\tau)$ and $\mathcal{H}$ and $\mathcal{N}$ are the Hamiltonian and particle number densities, respectively. Since in all the cases of interest to us, $\mathcal{H}-\mu\mathcal{N}$ is at most quadratic in $\pi$, the integration over the momentum fields is Gaussian and can be performed immediately. The result is
\begin{equation} Z(T,\mu,V) = \int\mathcal{D}\varphi\;\textrm{exp}\left(\int_0^\beta\textrm{d}\tau\int\textrm{d}^3\mathbf{x}\;\mathcal{L}'(\varphi,\dot{\varphi})\right), \label{eq:path_integral}
\end{equation}
where the integration is over all the fields satisfying (anti-)periodic boundary conditions in $\tau$ as discussed below. The Lagrangian $\mathcal{L}'(\varphi,\dot{\varphi})$ above is calculated from $\mathcal{H}-\mu\mathcal{N}$ and thus differs from the Lagrangian defining the vacuum theory when the chemical potentials are non-zero. More precisely, given $\mathcal{H}(\pi,\varphi)-\mu\mathcal{N}(\pi,\varphi)$, the associated Lagrangian will be
\begin{equation}
\mathcal{L}'(\varphi,\dot{\varphi}) = i\pi(\varphi,\dot{\varphi})\dot{\varphi} - \mathcal{H}\left(\pi(\varphi,\dot{\varphi}), \varphi\right) + \mu\mathcal{N}\left(\pi(\varphi,\dot{\varphi}),\varphi\right),
\end{equation}
where $\pi(\dot{\varphi},\varphi)$ is solved from
\begin{equation}
i\dot{\varphi} = \frac{\partial\left(\mathcal{H}-\mu\mathcal{N}\right)}{\partial\pi}.
\end{equation}

Due to the trace in the definition of the partition function, the fields in the path integral must satisfy appropriate boundary conditions in the compact imaginary time. It follows that bosonic fields must be periodic, $\phi(\mathbf{x},0) = \phi(\mathbf{x},\beta)$, while fermionic fields, being Grassmann variables, must be anti-periodic, $\psi(\mathbf{x},0) = -\psi(\mathbf{x},\beta)$. Hence it is convenient to expand the fields in Fourier series in imaginary time. From the boundary conditions above it follows that
\begin{equation}
\varphi(\mathbf{x},\tau) = T\sum_{n=-\infty}^\infty\;\textrm{e}^{i\omega_n\tau}\varphi_n(\mathbf{x}),
\end{equation}
where the so called Matsubara frequencies are given by
\begin{equation}
\omega_n = \left\{ \begin{array}{cl} 2n\pi T & \textrm{for bosons}, \\ & \\ (2n+1)\pi T & \textrm{for fermions}. \end{array} \right.
\end{equation}
With this Fourier expansion it then becomes possible to reinterpret the finite temperature theory as a three dimensional field theory with infinitely many fields $\varphi_n(\mathbf{x}),\;n\in \mathbb{N}$.

The gauge fields, containing non-physical degrees of freedom, must be treated with care (for a review, see \cite{Gross:1981br}). The gauge must be fixed in order to ensure that the integration is only over the configurations that are not gauge equivalent with respect to each other. Temporal gauge $A_0^a = 0$ is a typical choice (though it is not a sufficient condition). Also, since the Gauss law is not included in the Hamiltonian equations of motion, it should be considered as a constraint on $\pi_i^a$ (the canonically conjugate fields of $A_i^a$). In the absence of other fields, the Gauss law states that $D_i \pi_i^a = 0$. This constraint can be taken into account by means of Lagrange multiplier fields, for which we can impose periodic boundary conditions. In the end we can identify the Lagrange multipliers as the temporal components of the gauge fields and after integrating over the momentum fields, the resulting Lagrangian will have the conventional covariant form of a gauge field theory. The remaining gauge freedom, due to invariance under gauge transformations that are periodic in the imaginary time interval $\tau \in [0,\beta]$, can be removed by the Faddeev-Popov procedure, which introduces ghost fields to the theory. Although being Grassmann variables, they obey periodic boundary conditions since they are related to gauge transformations that are similarly periodic.

As a concrete example, consider an SU($2$) non-Abelian gauge field theory with a fermion and a scalar field transforming in the fundamental representation of the gauge group. The Lagrangian is given by (in Minkowski spacetime)
\begin{equation}
\mathcal{L} = D_\mu\Phi^\dagger D^\mu\Phi - \nu^2\Phi^\dagger\Phi - \frac{1}{4}G_{\mu\nu}^a G^{\mu\nu,a} + \bar{\psi}i D\!\!\!\!/\psi
\label{eq:exlagr}
\end{equation}
and the theory contains two linearly independent, mutually commuting conserved charges, the baryon number and the third component of isospin,
\begin{eqnarray}
B & = & \int\textrm{d}^3\mathbf{x}\; \bar{\psi}\gamma_0\psi, \nonumber \\
Q_3 & = & \int\textrm{d}^3\mathbf{x}\left[\frac{1}{2}\bar{\psi}\gamma_0 \tau^3\psi - \frac{i}{2}\left(\left(D_0\Phi\right)^\dagger \tau^3\Phi - \Phi^\dagger \tau^3 D_0\Phi\right) - \epsilon^{3bc}A^{\nu,b}G^c_{\mu\nu}\right].
\end{eqnarray}
Introducing chemical potentials $\mu_B$ and $\mu_Q$ for these charges and integrating over the canonically conjugate momentum fields we get for the partition function
\begin{equation}
\mathcal{Z} = \int\mathcal{D}\varphi\; \textrm{exp}\left[\int_0^\beta\textrm{d}\tau\int\textrm{d}^3\mathbf{x}\;\left(\mathcal{L}_E + \mu_B \overline{\psi}\gamma_0\psi\right)\right]
\end{equation}
where $\mathcal{L}_E$ is the Euclidean Lagrangian obtained from the original Lagrangian in Eq.~(\ref{eq:exlagr}) by going to imaginary time and by making a replacement \cite{Kapusta:1981aa,Kapusta:1990qc,Khlebnikov:1996vj}
\begin{equation}
A_0^3 \rightarrow A_0^3 - \frac{i\mu_Q}{g}.
\end{equation}
Gauge fixing and ghost terms are suppressed in the expression. We thus see that the chemical potentials related to the conserved gauge charges can be interpreted as background fields for the temporal components of the gauge fields. 

Although the relation between the temporal components of the gauge fields and the chemical potentials related to the corresponding gauge charges can be obtained by a straightforward calculation, there is a simple argument that suggests it as well. Heuristically, consider a system which contains an external charge density $q=Q/V=eN/V$ where $e$ is the elementary charge and $N$ is the number of external charged particles. This can be taken into account in the theory by adding a corresponding source term to the action:
\begin{eqnarray}
\mathcal{Z}(T,q) & = & \int\mathcal{D}\varphi\; \mathrm{exp}\left[\int_0^\beta\textrm{d}\tau\int\textrm{d}^3\mathbf{x}\;\left(\mathcal{L}_\mathrm{E} + iqA_0\right)\right].
\end{eqnarray}
The related chemical potential $\mu_Q$ will then be the conjugate variable to $N$, given by:
\begin{eqnarray}
\mu_Q & = & -\frac{T}{\mathcal{Z}(T,q)}\frac{\partial}{\partial N}\mathcal{Z}(T,q) \\
 & = & -\frac{ie}{\mathcal{Z}(T,q)}\int\mathcal{D}\varphi\,\left(\frac{T}{V}\int_0^\beta\!\mathrm{d}\tau\!\int\!\mathrm{d}^3\mathbf{x}\, A_0\right)\;\mathrm{exp}\left[\int_0^\beta\!\textrm{d}\tau\!\int\!\textrm{d}^3\mathbf{x}\left(\mathcal{L}_\mathrm{E} + iqA_0\right)\right]  = -ie\langle A_0\rangle, \nonumber
\end{eqnarray}
suggesting that these chemical potentials can be associated with the expectation values of the temporal components of the gauge fields \cite{Khlebnikov:1996vj}.

A condition for thermal equilibrium is that the free energy is stationary with respect to fluctuations around the expectation value of the gauge field $\langle A_0\rangle$. Thus, thermal equilibrium can be achieved only when the system is neutral with respect to gauge charges:
\begin{equation}
0 = \frac{\partial F}{\partial\langle A_0\rangle} \sim \frac{\partial F}{\partial \mu_Q} \sim Q.
\end{equation}
While the chemical potential for the baryon number (in general for any charge whose conservation is guaranteed by a global symmetry) can be chosen freely, we see that this is not the case for the chemical potentials for the gauge charges. Their values are determined by requiring the system to be neutral with respect to them. In general, their values will then depend on the chemical potential for the baryon number. Note that in the example above, the baryon number does not carry any gauge charge (\emph{i.e.} for any $\mu_B$, the number of baryons with third component of isospin being $+1/2$ is equal to those with $-1/2$) and thus the system is neutral when $\mu_Q = 0$. This is not the case in general. In the electroweak theory, global charges carry a non-zero (hyper)electric charge and the chemical potential for the (hyper)electric charge will then be non-zero.

\section{Dimensional reduction}
\label{sec:dimred}

Perturbative evaluation of the path integral in Eq.~(\ref{eq:path_integral}) at high temperatures is unreliable due to infrared divergences which arise when integrating over the bosonic zero modes $\phi_0(\mathbf{x})$ \cite{Linde:1980ts,Gross:1981br}. In loop expansion these modes behave as massless particles in three dimensions which makes the infrared behavior of the integrals worse. On the other hand, integration over the non-static (\emph{i.e.} $\omega_n \neq 0$) modes is infrared safe, since temperature, providing these modes a mass, acts as an effective infrared regulator. Since there is a clear scale hierarchy between the static and non-static modes, it is natural to consider developing an effective field theory describing the static modes by perturbatively integrating out all the other modes. This is called dimensional reduction \cite{Ginsparg:1980ef,Appelquist:1981vg}.

The dimensional reduction has many advantages. It essentially divides the problem of computing the path integral into two stages: 1. calculating the contribution from the non-static modes to the variable in question and constructing the effective theory, both tasks that can be carried out perturbatively, and 2. solving the effective theory, which one usually has to do with numerical computations. However, typically the effective theory is much better suited for numerical computations than the original four dimensional theory since it does not contain any fermions which are problematic to implement on a lattice. Also, since all the heavy scales have already been integrated out perturbatively, the lattice spacing can be taken to be larger (in physical units) which leads to larger lattice sizes and thus to a more precise treatment of the soft scales.



The effective theory is constructed to be the most general theory describing the relevant degrees of freedom which has the required symmetries. In general, any non-renormalizable coupling will be suppressed by powers of the large scale that was integrated out, in this case $\pi T$ and thus it normally is sufficient to consider just the renormalizable effective theories. The parameters of the effective theory are mapped to those of the full theory by requiring that the effective theory reproduces all the static, bosonic Green's functions to the desired accuracy. 

For a generic gauge field theory, the effective theory resulting from integrating out the non-static modes will be a three dimensional gauge field theory with fundamental (Higgs) and adjoint (temporal components of the gauge fields) scalars. The adjoint scalars correspond to the electrostatic modes of the original gauge field theory, $E_i \sim F_{0i} \sim \partial_i A_0$, and they obtain a thermal mass $m_D^2 \sim g^2 T^2$ in this effective theory due to Debye screening of electric fields. Consequently, this theory is often called the electrostatic effective theory. Also the fundamental scalar mass $m_3$ acquires temperature dependent corrections. In theories where the fundamental scalar is responsible for a spontaneous symmetry breaking, such as the Higgs scalar in the standard model, these corrections are the cause of the symmetry restoration at high temperatures, $m_3^2 \sim -\nu^2 + g^2 T^2 > 0$ when $g^2T^2\gtrsim \nu^2$. The three dimensional vector gauge boson, $A_i$, corresponding to the magnetic sector of the original theory ($B_i \sim \epsilon_{ijk}F_{jk} \sim \epsilon_{ijk}\partial_i A_k$) remains, however, massless, protected by gauge invariance. There nevertheless is a mass scale associated with it since in three dimensions the coupling becomes dimensionful, $g_3^2 \sim g^2T$. 

\begin{figure}[tb]
\begin{center}
\includegraphics[width=0.87\textwidth]{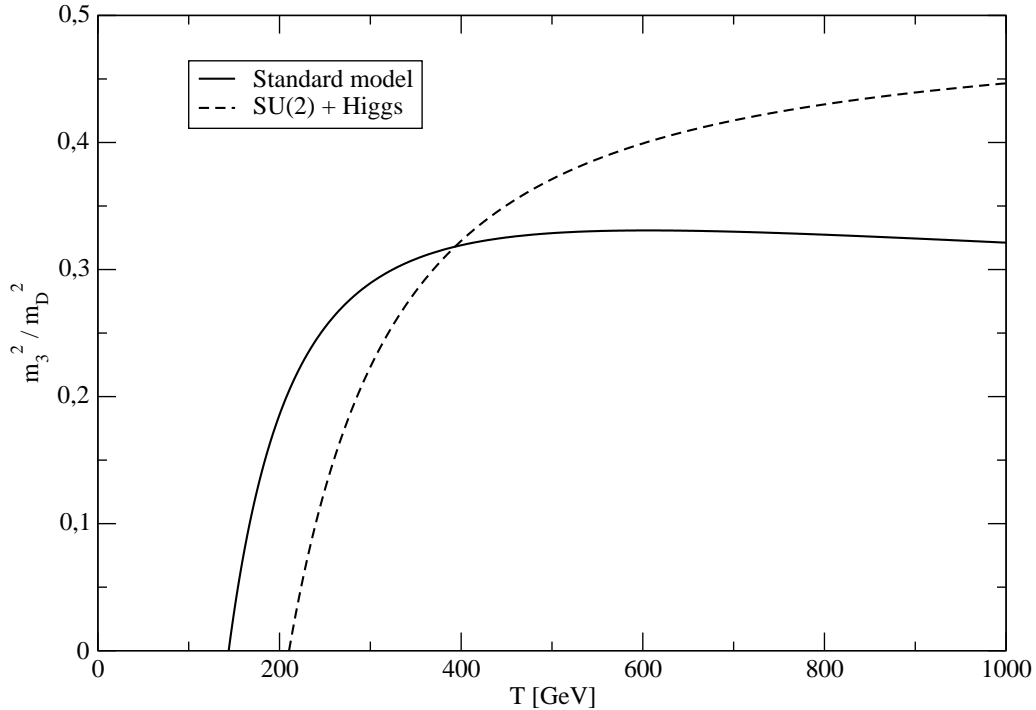}
\caption{The ratio of the two mass scales in the electrostatic effective theory for an SU($2$) + fundamental Higgs theory and for the standard model. The precise expressions for the masses are given in chapter \ref{cha:pressure} and in appendix \ref{app:presEW}.}
\label{fig:massratio}
\end{center}
\end{figure}

At very high temperatures the adjoint and fundamental scalars are parametrically equally heavy, $m_3\sim m_D\sim gT$, and thus both of them should be considered on an equal footing. However, if the fundamental scalar is responsible for a spontaneous symmetry breaking, then as the temperature is lowered, it becomes increasingly light compared with the adjoint scalar, as shown in Fig.~\ref{fig:massratio}. This indicates the onset of the phase transition in which the symmetry gets spontaneously broken. Near the phase transition point we then have again a mass hierarchy with the adjoint scalars being heavy and it is natural to integrate them out, leaving an effective theory containing a fundamental scalar and three dimensional gauge fields. This theory can then be used to compute the characteristics of the phase transition.

Even though the infrared problems related to screening of the electrostatic modes are tamed by the Debye mass induced for the electric modes, also the electrostatic effective theory suffers from infrared problems in perturbation theory. These are related to screening of the magnetic fields. This can be studied by integrating out the massive adjoint and fundamental scalars to obtain an effective theory describing the magnetostatic degrees of freedom, namely the three dimensional gauge bosons. The resulting theory is confining and it cannot be solved with perturbative calculations. This places a limit on how far the perturbative expansion of different variables can be extended.

We will consider the effect of introducing finite chemical potentials on the dimensional reduction in more detail in section \ref{sec:dimregatmu}, but let us briefly study some generic features of it. Chemical potentials for fermion numbers will change the form of the fermion propagators and thus the matching of fields and parameters of the different theories changes. This, however, will not change the structure of the effective theory. On the other hand, some of the symmetries of the four dimensional theory are explicitly broken by the finite chemical potentials, namely charge conjugation C, and this leads to new terms in the effective theory. More precisely, finite chemical potentials break C but preserve parity P and time reversal symmetry T and thus the theory is, in addition to being C breaking, also CP and CPT breaking (note that there is no controversy here since finite chemical potentials break Lorentz invariance). Depending on the field content of the system, the effective theory will then have a number of new terms.


\chapter{Pressure of the standard model}
\label{cha:pressure}

Pressure of a high temperature field theory is an important quantity both theoretically and phenomenologically. Defined by the path integral
\begin{eqnarray}
p(T) & = & \lim_{V\rightarrow\infty}\frac{T}{V} \ln \int\mathcal{D}A\mathcal{D}\psi\mathcal{D}\bar{\psi}\mathcal{D}\Phi\;\mathrm{exp}(-S),
\end{eqnarray}
it is the most basic quantity that one can compute in a high temperature quantum field theory and is thus of special interest. From a phenomenological point of view, knowing the pressure (and the other thermodynamic functions that can be evaluated when we know the pressure) is important, for example, in cosmology. Expansion of the universe, governed by the Einstein equations, depends on the pressure and energy density of the matter it is composed of. Assuming the universe to be flat (which it according to observations is) and having no cosmological constant (contribution of the present day cosmological constant on the expansion is negligible in the early universe), the evolution of the scale factor $R(t)$ is governed by
\begin{eqnarray}
H^2 = \frac{\dot{R}^2}{R^2} & = & \displaystyle \frac{8\pi G}{3}\epsilon(T), \\
\frac{\ddot{R}}{R} & = & -\frac{4\pi G}{3}\left(\epsilon(T)+3p(T)\right),
\end{eqnarray}
where $\epsilon(T)$ is the energy density. Thus many theoretical predictions in cosmology depend on how well we know the pressure, energy and also entropy densities in the early universe.

As an example, consider cold dark matter. Observations performed with the Wilkinson Microwave Anisotropy Probe (WMAP) suggest that as much as 23\% of the matter in the universe is cold dark matter \cite{Spergel:2003cb}. One possible candidate to explain cold dark matter is weakly interacting massive particles, or WIMPs, that might have been produced in the very early universe (for a recent review, see \cite{Bertone:2004pz}). In order to be able to predict the density of WIMPs in the universe today, we need to know the equation of state, \emph{i.e.} pressure, during the time when they decoupled from the thermal evolution of the universe \cite{Srednicki:1988ce}. It has been argued that differences of the order of 10\% in the equation of state can alter the relic density of WIMPs as much as 1\% \cite{Hindmarsh:2005ix}. It is possible that with the Planck satellite, designed to measure the power spectrum of the microwave background radiation with even higher precision than WMAP, we can measure the density of cold dark matter to this accuracy \cite{Balbi:2003en} and hence it is important to achieve the corresponding precision in theoretical calculations as well.

In this chapter we will review the calculation of the pressure in the electroweak sector of the standard model, assuming temperature to be so high that the system resides in the symmetric phase of the theory. The perturbative expansion is performed to order $g^5$ in the coupling constants, equivalent to three-loop order in loop expansion. In the next section, we analyze the structure of the expansion using a simpler SU($2$) + Higgs theory, after which we consider the numerical results for both the simpler theory as well as for the full electroweak theory.  

\section{Structure of the perturbative expansion of pressure}

For our purposes, full description of the electroweak theory requires five parameters: the gauge couplings $g$ and $g'$, the Higgs mass scale $\nu^2$ and self-coupling $\lambda$ and the top Yukawa coupling $g_Y$. In order to incorporate the QCD effects we have to add the strong coupling constant $g_s$. The perturbative expansion of pressure becomes then a very complex function of all these parameters and to study the structure of the expansion is strenuous. Also, since the Higgs sector of the theory carries only $4$ of the total of $106.75$ degrees of freedom, the contribution of the Higgs to the pressure is not easily visible.

In order to circumvent these problems, we will consider here a simpler theory that however has all the necessary properties, namely an SU($2$) + fundamental Higgs theory defined by the Euclidean Lagrangian
\begin{eqnarray}
\mathcal{L} & = & \frac{1}{4}G_{\mu\nu}^a G_{\mu\nu}^a + D_\mu\Phi^\dagger D_\mu\Phi - \nu^2\Phi^\dagger\Phi + \lambda\left(\Phi^\dagger\Phi\right)^2,
\end{eqnarray}
where $G_{\mu\nu}^a = \partial_\mu A_\nu^a - \partial_\nu A_\mu^a + g\epsilon^{abc}A_\mu^b A_\nu^c$ and $D_\mu\Phi = \partial_\mu\Phi - ig\tau^a A_\mu^a\Phi/2$. Characterized by only three couplings, the expansion for the pressure is easier to approach and the effects of the scalar sector are more evident. The expansion for the full electroweak theory is written down in appendix \ref{app:presEW} and the related numerical results are reviewed in section \ref{sec:numres}.

We will employ the framework of dimensional reduction, described in section \ref{sec:dimred}. The pressure can then be written as (we implicitly assume that the limit $V\rightarrow \infty$ is taken)
\begin{eqnarray}
p(T) & = & p_\mathrm{E}(T) + \frac{T}{V} \ln \int\mathcal{D}A_i\mathcal{D}A_0\mathcal{D}\Phi\;\mathrm{exp}(-S_\mathrm{E}),
\end{eqnarray}
where $p_\mathrm{E}(T)$ is the strict perturbative expansion of the pressure of the $4$d theory and $S_\mathrm{E}$ defines the effective theory describing the $n=0$ bosonic Matsubara modes,
\begin{eqnarray}
S_\mathrm{E} & = & \int\mathrm{d}^3\mathbf{x}\,\left(\frac{1}{4}G_{ij}^a G_{ij}^a + \frac{1}{2}\left(D_i A_0^a\right)^2 + \frac{1}{2}m_D^2 A_0^a A_0^a + \frac{1}{4}\lambda_A\left(A_0^a A_0^a\right)^2\right. \nonumber \\  
& & \left. \hspace{1.5cm}+ D_i\Phi^\dagger D_i\Phi + m_3^2\Phi^\dagger\Phi + \lambda_3\left(\Phi^\dagger\Phi\right)^2 + h_3 A_0^a A_0^a \Phi^\dagger\Phi\right),
\end{eqnarray}
where for the adjoint scalars $A_0^a$ we have $D_i A_0^a = \partial_i A_0^a + g_3 \epsilon^{abc}A_i^b A_0^c$. The matching of the parameters of this effective theory to those of the $4$d theory has been done in detail to the desired accuracy in \cite{Kajantie:1996dw,Gynther:2005dj} and we will just review the results here. 

The contribution of the non-zero Matsubara modes (that is, the scale $2\pi T$) to the pressure is contained in the parameter $p_\mathrm{E}(T)$ and in the couplings and masses of the effective theory $S_\mathrm{E}$. The goal is to evaluate the expansion of the pressure to order $g^5$ which determines the accuracy with which we need to know these parameters. For the couplings of the effective theory it is sufficient to use the tree level matching,
\begin{equation}
\begin{array}{rclrcl}
g_3^2 & = & g^2 T, & \lambda_3 & = & \lambda T, \vspace{0.2cm} \\
h_3 & = & {\displaystyle \frac{1}{4}g^2 T}, & \lambda_A & = & \mathcal{O}(g^4).
\end{array} \label{eq:coupl_match}
\end{equation}
Corrections of the order of $g^4$ to these would contribute to order $g^6$ in the pressure and thus we can neglect them now. The masses are needed to order $g^4$: 
\begin{eqnarray}
m_D^2 & = & T^2\left\{g^2\left[\frac{2}{3}+\frac{n_\mathrm{S}}{6} + \epsilon\left(\frac{4}{3}\ln\frac{\Lambda}{4\pi T} + \frac{4}{3}\frac{\zeta'(-1)}{\zeta(-1)} + \frac{n_\mathrm{S}}{3}\left(\ln\frac{\Lambda}{4\pi T} + \frac{1}{2} + \frac{\zeta'(-1)}{\zeta(-1)}\right)\right)\right]\right. \nonumber \\
& & + \frac{1}{(4\pi)^2}\left[g^4\left(\frac{88}{9}\ln\frac{\Lambda}{4\pi T} + \frac{20}{9} + \frac{88}{9}\gamma + n_\mathrm{S}\left(\frac{13}{6}\ln\frac{\Lambda}{4\pi T} + \frac{47}{72} + \frac{13}{6}\gamma\right)\right) \right. \nonumber \\
& & \left.\left.\hspace{1.6cm} + n_\mathrm{S}g^2\lambda - 2n_\mathrm{S}\frac{\nu^2}{T^2}\right]\right\}, \\
m_3^2(\Lambda) & = & -\nu^2\left[1 + \frac{1}{(4\pi)^2}\left(g^2\left(\frac{9}{2}\ln\frac{\Lambda}{4\pi T} + \frac{9}{2}\gamma\right) - \lambda\left(12\ln\frac{\Lambda}{4\pi T} + 12\gamma\right)\right)\right] \nonumber \\
& & + T^2\left[g^2\left(\frac{3}{16} + \epsilon\left(\frac{3}{8}\ln\frac{\Lambda}{4\pi T} + \frac{1}{4} + \frac{3}{8}\frac{\zeta'(-1)}{\zeta(-1)}\right)\right) \right. \nonumber \\
& & \left.\hspace{0.6cm} + \lambda\left(\frac{1}{2} + \epsilon\left(\ln\frac{\Lambda}{4\pi T} + 1 + \frac{\zeta'(-1)}{\zeta(-1)}\right)\right)\right. \nonumber \\
& & \hspace{0.6cm} + \frac{1}{(4\pi)^2}\left(g^4\left(-\frac{47}{16}\frac{\Lambda}{4\pi T} - \frac{19}{24} - \frac{13}{32}\gamma - \frac{81}{32}\frac{\zeta'(-1)}{\zeta(-1)}\right) \right. \\
& & \left.\left. \hspace{1.7cm} + 6\lambda^2\left(1 - \gamma + \frac{\zeta'(-1)}{\zeta(-1)}\right) + g^2\lambda\left(-\frac{9}{2}\ln\frac{\Lambda}{4\pi T} - \frac{15}{4} - \frac{9}{2}\frac{\zeta'(-1)}{\zeta(-1)}\right)\right) \right]. \nonumber
\end{eqnarray}
Here we have explicitly separated the contribution coming from the Higgs scalar to the Debye mass $m_D$ by keeping the number of fundamental scalars $n_\mathrm{S}$ general. This allows us to keep track of the effects of the Higgs sector. In the end one should set $n_\mathrm{S} = 1$. Order $\mathcal{O}(\epsilon)$ terms are required for the masses since there will be terms of the form $m^2/\epsilon$ in the pressure of the effective theory. Note also that the mass of the adjoint scalar is renormalization group invariant to the order needed while the mass of the fundamental scalar develops a pole at order $g_3^4$ in the effective theory that must be renormalized and therefore $m_3^2$ runs. The mass counterterm, which can be obtained by the matching procedure, is
\begin{eqnarray}
\delta m_3^2 & = & \frac{T^2}{(4\pi)^2\epsilon}\left(-\frac{81}{64}g^4 + 3\lambda^2 - \frac{9}{4}g^2\lambda\right).
\end{eqnarray}

The pressure $p_\mathrm{E}(T)$ is obtained from the $4$d theory by calculating its pressure in a strict perturbative expansion, \emph{i.e.} evaluating the diagrams depicted in Fig.~\ref{fig:pEdiags}. This will lead to infrared divergences at order $g^4$ (at three loops) which are not canceled even after renormalization. They will be canceled only when the pressure of the effective theory is taken into account. Since we assume the temperature to be so high that $\nu^2 \lesssim g^2 T^2$, we can expand the Higgs propagator in powers of $\nu^2$. Such expansion is possible since $p_\mathrm{E}(T)$ counts contributions only from the non-static modes and integration over them is infrared safe. Thus the expansion of $p_\mathrm{E}(T)$ in terms of $\nu^2$ is analytic.

Computing the diagrams in Fig.~\ref{fig:pEdiags} (done in \cite{Gynther:2005dj}), we get for $p_\mathrm{E}(T)$
\begin{eqnarray}
\frac{p_\mathrm{E}(T)}{T^4} & = & \frac{\pi^2}{90}\left(6 + 4n_\mathrm{S}\right) \label{eq:pEpres}\\
& & - g^2\left(\frac{1}{24} + \frac{5}{192}n_\mathrm{S}\right) - \frac{\lambda}{24}n_\mathrm{S} + \frac{1}{6}\frac{\nu^2}{T^2}n_\mathrm{S} \nonumber \\
& & +\frac{1}{(4\pi)^2}\left[g^4\left(\frac{1}{\epsilon} + \frac{97}{18}\ln\frac{\Lambda}{4\pi T} + \frac{29}{15} + \frac{1}{3}\gamma + \frac{55}{9}\frac{\zeta'(-1)}{\zeta(-1)} - \frac{19}{18}\frac{\zeta'(-3)}{\zeta(-3)} \right.\right. \nonumber \\
& & \left.\hspace{1.7cm} + \frac{n_\mathrm{S}}{12}\left(\frac{75}{16}\frac{1}{\epsilon} + \frac{195}{8}\ln\frac{\Lambda}{4\pi T} + \frac{8711}{960} + \frac{99}{32}\gamma + \frac{381}{16}\frac{\zeta'(-1)}{\zeta(-1)} - \frac{81}{32}\frac{\zeta'(-3)}{\zeta(-3)}\right)\right) \nonumber \\
& & \hspace{1.2cm} + \lambda^2 n_\mathrm{S}\left(\ln\frac{\Lambda}{4\pi T} + \frac{31}{60} + \frac{1}{2}\gamma + \frac{\zeta'(-1)}{\zeta(-1)} - \frac{\zeta'(-3)}{\zeta(-3)}\right) \nonumber \\
& & \hspace{1.2cm} + g^2\lambda n_\mathrm{S}\left(\frac{3}{8\epsilon} + \frac{15}{8}\ln\frac{\Lambda}{4\pi T} + \frac{11}{8} + \frac{3}{8}\gamma + \frac{3}{2}\frac{\zeta'(-1)}{\zeta(-1)}\right) \nonumber \\
& & \hspace{1.2cm} + \frac{\nu^2}{T^2}n_\mathrm{S}\left(-g^2\left(\frac{3}{4\epsilon} + \frac{9}{4}\ln\frac{\Lambda}{4\pi T} + \frac{5}{4} + \frac{3}{4}\gamma + \frac{3}{2}\frac{\zeta'(-1)}{\zeta(-1)}\right) - 2\lambda\left(\ln\frac{\Lambda}{4\pi T} + \gamma\right)\right) \nonumber \\
& & \hspace{1.2cm}\left. +\frac{\nu^4}{T^4}n_\mathrm{S}\left(2\ln\frac{\nu}{4\pi T} - \frac{3}{2} + 2\gamma\right) \right] + \mathcal{O}(g^6). \nonumber
\end{eqnarray}
Here we have again explicitly written down the contribution from the Higgs scalar. The pressure $p(T)$ is normalized so that $p(0)=0$ and the subtraction of the vacuum pressure, $p_\mathrm{vac.} \sim \nu^4$, is taken into account in the $\sim \nu^4$ term of the expansion of $p_\mathrm{E}(T)$. Note that the expression in Eq.~(\ref{eq:pEpres}) for $p_\mathrm{E}(T)$ is a high temperature expansion and thus is not suitable for studying the limit $T\rightarrow 0$ as such. The infrared divergences are manifest, corresponding to the terms $\sim 1/\epsilon$.

\begin{fmffile}{pE_diags}



\def\Ring#1{%
  \parbox{30\unitlength}{
  \begin{fmfgraph}(30,30)
	\fmfi{#1}{fullcircle scaled 1h shifted (.5w,.5h)}
  \end{fmfgraph}}}

\def\DiaEight#1#2{%
  \parbox{60\unitlength}{
  \begin{fmfgraph}(60,30)
	\fmfleft{i}
	\fmfright{o}
	\fmf{#1,right}{i,v,i}
	\fmf{#2,right}{o,v,o}
  \end{fmfgraph}}}

\def\Sunset#1#2{%
  \parbox{30\unitlength}{
  \begin{fmfgraph}(30,30)
	\fmfleft{i}
	\fmfright{o}
	\fmf{#1,right}{i,o,i}
	\fmf{#2}{i,o}
  \end{fmfgraph}}}

\def\FSunset#1#2#3{%
  \parbox{30\unitlength}{
  \begin{fmfgraph}(30,30)
	\fmfleft{i}
	\fmfright{o}
	\fmf{#1,right}{i,o}
        \fmf{#2,right}{o,i}
	\fmf{#3}{i,o}
  \end{fmfgraph}}}

\def\Mersu#1#2#3#4{%
  \parbox{30\unitlength}{
  \begin{fmfgraph}(30,30)
	\fmfipath{p}
	\fmfiset{p}{fullcircle scaled 30 rotated -30 shifted (15,15) }
	\fmfipair{v[]}
	\fmfiset{v1}{point 2length(p)/3 of p}
	\fmfiset{v2}{point 0 of p}
	\fmfiset{v3}{point length(p)/3 of p}
	\fmfiset{v4}{(v1+v2+v3)/3}
	\fmfi{#1}{subpath (0,length(p)/3) of p}
	\fmfi{#1}{subpath (length(p)/3,2length(p)/3) of p}
	\fmfi{#2}{subpath (2length(p)/3,length(p)) of p}
	\fmfi{#4}{v1--v4}
	\fmfi{#4}{v4--v2}
	\fmfi{#3}{v3--v4}
  \end{fmfgraph}}}

\def\FMersu#1#2#3#4{%
  \parbox{30\unitlength}{
  \begin{fmfgraph}(30,30)
	\fmfipath{p}
	\fmfiset{p}{fullcircle scaled 30 rotated -30 shifted (15,15) }
	\fmfipair{v[]}
	\fmfiset{v1}{point 2length(p)/3 of p}
	\fmfiset{v2}{point 0 of p}
	\fmfiset{v3}{point length(p)/3 of p}
	\fmfiset{v4}{(v1+v2+v3)/3}
	\fmfi{#1}{subpath (0,length(p)/3) of p}
	\fmfi{#1}{subpath (length(p)/3,2length(p)/3) of p}
	\fmfi{#2}{subpath (2length(p)/3,length(p)) of p}
	\fmfi{#4}{v4--v1}
	\fmfi{#4}{v2--v4}
	\fmfi{#3}{v3--v4}
  \end{fmfgraph}}}

\def\FFMersu#1#2#3#4{%
  \parbox{30\unitlength}{
  \begin{fmfgraph}(30,30)
	\fmfipath{p}
	\fmfiset{p}{fullcircle scaled 30 rotated -30 shifted (15,15) }
	\fmfipair{v[]}
	\fmfiset{v1}{point 2length(p)/3 of p}
	\fmfiset{v2}{point 0 of p}
	\fmfiset{v3}{point length(p)/3 of p}
	\fmfiset{v4}{(v1+v2+v3)/3}
	\fmfi{#1}{subpath (0,length(p)/3) of p}
	\fmfi{#1}{subpath (length(p)/3,2length(p)/3) of p}
	\fmfi{#2}{subpath (length(p),2length(p)/3) of p}
	\fmfi{#4}{v1--v4}
	\fmfi{#4}{v4--v2}
	\fmfi{#3}{v3--v4}
  \end{fmfgraph}}}

\def\DiaV#1#2#3#4{%
  \parbox{30\unitlength}{
  \begin{fmfgraph}(30,30)
	\fmfipath{p}
	\fmfiset{p}{fullcircle scaled 30 rotated -90 shifted (15,15) }
	\fmfipair{v[]}
	\fmfiset{v1}{point 0 of p}
	\fmfiset{v2}{point 3length(p)/8 of p}
	\fmfiset{v3}{point 5length(p)/8 of p}
	\fmfi{#1}{subpath (0,3length(p)/8) of p}
	\fmfi{#2}{subpath (3length(p)/8,5length(p)/8) of p}
	\fmfi{#1}{subpath (5length(p)/8,length(p)) of p}
	\fmfi{#3}{v1--v3}
	\fmfi{#4}{v2--v1}
  \end{fmfgraph}}}

\def\FDiaV#1#2#3#4{%
  \parbox{30\unitlength}{
  \begin{fmfgraph}(30,30)
	\fmfipath{p}
	\fmfiset{p}{fullcircle scaled 30 rotated 135 shifted (15,15) }
	\fmfipair{v[]}
	\fmfiset{v1}{point 3length(p)/8 of p}
	\fmfiset{v2}{point 3length(p)/4 of p}
	\fmfiset{v3}{point 0 of p}
	\fmfi{#2}{subpath (0,-length(p)/4) of p}
	\fmfi{#1}{subpath (0,3length(p)/4) of p}
	\fmfi{#3}{v1--v3}
	\fmfi{#4}{v2--v1}
  \end{fmfgraph}}}

\def\Basketball#1#2#3#4{%
  \parbox{45\unitlength}{
  \begin{fmfgraph}(45,30)
	\fmfipath{p[]}
	\fmfiset{p1}{fullcircle scaled 30 rotated -60 shifted (15,15) }
	\fmfiset{p2}{fullcircle scaled 30 rotated -120 shifted (30,15) }
	\fmfi{#1}{subpath (length(p1)/3,length(p1)) of p1}
	\fmfi{#2}{subpath (2length(p2)/3,length(p2)) of p2}
	\fmfi{#3}{subpath (0,length(p1)/3) of p1}
	\fmfi{#4}{subpath (0,2length(p2)/3) of p2}
  \end{fmfgraph}}}
  
  
\def\RingRing#1#2#3{%
  \parbox{60\unitlength}{
  \begin{fmfgraph}(60,40)
	\fmfi{#2}{fullcircle scaled .5h shifted (w/6,0.5h)}
	\fmfi{#3}{fullcircle scaled .5h shifted (5w/6,0.5h)}
	\fmfipath{p}
	\fmfiset{p}{fullcircle scaled 1h shifted (.5w,.5h)}
	\fmfi{#1}{subpath (2*angle(sqrt(15),1)*length(p)/360,(1/2-2*angle(sqrt(15),1)/360)*length(p)) of p}
	\fmfi{#1}{subpath ((1/2+2*angle(sqrt(15),1)/360)*length(p),(1-2*angle(sqrt(15),1)/360)*length(p)) of p}
  \end{fmfgraph}}}

\def\FlatFlat#1#2#3{%
  \parbox{60\unitlength}{
  \begin{fmfgraph}(60,40)
	\fmfipath{p[]}
	\fmfi{#1}{halfcircle scaled 1h shifted (.5w,.5h)}
	\fmfi{#1}{halfcircle scaled 1h rotated 180 shifted (.5w,.5h)}
	\fmfiset{p2}{fullcircle scaled .5h shifted (w/6,0.5h)}
	\fmfiset{p3}{fullcircle scaled .5h rotated 180 shifted (5w/6,0.5h)}
	\fmfi{#2}{subpath (angle(1,sqrt(15))*length(p2)/360,(1-angle(1,sqrt(15))/360)*length(p2)) of p2}
	\fmfi{#3}{subpath (angle(1,sqrt(15))*length(p3)/360,(1-angle(1,sqrt(15))/360)*length(p3)) of p3}
  \end{fmfgraph}}}

\def\LoopLoop#1#2#3{%
  \parbox{90\unitlength}{
  \begin{fmfgraph}(90,30)
	\fmfi{#1}{halfcircle scaled 1h shifted (.5w,.5h)}
	\fmfi{#1}{halfcircle scaled 1h rotated 180 shifted (.5w,.5h)}
	\fmfi{#2}{fullcircle scaled 1h shifted (w/6,.5h)}
	\fmfi{#3}{fullcircle scaled 1h rotated 180 shifted (5w/6,.5h)}
  \end{fmfgraph}}}

\def\RingLoop#1#2#3{%
  \parbox{70\unitlength}{
  \begin{fmfgraph}(70,40)
	\fmfi{#2}{fullcircle scaled .5h shifted (w/7,0.5h)}
	\fmfi{#3}{fullcircle scaled .5h rotated 180 shifted (6w/7,0.5h)}
	\fmfipath{p}
	\fmfiset{p}{fullcircle scaled 1h shifted (3w/7,.5h)}
	\fmfi{#1}{subpath (0,(1/2-2*angle(sqrt(15),1)/360)*length(p)) of p}
	\fmfi{#1}{subpath ((1/2+2*angle(sqrt(15),1)/360)*length(p),length(p)) of p}
  \end{fmfgraph}}}

\def\FlatLoop#1#2#3{%
  \parbox{70\unitlength}{
  \begin{fmfgraph}(70,40)
	\fmfi{#1}{halfcircle scaled 1h shifted (3w/7,.5h)}
	\fmfi{#1}{halfcircle scaled 1h rotated 180 shifted (3w/7,.5h)}
	\fmfipath{p}
	\fmfiset{p}{fullcircle scaled .5h shifted (w/7,0.5h)}
	\fmfi{#2}{subpath (angle(1,sqrt(15))*length(p)/360,(1-angle(1,sqrt(15))/360)*length(p)) of p}
	\fmfi{#3}{fullcircle scaled .5h rotated 180 shifted (6w/7,0.5h)}
  \end{fmfgraph}}}

\def\RingFlat#1#2#3{%
  \parbox{60\unitlength}{
  \begin{fmfgraph}(60,40)
	\fmfipath{p[]}
	\fmfiset{p}{fullcircle scaled 1h shifted (.5w,.5h)} 
	\fmfi{#1}{subpath (0,(1/2-2*angle(sqrt(15),1)/360)*length(p)) of p}
	\fmfi{#1}{subpath ((1/2+2*angle(sqrt(15),1)/360)*length(p),length(p)) of p}
	\fmfiset{p3}{fullcircle scaled .5h rotated 180 shifted (5w/6,0.5h)}
	\fmfi{#2}{fullcircle scaled .5h shifted (w/6,0.5h)}
	\fmfi{#3}{subpath (angle(1,sqrt(15))*length(p3)/360,(1-angle(1,sqrt(15))/360)*length(p3)) of p3}
  \end{fmfgraph}}}

  
\def\SelfenA#1#2{%
  \begin{fmfgraph}(60,30)
	\fmfipair{v[]}
	\fmfi{#2}{halfcircle scaled 1h shifted (0.5w,0.5h)}
	\fmfi{#2}{halfcircle scaled 1h rotated 180 shifted (0.5w,0.5h)}
	\fmfiset{v1}{(0,0.5h)}
	\fmfiset{v2}{(w,0.5h)}
	\fmfi{#1}{v1--(0.5w-0.5h,0.5h)}
	\fmfi{#1}{(0.5w+0.5h,0.5h)--v2}
  \end{fmfgraph}}

\def\SelfenAflat#1#2#3{%
  \begin{fmfgraph}(60,30)
	\fmfi{#2}{halfcircle scaled 1h shifted (0.5w,0)}
	\fmfi{#3}{(0,0)--(0.5w-0.5h,0)}
	\fmfi{#1}{(0.5w-0.5h,0)--(0.5w+0.5h,0)}
	\fmfi{#3}{(0.5w+0.5h,0)--(1w,0)}
  \end{fmfgraph}}
	
\def\SelfenB#1#2{%
  \begin{fmfgraph}(50,30)
	\fmfipair{v}
	\fmfi{#2}{fullcircle scaled 1h rotated -90 shifted (0.5w,0.5h)}
	\fmfiset{v}{(0.5w,0)}
	\fmfi{#1}{(0,0)--v}
	\fmfi{#1}{v--(w,0)}
  \end{fmfgraph}}


\fmfcmd{%
  style_def dash_dot expr p =
    save dpp, k;
    numeric dpp, k;
    dpp = ceiling (pixlen (p, 10) / (1.5*dash_len)) / length p;
    k=0;
    forever:
      exitif k+.33 > dpp*length(p);
      cdraw point k/dpp of p .. point (k+.33)/dpp of p;
      exitif k+.67 > dpp*length(p);
      cdrawdot point (k+.67)/dpp  of p;
      k := k+1;
    endfor
  enddef;}
  
\fmfcmd{%
  style_def dash_dot_arrow expr p =
    draw_dash_dot p;
    cfill (arrow p);
  enddef;}

\begin{figure}
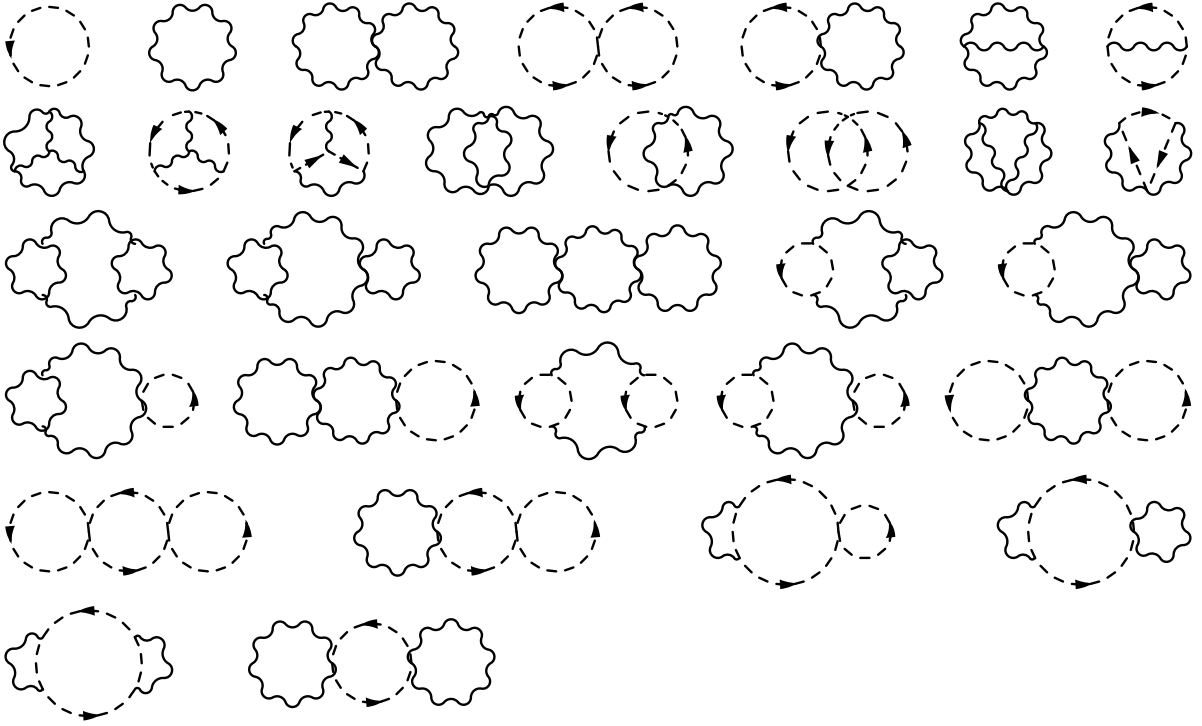

  \begin{minipage}{\textwidth}
        \fmfset{arrow_len}{2.5mm}
        \fmfset{dash_len}{2.5mm}
	\Ring{scalar}\hspace{\stretch{1}}
	\Ring{photon}\hspace{\stretch{1}}
	\DiaEight{photon}{photon}\hspace{\stretch{1}}
	\DiaEight{scalar}{scalar}\hspace{\stretch{1}}
	\DiaEight{scalar}{photon}\hspace{\stretch{1}}
	\Sunset{photon}{photon}\hspace{\stretch{1}}
	\Sunset{scalar}{photon}\\[3mm]
	\Mersu{photon}{photon}{photon}{photon}\hspace{\stretch{1}}
	\Mersu{scalar}{scalar}{photon}{photon}\hspace{\stretch{1}}
	\Mersu{scalar}{photon}{photon}{scalar}\hspace{\stretch{1}}
	\Basketball{photon}{photon}{photon}{photon}\hspace{\stretch{1}}
	\Basketball{scalar}{photon}{scalar}{photon}\hspace{\stretch{1}}
	\Basketball{scalar}{scalar}{scalar}{scalar}\hspace{\stretch{1}}
	\FDiaV{photon}{photon}{photon}{photon}\hspace{\stretch{1}}
	\FDiaV{photon}{scalar}{scalar}{scalar}\\[3mm]
	\RingRing{photon}{photon}{photon}\hspace{\stretch{1}}
	\RingLoop{photon}{photon}{photon}\hspace{\stretch{1}}
	\LoopLoop{photon}{photon}{photon}\hspace{\stretch{1}}
	\RingRing{photon}{scalar}{photon}\hspace{\stretch{1}}
	\RingLoop{photon}{scalar}{photon}\\[3mm]
	\RingLoop{photon}{photon}{scalar}\hspace{\stretch{1}}
	\LoopLoop{photon}{photon}{scalar}\hspace{\stretch{1}}
	\RingRing{photon}{scalar}{scalar}\hspace{\stretch{1}}
	\RingLoop{photon}{scalar}{scalar}\hspace{\stretch{1}}
	\LoopLoop{photon}{scalar}{scalar}\\[3mm]
	\LoopLoop{scalar}{scalar}{scalar}\hspace{\stretch{1}}
	\LoopLoop{scalar}{photon}{scalar}\hspace{\stretch{1}}
	\FlatLoop{scalar}{photon}{scalar}\hspace{\stretch{1}}
	\FlatLoop{scalar}{photon}{photon}\\[3mm]
	\FlatFlat{scalar}{photon}{photon}\hspace{1cm}
	\LoopLoop{scalar}{photon}{photon}
  \end{minipage}
\caption{Diagrams contributing to $p_\mathrm{E}(T)$ in the SU($2$) + fundamental Higgs theory. The dashed lines correspond to the fundamental scalar and the wavy lines to gauge bosons. Diagrams containing ghosts and counterterms are not drawn.}
\label{fig:pEdiags}
\end{figure}
\end{fmffile}

Contribution from the soft scale $gT$ to the pressure is contained in the pressure of the effective theory $S_\mathrm{E}$. Assuming that the adjoint and fundamental scalars are (parametrically) equally heavy (valid at very high temperatures), this is conveniently obtained by writing
\begin{eqnarray}
\frac{T}{V}\ln \int\mathcal{D}A_i\mathcal{D}A_0\mathcal{D}\Phi\;\mathrm{exp}(-S_\mathrm{E}) & = & p_\mathrm{M}(T) + \frac{T}{V} \ln \int\mathcal{D}A_i\;\mathrm{exp}(-S_\mathrm{M}) \label{eq:gtpres}
\end{eqnarray}
where $p_\mathrm{M}(T)$ is the contribution from the soft scales and the theory $S_\mathrm{M}$ describes the dynamics of the magnetic sector of the system,
\begin{eqnarray}
S_\mathrm{M} & = & \int\mathrm{d}^3\mathbf{x}\, \frac{1}{4}G_{ij}^a G_{ij}^a.
\end{eqnarray}
The theory is confining and cannot be solved using perturbative methods; however, it contributes to the pressure starting at order $g^6$ and can therefore be neglected now.

\begin{figure}
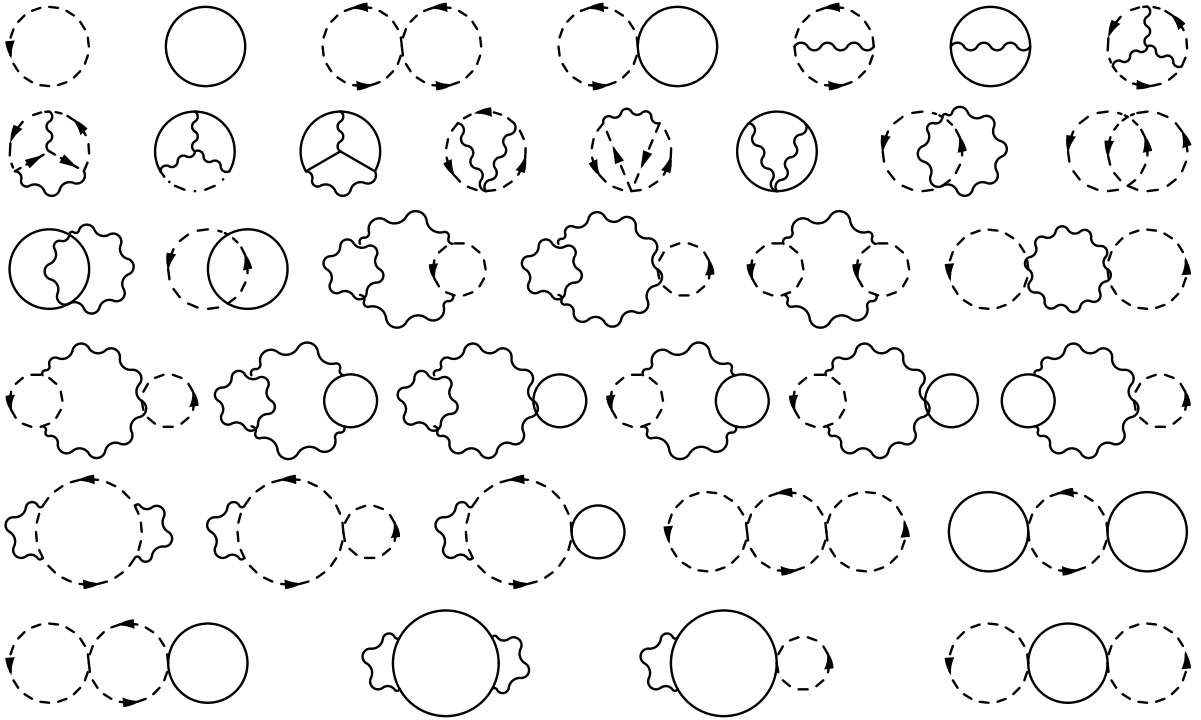

  \begin{minipage}{\textwidth}
        \fmfset{arrow_len}{2.5mm}
        \fmfset{dash_len}{2.5mm}
	\Ring{scalar}\hspace{\stretch{1}}
	\Ring{vanilla}\hspace{\stretch{1}}
	\DiaEight{scalar}{scalar}\hspace{\stretch{1}}
	\DiaEight{scalar}{vanilla}\hspace{\stretch{1}}
	\Sunset{scalar}{photon}\hspace{\stretch{1}}
	\Sunset{vanilla}{photon}\hspace{\stretch{1}}
	\Mersu{scalar}{scalar}{photon}{photon}\\[3mm]
	\Mersu{scalar}{photon}{photon}{scalar}\hspace{\stretch{1}}
	\Mersu{vanilla}{dash_dot}{photon}{photon}\hspace{\stretch{1}}
	\Mersu{vanilla}{photon}{photon}{vanilla}\hspace{\stretch{1}}
	\DiaV{scalar}{scalar}{photon}{photon}\hspace{\stretch{1}}
	\DiaV{scalar}{photon}{scalar}{scalar}\hspace{\stretch{1}}
	\DiaV{vanilla}{vanilla}{photon}{photon}\hspace{\stretch{1}}
	\Basketball{scalar}{photon}{scalar}{photon}\hspace{\stretch{1}}
	\Basketball{scalar}{scalar}{scalar}{scalar}\\[3mm]
	\Basketball{vanilla}{photon}{vanilla}{photon}\hspace{\stretch{1}}
	\Basketball{scalar}{vanilla}{scalar}{vanilla}\hspace{\stretch{1}}
	\RingRing{photon}{photon}{scalar}\hspace{\stretch{1}}
	\RingLoop{photon}{photon}{scalar}\hspace{\stretch{1}}
	\RingRing{photon}{scalar}{scalar}\hspace{\stretch{1}}
	\LoopLoop{photon}{scalar}{scalar}\\[3mm]
	\RingLoop{photon}{scalar}{scalar}\hspace{\stretch{1}}
	\RingRing{photon}{photon}{vanilla}\hspace{\stretch{1}}
	\RingLoop{photon}{photon}{vanilla}\hspace{\stretch{1}}
	\RingRing{photon}{scalar}{vanilla}\hspace{\stretch{1}}
	\RingLoop{photon}{scalar}{vanilla}\hspace{\stretch{1}}
	\RingLoop{photon}{vanilla}{scalar}\\[3mm]
	\FlatFlat{scalar}{photon}{photon}\hspace{\stretch{1}}
	\FlatLoop{scalar}{photon}{scalar}\hspace{\stretch{1}}
	\FlatLoop{scalar}{photon}{vanilla}\hspace{\stretch{1}}
	\LoopLoop{scalar}{scalar}{scalar}\hspace{\stretch{1}}
	\LoopLoop{scalar}{vanilla}{vanilla}\\[3mm]
	\LoopLoop{scalar}{scalar}{vanilla}\hspace{\stretch{1}}
	\FlatFlat{vanilla}{photon}{photon}\hspace{\stretch{1}}
	\FlatLoop{vanilla}{photon}{scalar}\hspace{\stretch{1}}
	\LoopLoop{vanilla}{scalar}{scalar}
  \end{minipage}
\caption{Diagrams contributing to $p_\mathrm{M}(T)$ in the SU($2$) + fundamental Higgs theory. The dashed lines correspond to the fundamental scalar, the wavy lines to the gauge bosons and the solid lines to the adjoint scalars.}
\label{fig:pMdiags}
\end{figure}

The pressure $p_\mathrm{M}(T)$ is obtained by computing the loop expansion of $S_\mathrm{E}$ to three loops. The required diagrams are listed in Fig.~\ref{fig:pMdiags} and the expansion is given by
\begin{eqnarray}
\frac{p_\mathrm{M}(T)}{T} & = & \frac{1}{4\pi}\left(\frac{4}{3}n_\mathrm{S}m_3^2 + m_D^3\right) \label{eq:pMpres}\\
&-&\hspace{-0.2cm} \frac{T}{(4\pi)^2}\left[n_\mathrm{S}m_3^2\left(g^2\left(\frac{3}{4\epsilon} + 3\ln\frac{\Lambda}{2m_3} + \frac{9}{4}\right) + 6\lambda\right)\right. \nonumber \\
& & \left. \hspace{1.2cm} + g^2m_D^2\left(\frac{3}{2\epsilon} + 6\ln\frac{\Lambda}{2m_D} + \frac{9}{2}\right) + \frac{3}{2}n_\mathrm{S}g^2 m_3 m_D\right] \nonumber \\
&+&\hspace{-0.2cm} \frac{T^2}{(4\pi)^3}\left[n_\mathrm{S}m_3\left(g^4\left(\frac{81}{8}\ln\frac{\Lambda}{2m_3} - \frac{15}{4}\ln\frac{m_3+m_D}{m_3} + 6\ln\frac{m_3+m_D}{m_D} - \frac{391}{32} - \frac{3\pi^2}{8} - \frac{5}{4}\ln 2  \right)\right.\right. \nonumber \\
& & \hspace{2.6cm} \left. + g^2\lambda\left(18\ln\frac{\Lambda}{2m_3} - \frac{9}{2} + 12\ln 2\right) - 24\lambda^2\left(\ln\frac{\Lambda}{2m_3} + \frac{23}{24} - \ln 2 \right)\right)\nonumber \\
& & \hspace{1.6cm} + m_D\left(g^4\left(-\frac{89}{2} - 2\pi^2 + 22\ln 2 + n_\mathrm{S}\left(\frac{9}{4}\ln\frac{m_3+m_D}{m_3} - \frac{43}{16}\right)\right) \right. \nonumber \\
& & \hspace{2.6cm} \left. + \frac{9}{2}n_\mathrm{S}g^2\lambda \right) \nonumber \\
& & \hspace{1.6cm} \left. + g^4\frac{m_3^2}{m_D}\left(\ln\frac{m_3+m_D}{m_3} + \frac{3}{8}\right) + g^4\frac{m_D^2}{m_3}\left(\ln\frac{m_3+m_D}{m_D} + \frac{9}{32}\right)\right] \nonumber
\end{eqnarray}
The couplings of the effective theory, $g_3^2,\;h_3\;\mathrm{and}\;\lambda_3$, are substituted with those of the original theory by using the matching given in Eq.~(\ref{eq:coupl_match}) in order to make the expression shorter.

\begin{fmffile}{pM_diags}

\end{fmffile}

The pressure of the theory can now be written as the sum of the individual parts,
\begin{eqnarray}
p(T) & = & p_\mathrm{E}(T) + p_\mathrm{M}(T) + \mathcal{O}(g^6). \label{eq:su2hpres}
\end{eqnarray}
with $p_\mathrm{E}(T)$ and $p_\mathrm{M}(T)$ given by Eqs.~(\ref{eq:pEpres}) and (\ref{eq:pMpres}), respectively. Substituting the expressions for the masses $m_D$ and $m_3$ to $p_\mathrm{M}(T)$, we see that the poles in $p_\mathrm{E}(T)$ and $p_\mathrm{M}(T)$ cancel each other. However, a physical effect remains in the form of terms $\sim g^4\ln\left(m/T\right) \sim g^4\ln g$ where $m$ refers to masses in the effective theory, $m_3$ and $m_D$. Additionally, the fact that the fundamental scalar mass runs in the effective theory leads to terms of the order $g^5\ln\left(\Lambda/m_3\right)$ in the expansion. Dependence on the scale $\Lambda$ vanishes order by order, as it should for a physical quantity, when running of the parameters is taken into account, but will not vanish completely unless an all-orders result is considered. The scale must therefore be fixed and we choose $\Lambda = 2\pi T$. Sensitivity to changing the scale from this is studied later. 

\subsection{Approaching the crossover transition}

Above we implicitly assumed that the fundamental and adjoint scalar masses are of the same order of magnitude (parametrically). However, as the temperature of the system is decreased, the fundamental scalar becomes increasingly light compared to the adjoint scalar (see Fig.~\ref{fig:massratio}). Near the crossover transition in which the symmetry of the theory becomes spontaneously broken, the mass of the fundamental scalar will effectively vanish, $m_3^2 \sim 0$. This renders the computation of $p_\mathrm{M}(T)$ unreliable near the crossover, as especially evidenced by the terms $\sim m_D^2/m_3$ in the expansion. Thus $p_\mathrm{M}(T)$ must be recalculated near the crossover. This is conveniently done by constructing a new effective theory. Since we have a scale hierarchy, $m_3 < m_D$, we can integrate out the adjoint scalars to obtain an effective theory for the fundamental scalars and $3$d gauge fields,
\begin{eqnarray}
S_{\mathrm{E}'} & = & \int \mathrm{d}^3\mathbf{x}\,\left(\frac{1}{4}G_{ij}^a G_{ij}^a + D_i\Phi^\dagger D_i\Phi + \widetilde{m}_3^2\Phi^\dagger\Phi + \widetilde{\lambda}_3\left(\Phi^\dagger\Phi\right)^2\right). 
\end{eqnarray}
The couplings of this theory will to the required order be the same as the couplings of $S_\mathrm{E}$, $\widetilde{g_3}^2 = g_3^2,\;\widetilde{\lambda}_3 = \lambda_3$. The fundamental scalar mass, on the other hand, does receive a correction,
\begin{eqnarray}
\widetilde{m}_3^2 & = & m_3^2 - \frac{3h_3 m_D}{4\pi}\left[1 + 2\left(1+\ln\frac{\Lambda}{2 m_D}\right)\epsilon\right] + \mathcal{O}(g^4). \label{eq:ptmass}
\end{eqnarray}
We use a powercounting rule $m_3^2 \sim \widetilde{m}_3^2 \sim g^3 T^2$ (valid only near the crossover) and then the order $g^4$ corrections to the mass will contribute to the pressure at order $g^{5.5}$ and will be neglected now. Note that $\widetilde{m}_3^2$ is renormalization group invariant to this order.

The contribution from the effective theories to the pressure is now reorganized so that, instead of Eq.~(\ref{eq:gtpres}), we have
\begin{eqnarray}
\frac{T}{V} \ln \int\mathcal{D}A_i\mathcal{D}A_0\mathcal{D}\Phi\;\mathrm{exp}(-S_\mathrm{E}) & = &
p_{\mathrm{M}1}(T) + \frac{T}{V} \ln \int\mathcal{D}A_i\mathcal{D}\Phi\;\mathrm{exp}(-S_{\mathrm{E}'}) \nonumber \\
& = & p_{\mathrm{M}1}(T) + p_{\mathrm{M}2}(T) + \frac{T}{V} \ln \int\mathcal{D}A_i\;\mathrm{exp}(-S_{\mathrm{M}'}).
\end{eqnarray}
The theory $S_{\mathrm{M}'}$ will differ from the theory $S_\mathrm{M}$ only by the matching of the parameters, the structure of the theories is the same. Its contribution to the pressure, of the order $\mathcal{O}(g^6)$, can again be neglected.

\begin{figure}
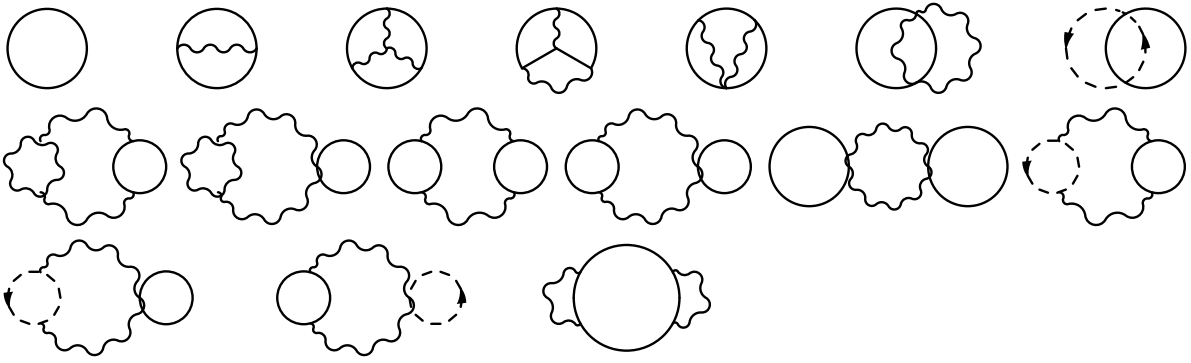

  \begin{minipage}{\textwidth}
        \fmfset{arrow_len}{2.5mm}
        \fmfset{dash_len}{2.5mm}
        \Ring{vanilla}\hspace{\stretch{1}}
        \Sunset{vanilla}{photon}\hspace{\stretch{1}}
        \Mersu{vanilla}{vanilla}{photon}{photon}\hspace{\stretch{1}}
        \Mersu{vanilla}{photon}{photon}{vanilla}\hspace{\stretch{1}}
        \DiaV{vanilla}{vanilla}{photon}{photon}\hspace{\stretch{1}}
        \Basketball{vanilla}{photon}{vanilla}{photon}\hspace{\stretch{1}}
        \Basketball{scalar}{vanilla}{scalar}{vanilla}\\[3mm]
        \RingRing{photon}{photon}{vanilla}\hspace{\stretch{1}}
        \RingLoop{photon}{photon}{vanilla}\hspace{\stretch{1}}
        \RingRing{photon}{vanilla}{vanilla}\hspace{\stretch{1}}
        \RingLoop{photon}{vanilla}{vanilla}\hspace{\stretch{1}}
        \LoopLoop{photon}{vanilla}{vanilla}\hspace{\stretch{1}}
        \RingRing{photon}{scalar}{vanilla}\\[3mm]
        \RingLoop{photon}{scalar}{vanilla}\hspace{1cm}
        \RingLoop{photon}{vanilla}{scalar}\hspace{1cm}
        \FlatFlat{vanilla}{photon}{photon}
  \end{minipage}
\caption{Diagrams contributing to $p_{\mathrm{M}1}(T)$. The solid lines correspond to the adjoint scalars, the dashed lines to the fundamental scalar and the wavy lines to the gauge bosons.}
\label{fig:pM1diags}
\end{figure}

The diagrams needed to calculate $p_{\mathrm{M}1}(T)$ are given in Fig.~\ref{fig:pM1diags}. Those are computed in \cite{Gynther:2005av} and the result reads
\begin{eqnarray}
\frac{p_{\mathrm{M}1}(T)}{T} & = & \frac{1}{4\pi}m_D^3 - \frac{T}{(4\pi)^2}6g^2m_D^2\left(\frac{1}{4\epsilon} + \frac{3}{4} + \ln\frac{\Lambda}{2m_D}\right) \nonumber \\
& & + \frac{T^2}{(4\pi)^3}g^4m_D\left[-\frac{89}{2}-2\pi^2+22\ln 2 -n_\mathrm{S}\left(\frac{9}{16\epsilon} + \frac{27}{8}\ln\frac{\Lambda}{2m_D} + 3\right)\right],
\end{eqnarray}
where we have again substituted the matching of the couplings already to the expression. Note that, in addition to poles at order $g^2m_D^2$ that are canceled against the poles in $p_\mathrm{E}(T)$ and which correspond to infrared divergences related to screening of electric fields, there are also poles at order $g^4m_D$ that are only canceled when the pressure $p_{\mathrm{M}2}(T)$ is taken into account. These poles are related to the infrared divergences that appear in the theory given by $S_\mathrm{E}$ when the mass of the fundamental scalar is taken to the limit $m_3 \rightarrow 0$.

\begin{fmffile}{pM1_diags}

\end{fmffile}

\begin{figure}
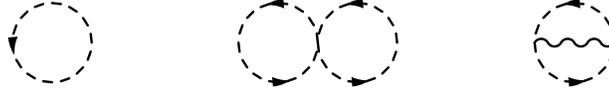

\begin{center}
  \begin{minipage}{8cm}
        \fmfset{arrow_len}{2.5mm}
        \fmfset{dash_len}{2.5mm}
	\Ring{scalar} \hspace{\stretch{1}}
	\DiaEight{scalar}{scalar}\hspace{\stretch{1}}
	\Sunset{scalar}{photon}
  \end{minipage}
\caption{Diagrams contributing to $p_{\mathrm{M}2}$. The dashed lines correspond to the fundamental scalar and the wavy lines to the gauge bosons.}
\label{fig:pM2diags}
\end{center}
\end{figure}

Since the fundamental scalar mass is small near the crossover, the pressure from $S_{\mathrm{E}'}$ is only needed to two loops. The diagrams needed are shown in Fig.~\ref{fig:pM2diags} and the result for the pressure reads
\begin{eqnarray}
\frac{p_{\mathrm{M}2}(T)}{T} & = & \frac{1}{4\pi}\frac{4}{3}\widetilde{m}_3^3 - \frac{T}{(4\pi)^2}\left[\frac{3}{4}g^2\widetilde{m}_3^2\left(\frac{1}{\epsilon} + 3 + 4\ln\frac{\Lambda}{2\widetilde{m}_3}\right) + 6\lambda\widetilde{m}_3^2\right]. \label{eq:pm2pres}
\end{eqnarray}
The poles here cancel against those coming from $p_\mathrm{E}(T)$ and $p_{\mathrm{M}1}(T)$.

Collecting the results together, the pressure near the crossover is given by
\begin{eqnarray}
p(T) & = & p_\mathrm{E}(T) + p_{\mathrm{M}1}(T) + p_{\mathrm{M}2}(T) + T^4\mathcal{O}(g^{5.5}). \label{eq:su2hprespt}
\end{eqnarray}
This expression is well behaved when $\widetilde{m}_3 \rightarrow 0$ as it should. Also, all the $1/\epsilon$ poles are canceled in the final result just like for the previous calculation and similar types of terms, of orders $g^4 \ln g$ and $g^5 \ln g$ are found in the expansion. Note that in this case, the presence of the terms $\sim g^5\ln g$ is related to the infrared divergences that we encounter at the limit $m_3 \rightarrow 0$, not to the renormalization of the fundamental scalar mass in the effective theories.

\begin{fmffile}{pM2_diags}

\end{fmffile}

\section{Numerical results}
\label{sec:numres}

In this section we will analyze the numerical consequences of the expansion derived in the previous section. We will start with the simpler SU($2$) + Higgs theory, since the effects of the Higgs sector should be more evident in it. We will then turn to the full standard model for which the perturbative expansion is given in appendix \ref{app:presEW}. The numerical values of all the relevant parameters are as given in section \ref{sec:notation}, with the Higgs mass chosen to be $m_H = 130$~GeV. The scale is chosen so that $\Lambda = 2\pi T$.

\begin{figure}[!hptb]
\begin{minipage}[t]{.9\textwidth}
\begin{center}\vspace{-3cm}  
\includegraphics[width=0.8\textwidth]{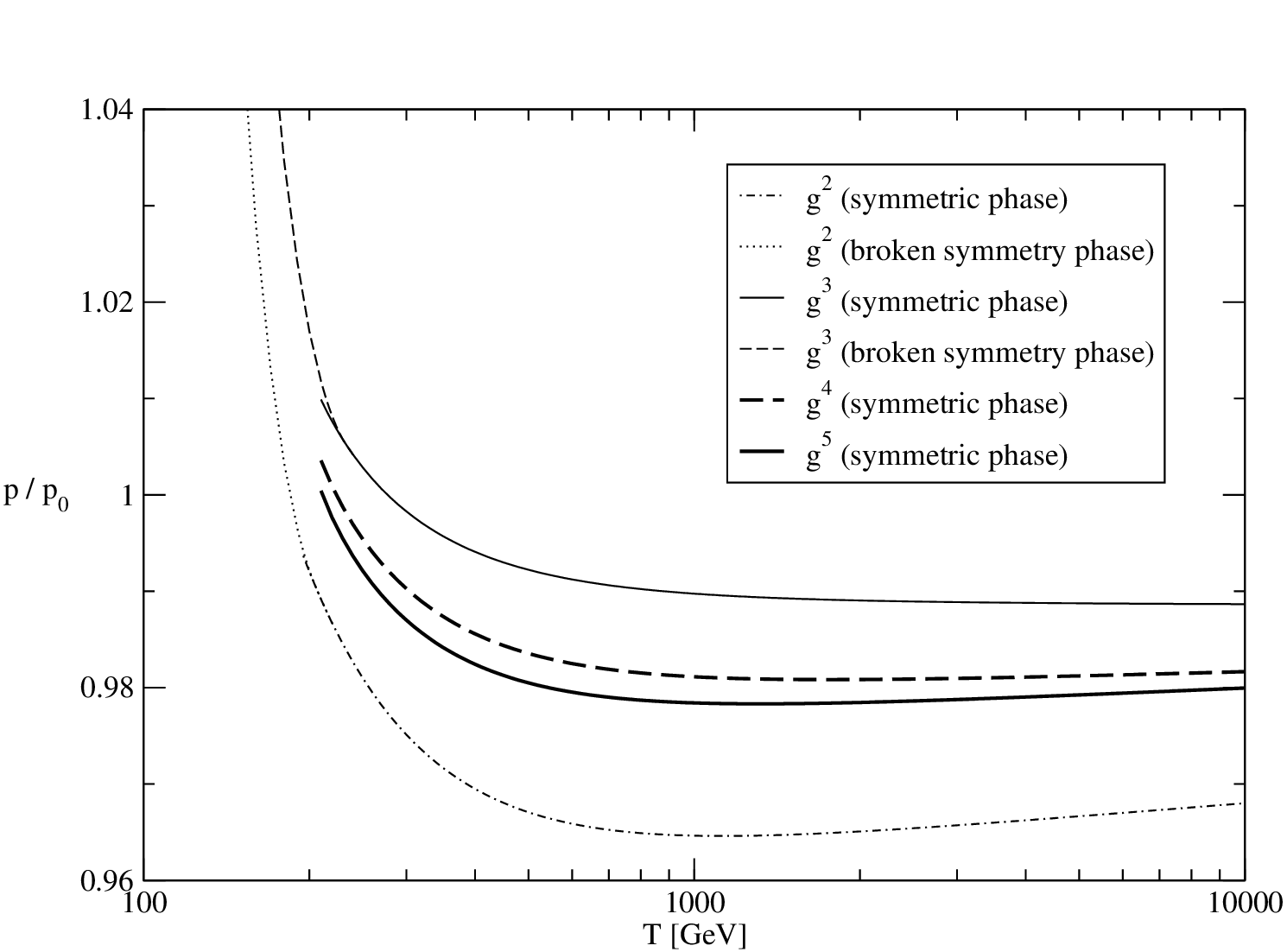}
\caption{The pressure of the SU($2$) + Higgs theory at different orders.}
\label{fig:su2hpres}
\end{center}
\end{minipage}
\begin{minipage}[t]{.9\textwidth}
\begin{center}\vspace{3cm}  
\includegraphics[width=0.8\textwidth]{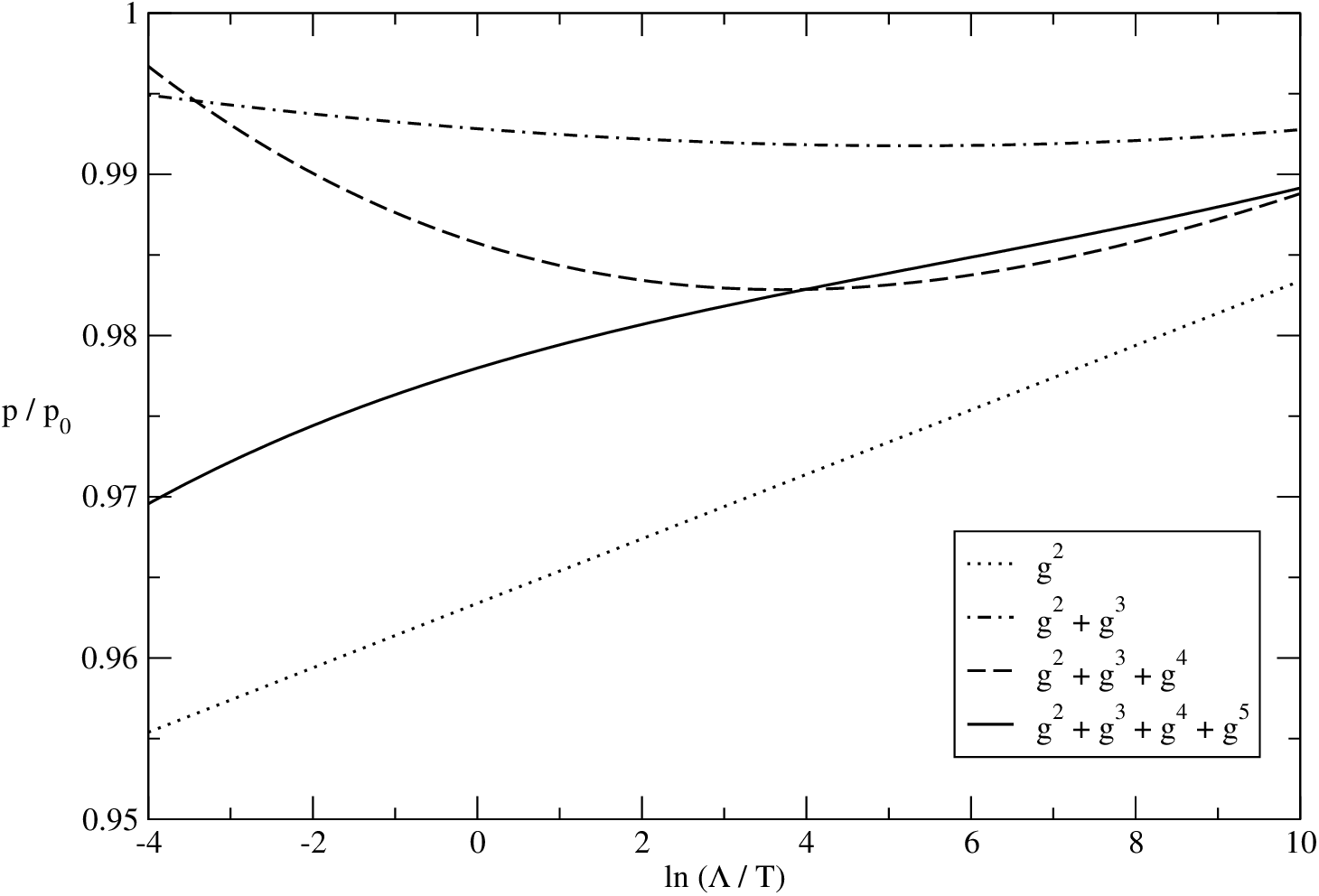}
\caption{Scale dependence of the pressure of SU($2$) + Higgs theory at different orders. Temperature is fixed to $500$~GeV.}
\label{fig:su2hscale}
\end{center}
\end{minipage}
\end{figure}

The pressure of the SU($2$) + Higgs theory is plotted in Fig.~\ref{fig:su2hpres} at different orders of the perturbative expansion. Using two-loop effective potential calculations \cite{Arnold:1992rz,Fodor:1994bs} we have added the pressure of the broken phase to order $g^3$ to the picture to indicate where the phase transition (crossover) takes place. The result is normalized to the ideal gas pressure of massless particles $p_0(T)$, given by
\begin{eqnarray}
p_0(T) & = & \frac{\pi^2}{9}T^4.
\end{eqnarray}

As can be seen from the figure, the pressure does not differ from the pressure of ideal gas by more than 2\%. This reflects the fact that the theory is weakly coupled and thus the effect of interactions remains small. This is confirmed by the fact that the perturbative expansion seems to converge well: contribution of each new term in the expansion is smaller than that of the previous terms. 

Convergence of the expansion can also be studied by considering the scale dependence of the result. Scale dependence should be reduced with each new order in the expansion if perturbation theory is valid. As can be seen in Fig. \ref{fig:su2hscale}, this is partially so in this theory. Running the scale by $6$ orders magnitude at a fixed temperature, the pressure changes only by about 2\%. Also, scale dependence seems to be largest at order $g^2$. However, the change in the pressure induced by running the scale is as big as is the general effect from the interactions. Also, the difference between the scale dependence at orders $g^2$ and $g^5$ is not large. In fact, the result is less dependent on the scale at orders $g^3$ and $g^4$ than at order $g^5$. This is in accord with results obtained in \cite{Arnold:1995eb,*Arnold:1994ps} for the pressure of QCD. We can thus deduce that although the perturbative expansion seems to converge well, it can not be ruled out that higher order terms have as large a contribution to the pressure as the presently calculated terms.

\begin{figure}[!tb]
\begin{center}
\includegraphics[width=0.87\textwidth]{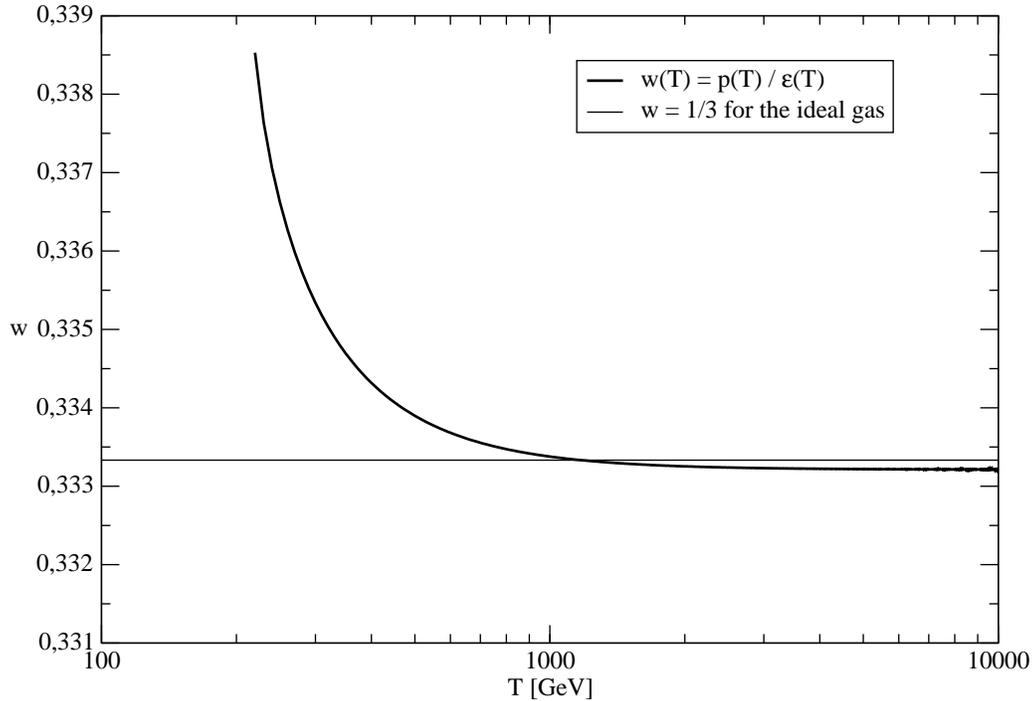}
\caption{The equation of state parameter $w(T)$ plotted for SU($2$) + Higgs theory. The ideal gas result for radiation is given for reference.}
\label{fig:su2hw}
\end{center}
\end{figure}

Since there is another explicit mass scale in the theory, $\nu^2$, we can also expect the equation of state of this system to differ from that of massless particles, at least when the temperature is not too much above the scale $\nu$. This can be conveniently studied by computing the equation of state parameter $w(T)$, defined by $p(T) = w(T)\epsilon(T)$. For radiation $w(T) = 1/3$ while for non-relativistic matter we have $w(T) = 0$ (pressureless dust) and for cosmological constant $w(T)=-1$. In Fig.~\ref{fig:su2hw} we have plotted $w(T)$ for matter described by the theory in question. As can be seen, for high temperatures the system behaves very much as radiation, but as the temperature decreases, we see deviation from $w(T) = 1/3$. This is to be expected since the terms $\nu^2 T^2$ and $\nu^4$ become increasingly important as temperature is lowered, making the pressure to deviate from the form $p\sim T^4$. The difference to $w(T)=1/3$ of radiation is, however, still small. Although $w(T)$ seems to grow near the crossover transition, one expects it to decrease soon after the transition since the system will contain a number of massive fields whose contribution to the pressure is negligible compared with their contribution to the energy density, $p(T)/\epsilon(T) = T/m$, $m \gg T$. 

\begin{figure}[!tb]
\begin{center}
\includegraphics[width=0.87\textwidth]{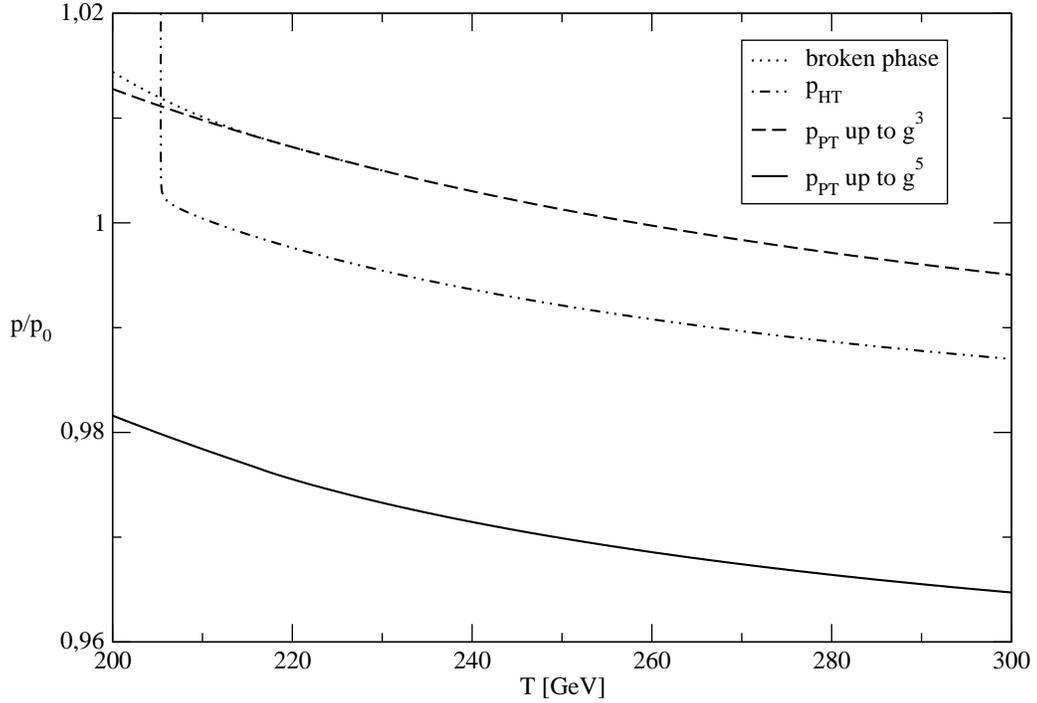}
\caption{The pressure of SU($2$) + Higgs theory as plotted all the way down to the crossover temperatures. Unphysical singularity in the high temperature result $p_\mathrm{HT}(T)$ is manifest while the consistent calculation taking into account the lightness of the fundamental scalar, denoted by $p_\mathrm{PT}(T)$ in the graph, is seen to behave well.}
\label{fig:su2hprespt}
\end{center}
\end{figure}

As we decrease the temperature, we eventually reach the crossover transition and the expression used for pressure at high temperatures (denoted now as $p_\mathrm{HT}(T)$) ceases to be valid. This is seen in Fig.~\ref{fig:su2hprespt} in which we see an unphysical singularity for the pressure when plotted using the high temperature expression given in Eq.~(\ref{eq:su2hpres}). The correct result for the pressure in that region, $p_\mathrm{PT}(T)$, defined by Eq.~(\ref{eq:su2hprespt}), is given by the solid curve in Fig.~\ref{fig:su2hprespt}. However, the singular behavior of $p_\mathrm{HT}(T)$ is isolated to a narrow range of temperatures. This can be understood by considering the leading order terms responsible for the singularity, the terms of the form $\sim m_D^2/m_3$ in the expansion of pressure. We get for the singular terms
\begin{eqnarray}
\frac{p_\mathrm{singular}(T)}{p_0(T)} & = & \frac{135}{4096\pi^5}\frac{g^6}{\sqrt{\frac{3}{8}g^2+\lambda}}\sqrt{\frac{T_0}{\delta T}}, \label{eq:singterms}
\end{eqnarray}
where $T_0$ is the temperature such that $m_3(T_0) = 0$ and $\delta T$ is the deviation from that, $\delta T = T - T_0$. Since the numerical factor of this term is very small, the temperature must be very close to $T_0$ for any effects from this term to be manifest.

\begin{figure}
\begin{center}
\begin{fmfgraph}(80,80)
        \fmfi{dashes}{fullcircle scaled 40 shifted (40,40) }
        \fmfi{vanilla}{fullcircle scaled 20 shifted (40,10) }
        \fmfi{vanilla}{fullcircle scaled 20 shifted (40,70) }
        \fmfi{vanilla}{fullcircle scaled 20 shifted (10,40) }
        \fmfi{vanilla}{fullcircle scaled 20 shifted (70,40) }
\end{fmfgraph}
\caption{The type of diagrams that are resummed to the pressure $p_\mathrm{PT}$. The dashed line corresponds to the fundamental scalar and the solid lines to the adjoint scalars.}
\label{fig:resumdiags}
\end{center}
\end{figure}
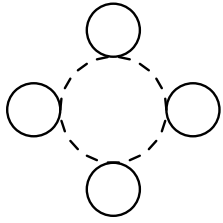

We can also see that $p_\mathrm{PT}(T)$ runs consistently somewhat below the high temperature result. This is due to the fact that when inserting the mass $\widetilde{m}_3^2$ from Eq.~(\ref{eq:ptmass}) to the expansion of the pressure $p_{\mathrm{M}2}(T)$ in Eq.~(\ref{eq:pm2pres}), we effectively resum the class of diagrams in Fig.~\ref{fig:resumdiags} to the pressure. The high temperature result, instead, would correspond to expanding the mass $\widetilde{m}_3^2$ in powers of $h_3 m_D/m_3^2$, the leading order given by $\widetilde{m}_3^2 = m_3^2$, when inserted to the expansion of pressure. At high temperatures, this expansion can be carried out since then $h_3 m_D/m_3^2 \sim g \ll 1$, but near the crossover $h_3 m_D/m_3^2 \sim 1$ according to our power counting rules and such expansion cannot be performed then.

\begin{fmffile}{resum_diags}

\end{fmffile}

\begin{figure}[!htb]
\begin{center}
\vspace{-1.5cm}
\includegraphics[width=0.8\textwidth]{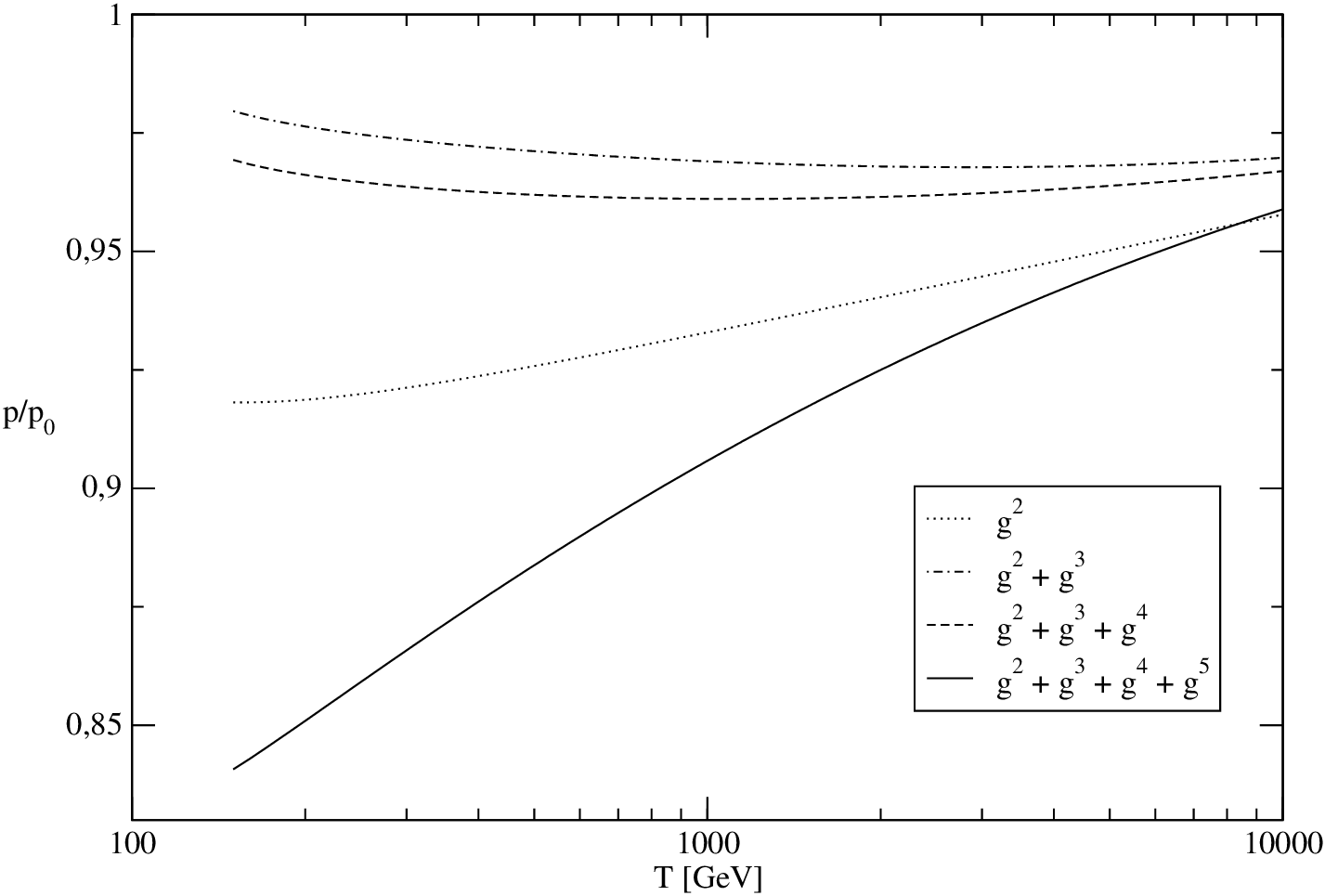}
\caption{Pressure of the standard model at different orders of perturbation theory.}
\label{fig:presSM}
\vspace{2.5cm}
\includegraphics[width=0.8\textwidth]{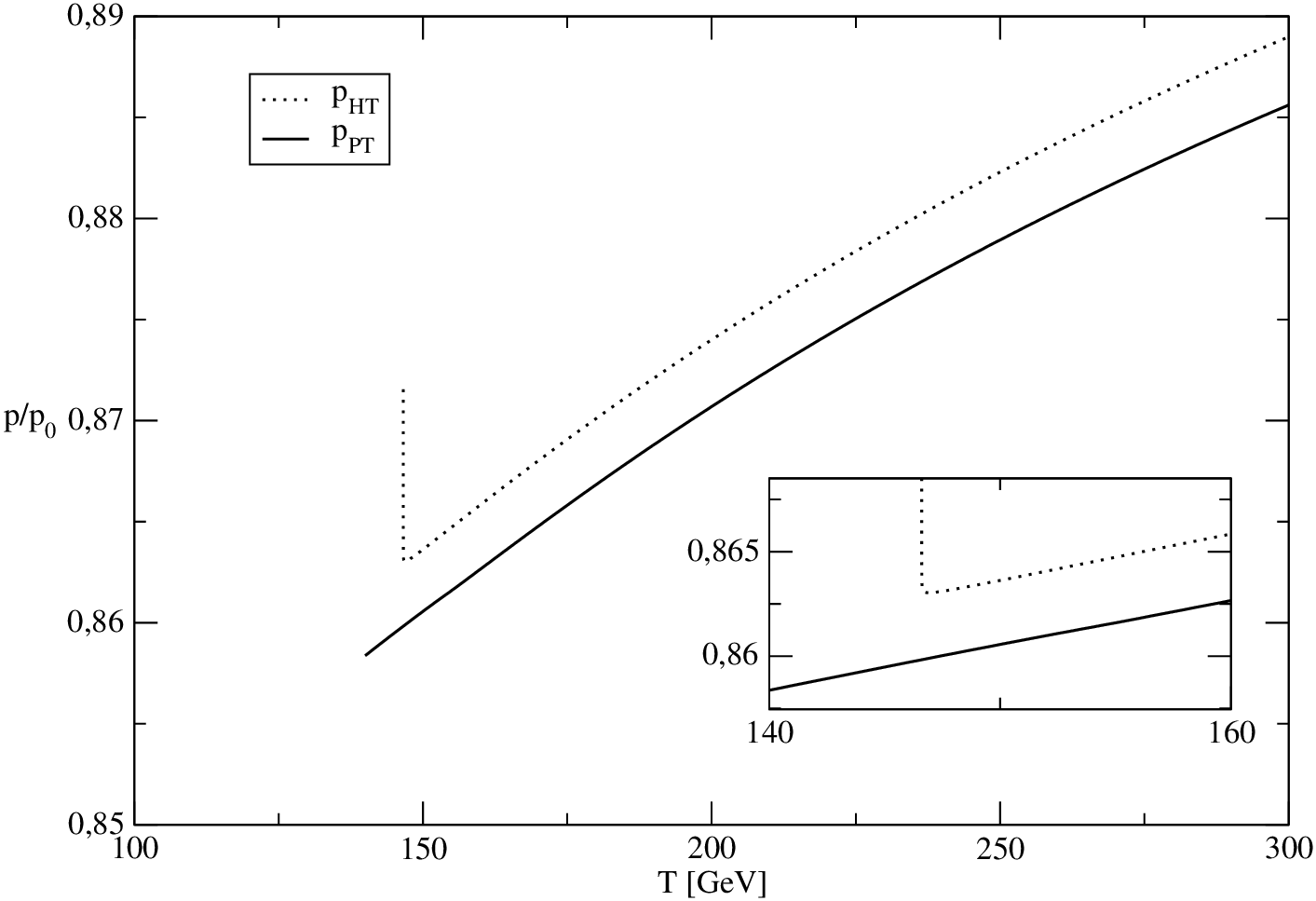}
\caption{Pressure of the standard model near the electroweak crossover.}
\label{fig:prescrossover}
\end{center}
\end{figure}

Let us next consider the pressure of the full standard model. The expansion in powers of the coupling constants is given in appendix \ref{app:presEW}. The pressure, including the QCD contribution, is plotted at different orders of perturbation theory in Fig.~\ref{fig:presSM}, normalized again to the pressure of ideal gas of massless particles, 
\begin{equation}
p_0(T) = 106.75 \frac{\pi^2}{90} T^4. 
\end{equation}
It is seen to significantly deviate from the ideal gas result, by up to $15$\%.  The convergence of the expansion is also much worse than that of the SU($2$) + Higgs theory. This was to be expected since there are two large couplings in the theory, the strong coupling constant $g_s^2(m_Z) \approx 1.5$ and the top Yukawa coupling $g_Y^2(m_Z) \approx 1.4$. The scale dependence of the result is not plotted here since one does not expect it to differ from the QCD results \cite{Arnold:1995eb,*Arnold:1994ps,Braaten:1996jr}: QCD degrees of freedom constitute the majority of all the degrees of freedom in the standard model and, moreover, the scale dependence coming from running of $g_s$ has by far the largest effect.

The pressure near the electroweak crossover is plotted in Fig.~\ref{fig:prescrossover}, similarly normalized. The unphysical singularity in the pressure when plotted using the expression valid at high temperatures is again manifest, but the calculation that takes into account that the fundamental scalar is light around the crossover behaves smoothly. Note that the singular behavior is again isolated to a very narrow temperature range, even more so than in the SU($2$) + Higgs theory. The reason is that this effect stems from the fundamental scalar sector which, as already noted, carries just a tiny fraction of all the degrees of freedom and thus the numerical factor in front of the singular terms (normalized to the ideal gas pressure) is even smaller (see Eq.~(\ref{eq:singterms})).

\begin{figure}[!tb]
\begin{center}
\includegraphics[width=0.65\textwidth]{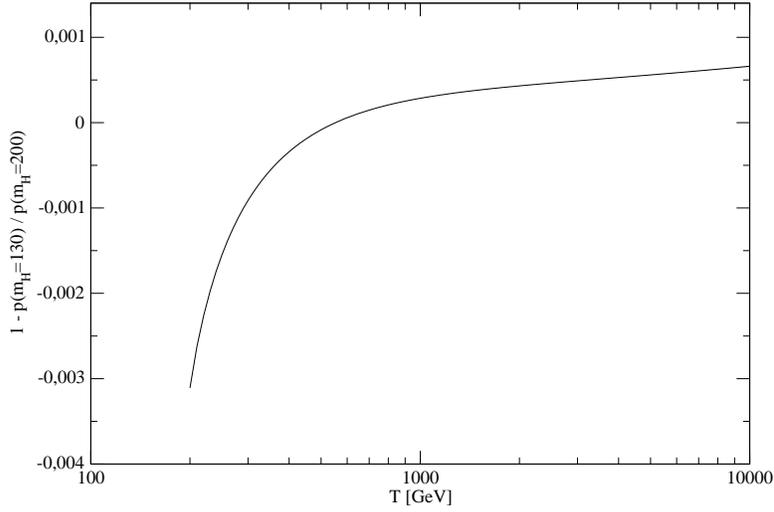}
\caption{Relative difference between pressures at $m_H = 130$~Gev and $m_H = 200$~GeV.}
\label{fig:higgsmassdep}
\end{center}
\hspace{1cm}
\end{figure}

We can study the effects that the scalar sector has on the pressure by altering the mass of the Higgs boson. Plotted in Fig.~\ref{fig:higgsmassdep}, we see that changing the Higgs mass from $130$~GeV to $200$~GeV has a minimal effect on the pressure, again reflecting the small relative number of degrees of freedom. It is then not necessary to repeat the numerical analysis here for a number of different Higgs masses.

\begin{figure}[!tb]
\begin{center}
\includegraphics[width=0.65\textwidth]{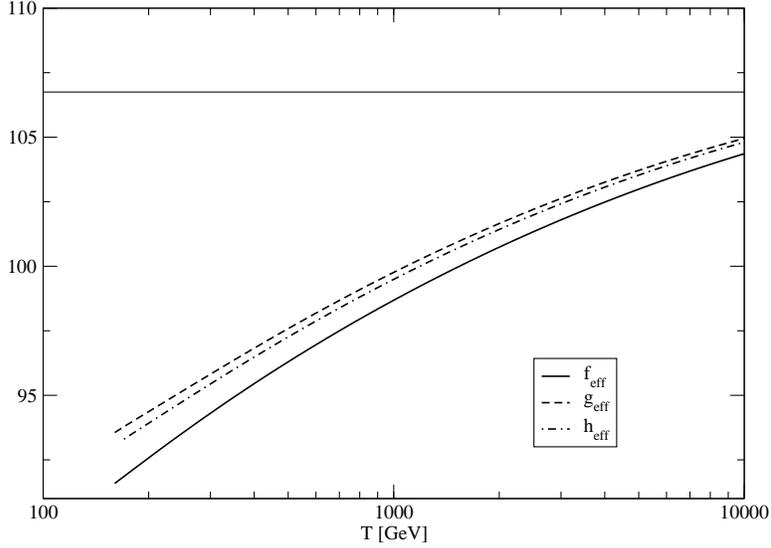}
\caption{The effective number of degrees of freedom in terms of pressure, energy density and entropy density, as defined in the text. For ideal gas each of these would be equal to $106.75$.}
\label{fig:otherplots}
\end{center}
\end{figure}

Using the result for the pressure, we can compute also other thermodynamic variables. In Fig.~\ref{fig:otherplots} we have plotted the effective number of bosonic degrees of freedom in terms of pressure ($f_\mathrm{eff}$), energy density ($g_\mathrm{eff}$) and entropy density ($h_\mathrm{eff}$). Those are defined by
\begin{equation}
\begin{array}{rclrclrcl}
p(T) & = & {\displaystyle f_\mathrm{eff} \frac{\pi^2}{90}T^4}, & \epsilon(T) & = & {\displaystyle g_\mathrm{eff} \frac{\pi^2}{30}T^4}, & s(T) & = & {\displaystyle h_\mathrm{eff} \frac{2\pi^2}{45}T^3}
\end{array}
\end{equation} 
and for ideal gas their values would be equal, $f_\mathrm{eff}^\mathrm{id.} = g_\mathrm{eff}^\mathrm{id.} = h_\mathrm{eff}^\mathrm{id.} = 106.75$. As can be seen, the effect of interactions on the energy and entropy densities is comparable to that they have on the pressure. 

\begin{figure}[!tb]
\begin{center}
\includegraphics[width=0.8\textwidth]{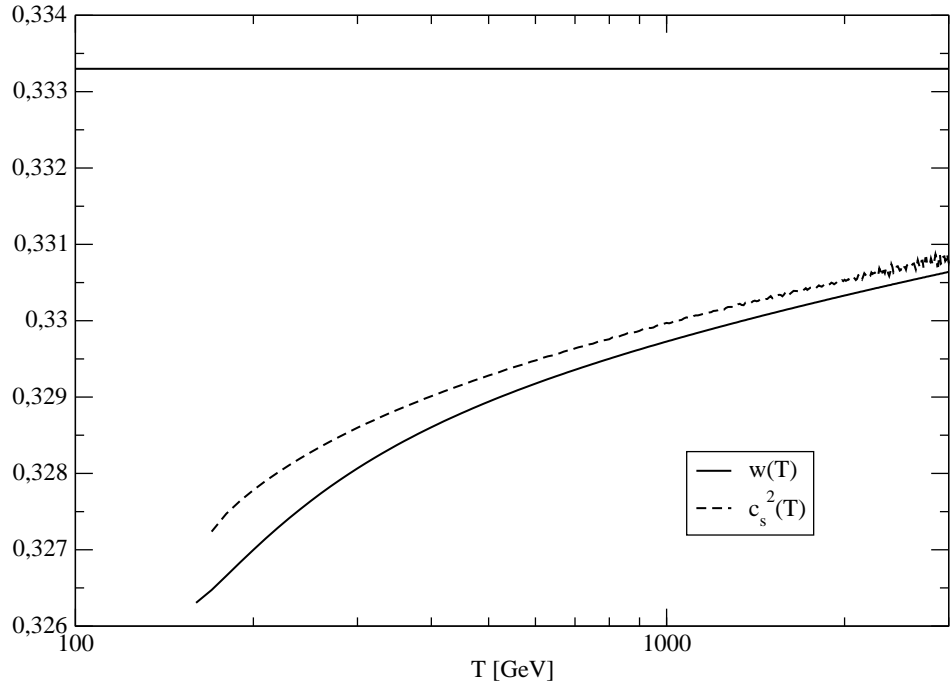}
\caption{The equation of state parameter $w(T)$ and the speed of sound $c_s^2(T)$ for the standard model, along with the ideal gas result $w(T) = c_s^2 = 1/3$. }
\label{fig:eqofstate}
\end{center}
\end{figure}

Finally, in Fig.~\ref{fig:eqofstate} we plot the equation of state parameter $w(T)$ and the speed of sound, $c_s^2(T) = p'(T)/\epsilon'(T)$ (where $'$ refers to derivative with respect to temperature). They both equal $1/3$ for ideal gas. Unlike for pressure, energy density and entropy density, the effect of interactions on these variables is observed to be rather weak. We do see deviation from the ideal gas results, but the effect is not much more than 2\%. We can thus deduce that though the thermodynamic potentials do significantly deviate from the ideal gas results, the matter still seems to behave very much like radiation.

In this thesis, the computation of the pressure is carried out assuming the system resides in the symmetric phase of the theory. However, it is of phenomenological interest to evaluate the pressure also in the broken symmetry phase. This can be done by extending the effective potential calculations to the corresponding precision, but it is a highly non-trivial task due to the complicated structure of the theory in the broken phase. Phenomenological estimates for the pressure at temperatures between the electroweak scale and the QCD scale have, however, been presented \cite{Laine:2006cp}.

\chapter{The electroweak phase diagram at finite chemical potentials}
\label{cha:phase_diagram}

As discussed in the introduction, much of the research on electroweak thermodynamics has concentrated on studying the properties of the electroweak phase transition. In this chapter, we will review the effect of finite chemical potentials related to conserved fermion numbers on the phase diagram. The motivation is to understand the phase structure of the electroweak theory in more general terms, but the computation may have phenomenological interest as well. Although the baryon number asymmetry in the universe can be estimated with observations (baryon-to-photon ratio is of the order of $10^{-10}$), measuring the lepton number asymmetry in the universe is more difficult since neutrinos interact only weakly with other particles. Consequently, a large lepton number asymmetry, residing in neutrinos, has presently not been ruled out. For this reason, we will in the next section briefly review some aspects of neutrino cosmology. We then move on to consider the computation of the phase diagram of the electroweak theory in presence of finite chemical potentials.

\section{Lepton asymmetry in the universe}

Neutrinos interact with other particles only via weak interactions and for this reason they decouple from thermal evolution of the rest of the universe at an early stage. Assuming the neutrinos are massless, the interaction rate of processes keeping the neutrinos in thermal contact with other particles (such as $\nu e \leftrightarrow \nu e$ and $\bar{\nu}\nu \leftrightarrow \bar{e}e$) is roughly given by $\Gamma = n\sigma \sim G_\mu^2 T^5$ where $n$ is the number density of neutrinos and we have estimated the cross section $\sigma$ to behave as $\sigma \sim G_\mu^2 T^2$. This rate is of the same order as the expansion rate of the universe when the temperature is about $1$~MeV, assuming that there is no large asymmetry between the numbers of neutrinos and antineutrinos. After decoupling, neutrinos retain a thermal distribution, but the temperature $T_\nu$ of the neutrinos, decreasing as the universe expands according to $T_\nu \sim 1/R$, may differ from the temperature of the rest of the universe (photons). Indeed, shortly after the neutrino decoupling, the temperature of the universe drops below the electron mass threshold and entropy carried by the electron-positron pairs is transferred to photons. Consequently, the temperature of photons will be boosted by a factor of $(11/4)^{1/3}$ compared to the temperature of the neutrino background. Since the temperature of the microwave background radiation today is about $2.735$~K, we get that the corresponding temperature of the neutrino background is about $1.95$~K. This corresponds to a neutrino background of about $n_\nu \approx n_{\bar{\nu}} \approx 56$ neutrinos per $\mathrm{cm}^3$ in the universe today.

However, there are no direct observations concerning the neutrino degeneracy, \emph{i.e.} asymmetry between neutrinos and antineutrinos, in the universe. Upper limits for the neutrino degeneracy can be obtained by considering big bang nucleosynthesis (BBN) which is sensitive to a degeneracy in electron neutrinos $\nu_e$. Excess of $\nu_e$ with respect to $\bar{\nu}_e$ will induce changes in the beta reactions leading to lower neutron to proton ratio on which the abundances of primordially produced light elements depend. This constrains the chemical potential of $\nu_e$ to lie between $-0.01 < \mu_{\nu_e}/T_{\nu_e} < 0.22$, where $T_{\nu_e}$ is the temperature of electron neutrino background. Asymmetry in the $\mu$ and $\tau$ neutrinos affects the BBN by hastening the expansion of the universe and this bounds the corresponding chemical potentials by $|\mu_{\nu_{\mu,\tau}}/T_{\nu_{\mu,\tau}}| < 2.6$ \cite{Hansen:2001hi}. One can also obtain limits for the neutrino degeneracy by considering the power spectrum of the microwave background radiation. Based on the data from WMAP, Lattanzi, Ruffini and Vereshchagin \cite{Lattanzi:2005qq} find that the neutrino degeneracy is constrained to be $0 < |\mu/T| < 1.1$ and claim that statistical fits, in fact, prefer a large neutrino degeneracy, $\mu/T \simeq 0.6$

There are a number of models that could explain a presence of a large lepton number asymmetry in the universe without leading to a comparable baryon number asymmetry, see for example \cite{Casas:1997gx,McDonald:1999in}. An obvious problem is to avoid the transformation of any generated net lepton number to a comparable net baryon number via sphaleron mediated processes that are unsuppressed when the electroweak symmetry is unbroken. The common argument is that a large lepton number asymmetry would prevent the restoration of the electroweak symmetry even at high temperatures \cite{Linde:1976kh} and thus the sphaleron processes would in fact be suppressed. One can also consider a situation in which there are large net lepton numbers $L_i$ within each family, but in such a way that the sum of the lepton numbers is small, $L_e + L_\mu + L_\tau \ll L_{e,\mu,\tau}$. This requires fine tuning and thus is not natural.  

\section{Dimensional reduction at finite chemical potentials}
\label{sec:dimregatmu}

As discussed previously, the exactly conserved global charges in the standard model are the baryon - lepton number charges,
\begin{eqnarray}
X_i & = & \frac{1}{3}B - L_i,\quad i = e,\mu,\tau,
\end{eqnarray}
where $i$ refers to different families and $B$ and $L_i$ are defined by
\begin{eqnarray}
B & = & \frac{1}{3}\sum_{c,i}\int\mathrm{d}^3\mathbf{x}\;\bar{q}_{c,i}\;\gamma_0\; q_{c,i}, \\
L_i & = & \int\mathrm{d}^3\mathbf{x}\; \left(\bar{e}_i \gamma_0\; e_i + \bar{\nu}_i \gamma_0\; a_L \nu_i\right),\quad a_L = \frac{1}{2}(1-\gamma_5).
\end{eqnarray}
Here $c$ refers to color, $q_{c,i}$ are the quarks and $e_i$ and $\nu_i$ are the electron type lepton and the corresponding neutrino of each family. We assign chemical potentials $\mu_i$ to these conserved charges, but for computational reasons, it is convenient to introduce separate chemical potentials for the baryon number $B$ and for the lepton numbers $L_i$, namely $\mu_B$ and $\mu_{L_i}$. These are then related so that 
\begin{equation}
\begin{array}{rclrcl}
\mu_B & = & {\displaystyle \frac{1}{3}\sum_{i=e,\mu,\tau}\mu_i}, & \mu_{L_i} & = & -\mu_i.
\end{array}
\end{equation}
For this reason, we will refer to the chemical potentials $\mu_i$ as ``leptonic chemical potentials''. Note that each colored quark carries a baryonic chemical potential $\mu_B / 3$.

Additionally, there are conserved gauge charges, but as discussed in chapter \ref{cha:ft}, the corresponding chemical potentials can be absorbed to the temporal components of the gauge fields and are thus not manifest in the path integral \cite{Kapusta:1981aa,Kapusta:1990qc,Khlebnikov:1996vj}. The equilibrium properties of the standard model in the presence of the chemical potentials $\mu_i$ are then described by the partition function
\begin{eqnarray}
\mathcal{Z} & = & \int\mathcal{D}\varphi\exp\left[-S + \int_0^\beta\mathrm{d}\tau\left(\mu_B B + \sum_i\mu_{L_i} L_i\right)\right],
\end{eqnarray}
where $S$ is the Euclidean action of the electroweak theory.

We are now interested in computing the phase diagram of the electroweak theory when the chemical potentials $\mu_i$ are non-zero. A proper approach at high temperatures is again the framework provided by dimensional reduction. As discussed in the previous sections, close to the phase transition the relevant light degrees of freedom are the fundamental scalar $\Phi$ and the magnetostatic gauge bosons $A_i^a,\;B_i$. Construction of the effective theory describing these fields is performed in detail in \cite{Gynther:2003za} and we will merely review the generic features of the computation here.

As pointed out in section \ref{sec:dimred}, introducing chemical potentials to the system will reduce the number of discrete symmetries the theory has (or at least may have) otherwise. Namely, the theory becomes CP and CPT breaking. The effective theory may then contain terms that break these symmetries but which still respect three dimensional gauge- and rotational invariances (for a review on how different fields transform under the discrete symmetries, see \cite{Kajantie:1997ky}). The most general renormalizable theory describing the three dimensional gauge fields and the fundamental scalar in question can then be written as
\begin{eqnarray}
\mathcal{L}_\mathrm{eff.} & = & \frac{1}{4}G_{ij}^a G_{ij}^a + \frac{1}{4}F_{ij}F_{ij} + D_i\Phi^\dagger D_i\Phi + \widetilde{m}_3^2\Phi^\dagger\Phi + \widetilde{\lambda}_3\left(\Phi^\dagger\Phi\right)^2 \nonumber \\
& & + \widetilde{\alpha}\epsilon_{ijk}\left(A_i^a G_{jk}^a - \frac{1}{3}g_3\epsilon^{abc}A_i^a A_j^b A_k^c\right) + \widetilde{\alpha}'\epsilon_{ijk}B_i F_{jk}.
\end{eqnarray}
Here the Chern-Simons terms (terms proportional to $\widetilde{\alpha}$ and $\widetilde{\alpha}'$) break CP and CPT and thus vanish automatically when chemical potentials are not present. However, they appear in the effective theory if we have a finite chemical potential for the non-conserved $B+L$ \cite{Redlich:1985md}. Since we consider only the strictly conserved $B-L$ numbers, the Chern-Simons terms will not be present in the effective theory, \emph{i.e.} $\widetilde{\alpha} = \widetilde{\alpha}' \equiv 0$. The structure of the effective theory is thus the same as it is when the chemical potentials vanish, only the matching of the parameters $\widetilde{g}_3^2,\,\tilde{g}_ 3'^2,\widetilde{\lambda}_ 3$ and $\widetilde{m}_3^2$ to the physical variables changes. The matching is given in \cite{Gynther:2003za}.

\begin{table}[tb]
\begin{centering}
\begin{tabular}{cl}
Dimensionality & Terms \\
\hline
${\displaystyle \textrm{GeV}^{1/2}}$ & ${\displaystyle iB_0}$ \\
${\displaystyle \textrm{GeV}^{3/2}}$ & ${\displaystyle iB_0^3,\;i\Phi^\dagger A_0^a\tau^a\Phi,\;i\Phi^\dagger B_0\Phi,\;i B_0 A_0^a A_0^a}$ \\
${\displaystyle \textrm{GeV}^2}$     & ${\displaystyle \epsilon_{ijk}B_i F_{jk},\;\epsilon_{ijk}\left(A_i^a G_{jk}^a - \frac{1}{3}g\epsilon^{abc}A_i^a A_j^b A_k^c\right)}$ \\
\hline
\end{tabular}
\caption{The dimensionally lowest order terms violating CP and CPT invariances in the electroweak theory.}
\label{tab:cpterms}
\end{centering}
\end{table}

When constructing the theory describing the light degrees of freedom, we again need to first develop an effective theory containing the electrostatic modes $A_0^a$ and $B_0$ as well. Unlike the theory describing the light degrees of freedom, the structure of the electrostatic effective theory does change when chemical potentials are introduced. The dimensionally lowest order terms (that are not present when $\mu_i=0$) that must be taken into account in this theory are listed in table \ref{tab:cpterms} for the standard model. The Lagrangian of the electrostatic effective theory is then
\begin{eqnarray}
\mathcal{L}_\mathrm{E} &=& \frac{1}{4}G_{ij}^a G_{ij}^a +\frac{1}{4}F_{ij}F_{ij} +(D_i\Phi)^\dagger(D_i\Phi)
        +m_3^2\Phi^\dagger\Phi +\lambda_3(\Phi^\dagger\Phi)^2 \nonumber \\
        &&+\half(D_i A_0^a)^2 +\half\mD^2 A_0^a A_0^a +\frac{1}{4}\lambda_A (A_0^a A_0^a)^2 +\half(\partial_i B_0)^2
        +\half\mD'^2 B_0 B_0 \nonumber \\
        && +h_3\Phi^\dagger\Phi A_0^a A_0^a +h_3'\Phi^\dagger\Phi B_0 B_0
        -\half g_3 g_3' B_0\Phi^\dagger A_0^a \tau^a\Phi \nonumber \\
	&& + \kappa_1 B_0 +\kappa_3 B_0^3 + \rho\Phi^\dagger A_0^a\tau^a \Phi + \rho'\Phi^\dagger\Phi B_0 + \rho_G B_0 A_0^a A_0^a \nonumber \\
& & + \alpha\epsilon_{ijk}\left(A_i^a G_{jk}^a - \frac{1}{3}g_3\epsilon^{abc}A_i^a A_j^b A_k^c\right) + \alpha'\epsilon_{ijk}B_i F_{jk}
\end{eqnarray}
The coefficients for the Chern-Simons terms vanish identically again. Of special importance is the term $\kappa_1 B_0$. Having such linear terms in the Lagrangian leads, in equilibrium, to condensates of the corresponding field, to leading order 
\begin{eqnarray}
\langle B_0\rangle & = & -\frac{\kappa_1}{\mD'^2} \neq 0.
\end{eqnarray} 
As discussed before, this implies a finite chemical potential for the hypercharge. The linear term $\kappa_1 B_0$ in the Lagrangian thus ensures neutrality of the system with respect to hypercharge. Performing the matching (see \cite{Gynther:2003za} for details) we get to leading order for $\kappa_1$:
\begin{eqnarray}
\kappa_1 & = & -\frac{i\pi}{3}g' T^{5/2}\left[\sum_{i}\frac{\mu_{L_i}}{\pi T}\left(1+\frac{\mu_{L_i}^2}{\pi^2T^2}\right) - \frac{\mu_B}{\pi T}\left(1+\frac{1}{9}\frac{\mu_B^2}{\pi^2T^2}\right)\right].
\end{eqnarray} 
Matching of all the other terms to physical parameters is given to high accuracy in \cite{Gynther:2003za}.

\section{The phase diagram}

The electroweak phase transition is described by the effective theory $\mathcal{L}_\mathrm{eff.}$ (with $\widetilde{\alpha} = \widetilde{\alpha}' = 0$) introduced in the previous section. Due to infrared divergences encountered in perturbative calculations, this theory must be solved with numerical computations. To this end, lattice Monte Carlo studies have been carried out \cite{Kajantie:1995kf,Kajantie:1996mn,Kajantie:1996qd,Karsch:1996yh,Gurtler:1997hr} and the phase diagram of the theory has been reliably solved.\footnote{We can apply the same numerical computations now, since the structure of the effective theory remains the same when finite chemical potentials are introduced.} In practice, one can neglect the U($1$) subgroup and just consider SU($2$) + Higgs theory \cite{Kajantie:1996qd}.

It is convenient to introduce dimensionless quantities
\begin{equation}
\begin{array}{rclrclrcl}
x & = & {\displaystyle \frac{\widetilde{\lambda}_3}{\widetilde{g}_3^2}}, & y & = & {\displaystyle \frac{\widetilde{m}_3^2(\widetilde{g}_3^2)}{\widetilde{g}_3^4}}, & z & = & {\displaystyle \frac{\widetilde{g}_3'^2}{\widetilde{g}_3^2}}
\end{array}
\end{equation}
and let the coupling $\widetilde{g}_3^2$ to provide the dimensions. Of these, $z$ is essentially constant, $z\approx 0.3$ ($=0$ for SU($2$) + Higgs theory), and $y$ determines the phase of the system: at tree level, if $y > 0$ the system resides in the symmetric phase while if $y < 0$ the symmetry is spontaneously broken. Thus, at tree level the critical temperature is determined by setting $y=0$. The remaining variable, $x$, depending on the value of the Higgs mass, parametrizes the theory. 

\begin{figure}[!tb]
\begin{center}
\includegraphics[width=0.87\textwidth]{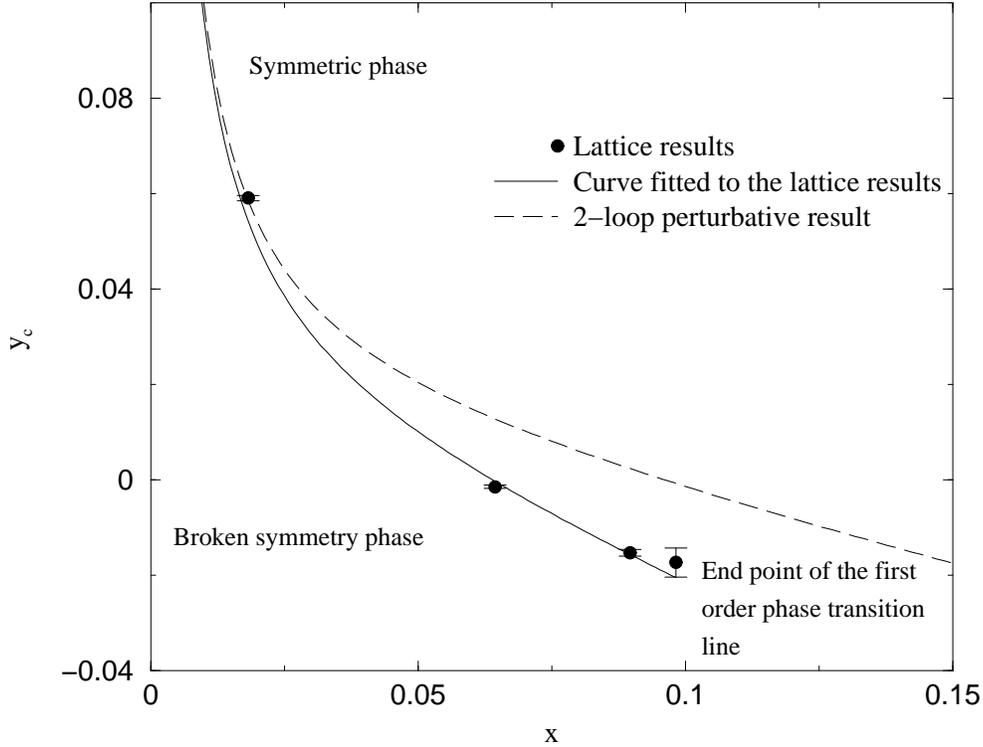}
\caption{The phase diagram of the effective theory in terms of the dimensionless variables $x$ and $y$. The dashed line corresponds to the perturbative result based on two-loop effective potential calculations.}
\label{fig:xyphdg}
\end{center}
\end{figure}

The phase diagram of the theory in terms of the dimensionless variables $x$ and $y$ is plotted in Fig.~\ref{fig:xyphdg}. Both the perturbative result (based on two-loop effective potential calculations in the effective theory, \cite{Farakos:1994kx}) and numerical results are presented. The perturbative result and the numerical results agree well for small values of $x$ (which amounts to small Higgs masses and chemical potentials, as will be discusses shortly), but while perturbative computations suggest that there is a first order phase transition for all values of $x$, numerical studies show that there is an endpoint to the critical curve at $x\approx 0.1$ above which there is only a crossover transition. 

In order to map the phase diagram in Fig.~\ref{fig:xyphdg} to a phase diagram in terms of temperature, chemical potentials and the Higgs mass, we need to apply the matching between the parameters of the effective theory and the physical $4$d theory, given in detail in \cite{Gynther:2003za}. The critical temperature $T_c = T_c(\mu_i,m_H)$ is then given by $y(T_c,\mu_i,m_H) = y_c(x(T_c,\mu_i,m_H))$ using $y_c(x)$ in Fig.~\ref{fig:xyphdg}. In order to simplify the study, we assume all the leptonic chemical potentials to be equal to each other, $\mu_i = \mu\;\forall i,\;i= e,\mu,\tau$. Then, to leading order
\begin{eqnarray}
x(T,\mu,m_H) & = & \frac{m_H^2}{8m_W^2} + \frac{1}{g^2}\frac{96}{1331}\frac{\mu^2}{T^2}, \\
y(T,\mu,m_H) & = & -\frac{m_H^2}{2g^4T^2} + \frac{1}{g^2}\left(\frac{m_H^2}{16m_W^2} + \frac{3}{16} + \frac{g'^2}{16g^2} + \frac{g_Y^2}{4g^2}\right) - \frac{1}{g^4}\frac{16}{121}\frac{\mu^2}{T^2}.
\end{eqnarray}
These expressions can be used to qualitatively understand the effect of the chemical potentials on the phase diagram. As can be seen, the chemical potentials lead to decreasing the value of $y$ and increasing the value of $x$. Thus, chemical potentials tend to break the symmetry of the theory and, consequently, the critical temperature will increase. This is in accord with results already obtained by Linde \cite{Linde:1976kh}. Moreover, the increasing value of $x$ implies that the transition becomes weaker and that for sufficiently large chemical potentials there will be no phase transition at all, regardless of the value of the Higgs mass.

\begin{figure}[!tb]
\begin{center}
\includegraphics[width=0.87\textwidth]{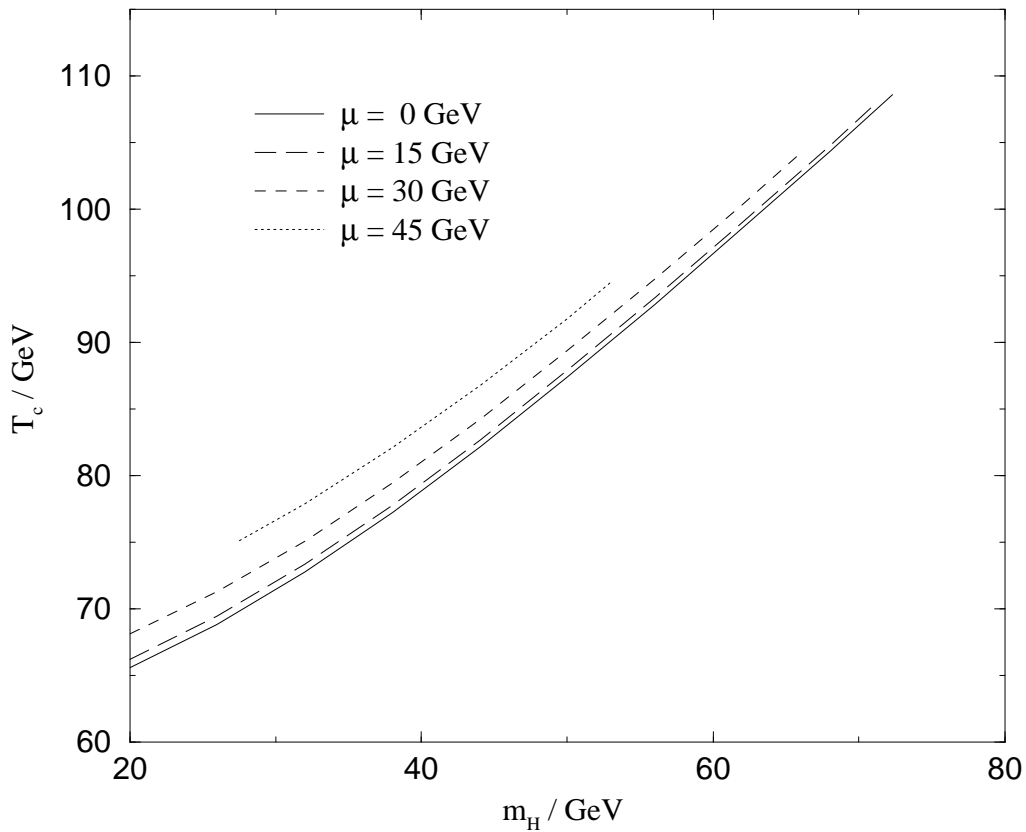}
\caption{The phase diagram of the electroweak theory in terms of the Higgs mass and temperature.}
\label{fig:phdgrmt}
\end{center}
\end{figure}

\begin{figure}[!tb]
\begin{center}
\vspace{-1.5cm}
\includegraphics[width=0.87\textwidth]{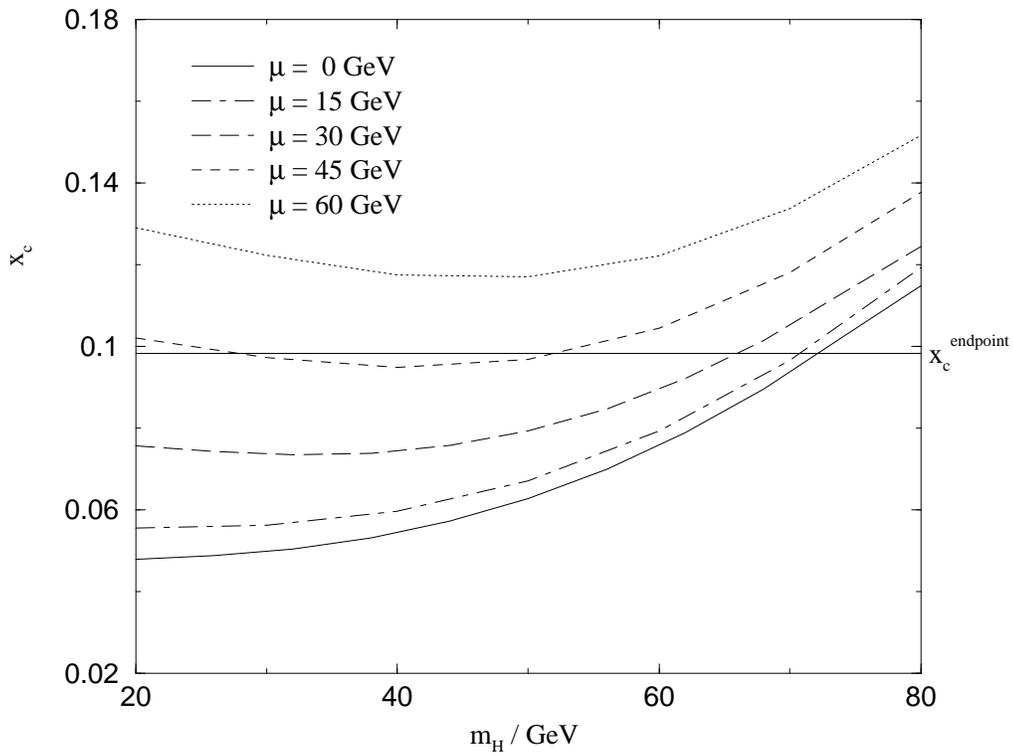}
\caption{The value of $x_c = x(T_c,\mu,m_H)$ as a function of $m_H$ for a number of different values of the chemical potential $\mu$.}
\label{fig:xcmh}
\end{center}
\end{figure}

These qualitative considerations are verified by the more accurate treatment of matching the phase diagram of the effective theory in terms of physical parameters. The electroweak phase diagram in terms of the Higgs mass and temperature is plotted in Fig.~\ref{fig:phdgrmt} for a number of different values of the chemical potential $\mu$. The location of the endpoint of the first order phase transition line is seen to move to smaller values of the Higgs mass as the chemical potentials are increased, indicating that the chemical potentials make the transition weaker. At the same time, the critical temperature is slightly increased. Note also that, for some values of $\mu$, there is a first order phase transition only if the Higgs mass is within some range $m_H^\textrm{lower limit} < m_H < m_H^\textrm{upper limit}$. The reason is that, for a fixed and sufficiently large value of the chemical potential, the value of $x$ along the first order phase transition line\footnote{Of course, there is no first order phase transition line if $x > x_\mathrm{endpoint} \approx 0.1$. We can, however, formally continue the curve fitted to the lattice results in Fig.~\ref{fig:xyphdg} to values $x>x_\mathrm{endpoint}$ and we refer to that continuation as the critical curve in that region of the parameter space. One should not attach any physical meaning to such a continuation, it merely serves as a convenient tool for analysis.}, $x_c=x(T_c,\mu,m_H)$, is actually a decreasing function of $m_H$ for small values of $m_H$, see Fig.~\ref{fig:xcmh}. Thus, although $x(T_c,\mu,m_H) > x_\mathrm{endpoint} \approx 0.1$ for small values of $m_H$ when $\mu$ is large enough, it may decrease so much as $m_H$ is increased that it becomes smaller than $x_\mathrm{endpoint}$ for some values of $m_H$. 

\begin{figure}[!tb]
\begin{center}
\includegraphics[width=0.87\textwidth]{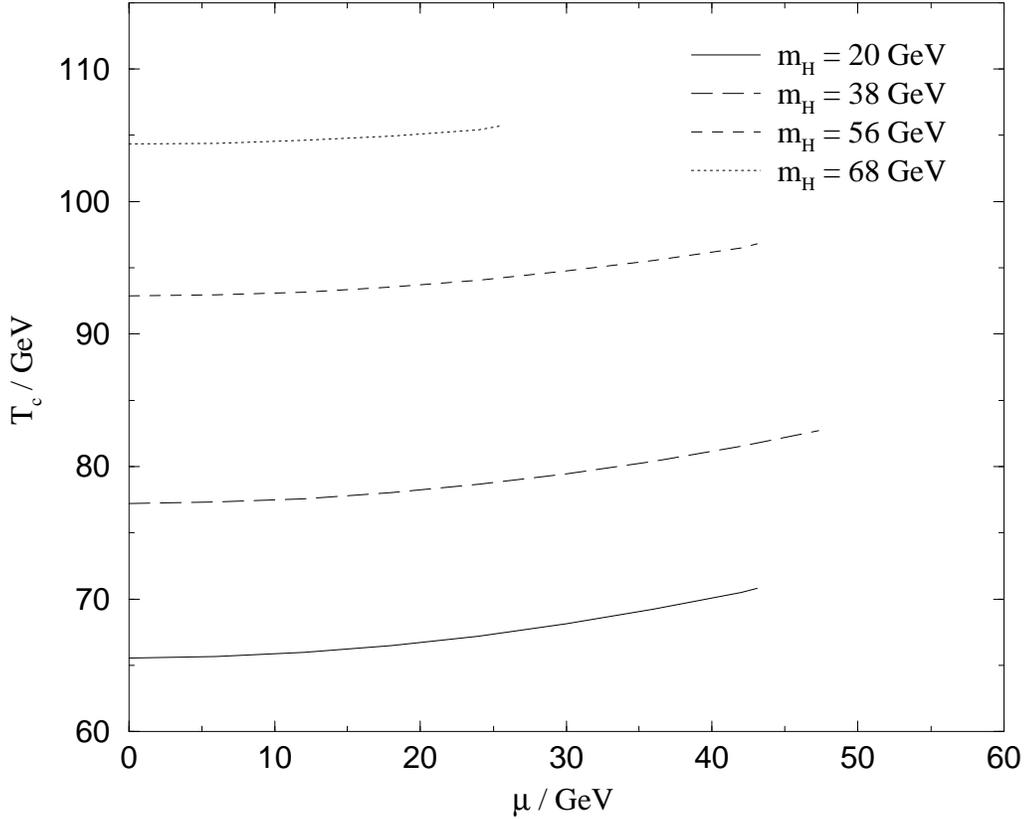}
\caption{The phase diagram of the electroweak theory in terms of the chemical potential and temperature.}
\label{fig:phdgrmut}
\end{center}
\end{figure}

In Fig.~\ref{fig:phdgrmut} we plot the phase diagram in terms of the chemical potential and temperature. As can be seen, there is again an endpoint to the first order phase transition lines, regardless of the value of the Higgs mass. If the chemical potential has an absolute value of $\mu \gtrsim 50$~GeV, there is no phase transition in the electroweak theory for any value of the Higgs mass, just a crossover transition. Finally, in table \ref{tab:endpoint} we list the values of the Higgs mass and critical temperature corresponding to the endpoint of the first order phase transition line for a number of different values of the chemical potential.

\begin{table}[!tb]
\begin{center}
\begin{tabular}{|l|l|l|}
\hline
$\mu$ & $m_H^\mathrm{endpoint}$ & $T_c^\mathrm{endpoint}$ \\ \hline
$0$~GeV & $72$~GeV & $109$~GeV \\ \hline
$15$~GeV & $71$~GeV & $108$~GeV \\ \hline
$30$~GeV & $66$~GeV & $104$~GeV \\ \hline
$45$~GeV & $52$~GeV & $94$~GeV \\ \hline
\end{tabular}
\caption{The location of the endpoint of the first order phase transition line for a number of different values for the chemical potential.}
\label{tab:endpoint}
\end{center}
\end{table}

There is a simple physical interpretation of the results. Introducing a finite chemical potential for the conserved fermion numbers will in general lead to a non-zero (hyper)charge density in the system. However, the system must be neutral with respect to gauge charges in equilibrium and thus a finite chemical potential for the hypercharge is generated to guarantee this. This effectively induces a finite chemical potential for the scalar doublet as well since it carries a hypercharge. As well known, a finite chemical potential for bosons leads to Bose-Einstein condensation\footnote{For relativistic bosons, condensation happens when $|\mu| = m$, where $\mu$ is the chemical potential and $m$ is the mass of the bosons. For non-relativistic bosons $\mu \leq 0$ and condensation happens in the limit $\mu \rightarrow 0$.} and thus, in this case, to symmetry breaking. To overcome this, even larger temperatures are needed to restore the symmetry of the theory.

The computation of the phase diagram is by construction limited to small chemical potentials. The reason is that, in principle, dimensional reduction requires there to be a large scale hierarchy and the temperature to be the largest scale in the system. If there are other comparable scales, then the effective theories may contain non-renormalizable terms that are not suppressed by any large scale hierarchy and which therefore cannot be excluded from the theory. This would spoil dimensional reduction. 

Within QCD, it has been studied in more detail how large chemical potentials one can consider in this framework. Perhaps surprisingly, it turns out that one can allow the chemical potentials to be as large as $\mu \lesssim 4T$ \cite{Hart:2000ha}. Vuorinen \cite{Vuorinen:2003fs}, computing the QCD pressure and comparing calculations in the limit $\mu/T \ll 1$ and at $T = 0$, also concludes that dimensional reduction seems to work well for very large values of $\mu/T$. Parametrically, dimensional reduction works for $T\ll \mu$ if the mass scales of the effective theories are smaller than the scales that were integrated out, $\pi^2 T^2 \gg m_D^2 \sim g_s^2\left(T^2 + \mu^2/\pi^2\right) \sim g_s^2\mu^2/\pi^2$, \emph{i.e.}, for $\mu/T \ll \pi^2/g_s$. However, whether or not similar conclusions can be drawn within electroweak physics is an open question.

\chapter{Comparison of electroweak and QCD thermodynamics}

In this thesis we have concentrated on electroweak thermodynamics. However, also finite temperature QCD is relevant, both from theoretical and phenomenological viewpoints: with ongoing experiments at RHIC in Brookhaven (and in near future at LHC at CERN) we are in a position to create and study strongly interacting matter \cite{Ritter:2004xj}. In this chapter, we will briefly consider the similarities and differences between the QCD and electroweak thermodynamics. 

\section{Comparison of the pressures}

Perturbative evaluation of the pressure of QCD with massless quarks has a long history \cite{Braaten:1996jr,Shuryak:1977ut,Chin:1978gj,Kapusta:1979fh,Toimela:1982hv,Arnold:1995eb,*Arnold:1994ps,Zhai:1995ac,Kajantie:2002wa,Kajantie:2003ax} and a number of theoretical breakthroughs has been achieved while extending the expansion further (for a brief review on the history of this topic, see \cite{Vuorinen:2004rd}). Today, the expansion is known to the last perturbatively calculable term \cite{Kajantie:2002wa,Vuorinen:2003fs}. The generic structure of the expansion, as far as perturbative computations can extend, is
\begin{eqnarray}
\frac{p_\mathrm{QCD}(T)}{T^4} & = & c_0 + c_2 g_s^2 + \left(c_4 + c_4'\ln g_s\right)g_s^4 + c_5 g_s^5 + c_6' g_s^6\ln g_s  + \mathcal{O}\left(g_s^6\right).
\end{eqnarray}
Computation of the order $g_s^6$ term cannot anymore be carried out within perturbation theory \cite{Linde:1980ts,Gross:1981br} and to obtain it requires numerical input. First steps in computing it have already been taken \cite{Hietanen:2004ew}.

Some general observations can now be made of the expansion. Although, in principle, there is a mass scale inherent to QCD, the Landau pole $\Lambda_\mathrm{QCD}$,
\begin{eqnarray}
\frac{1}{g_s^2(\Lambda)} & = & \frac{1}{16\pi^2}\left(22-\frac{4}{3}N_f\right)\ln\frac{\Lambda}{\Lambda_\mathrm{QCD}}\quad\textrm{(to one-loop)},
\end{eqnarray}
it only arises in quantum theory. At tree level the only mass scale is provided by temperature. Hence, the expression for the perturbative expansion of pressure deviates from the form $p(T) \sim T^4$ only by logarithmic terms. This should be compared to the expression for the pressure in the full standard model. There we have another explicit mass scale at tree level and, consequently, the expansion of the pressure has a different polynomial structure altogether, to leading order $p(T) \sim T^4 + \nu^2 T^2 + \nu^4 + \mathcal{O}(\nu^6)$. The complete expansion has an even richer structure, containing roots and logarithms of linear combinations of $\nu^2$ and $T^2$, as seen in chapter \ref{cha:pressure}. Similar structure applies to the couplings as well: in QCD with massless quarks there is only one coupling constant, $g_s$, and thus the expansion is a polynomial in powers and logarithms of $g_s$. In the electroweak theory there are multiple couplings, leading to non-trivial combinations of them and thus the expansion is not a simple polynomial in the couplings and their logarithms. As a more subtle difference, there are no terms of the order $g_s^5\ln g_s$ in the expansion of QCD pressure. This is related to the fact that the emerging effective theories are finite to the order needed and thus require no renormalization. In the electroweak case, the fundamental scalar mass runs in the effective theories at order $g_3^4$ and this leads to terms of the order $g^5\ln g$ in the expansion.

The perturbative expansion of QCD pressure converges poorly due to the magnitude of the strong coupling. Especially, evaluating the pressure near the QCD phase transition is in principle impossible by using perturbative methods since the critical temperature of the transition is close to the Landau pole of the coupling. The expansion of the electroweak pressure behaves much better in this sense: the theory is weakly coupled and the expansion appears to converge well. Moreover, electroweak pressure can be studied with perturbative computations even near the electroweak phase transition since the transition (crossover) is driven by the Higgs scalar and not by the theory becoming confining. However, as was seen previously, even in the electroweak case it cannot be ruled out that higher order corrections have a comparable contribution to the pressure.

\section{The phase diagrams}

Both QCD and the electroweak theory contain a phase transition. The exact properties and critical temperatures of the transitions depend on the chemical potentials and the values of parameters of the theories. In this section, we will compare the phase diagrams of these theories. The electroweak phase diagram is considered in terms of the leptonic chemical potentials and the theory is parametrized by the Higgs mass, while the QCD phase diagram \cite{Fodor:2001pe,*Fodor:2004nz,deForcrand:2002ci,Karsch:2001vs} is considered in terms of the baryonic chemical potential and the theory is parametrized in terms of the strange quark mass (\emph{i.e.} whether the system behaves as $N_f=2$ of as $N_f=3$). 

\begin{figure}[!tb]
\begin{center}
\includegraphics[width=0.87\textwidth]{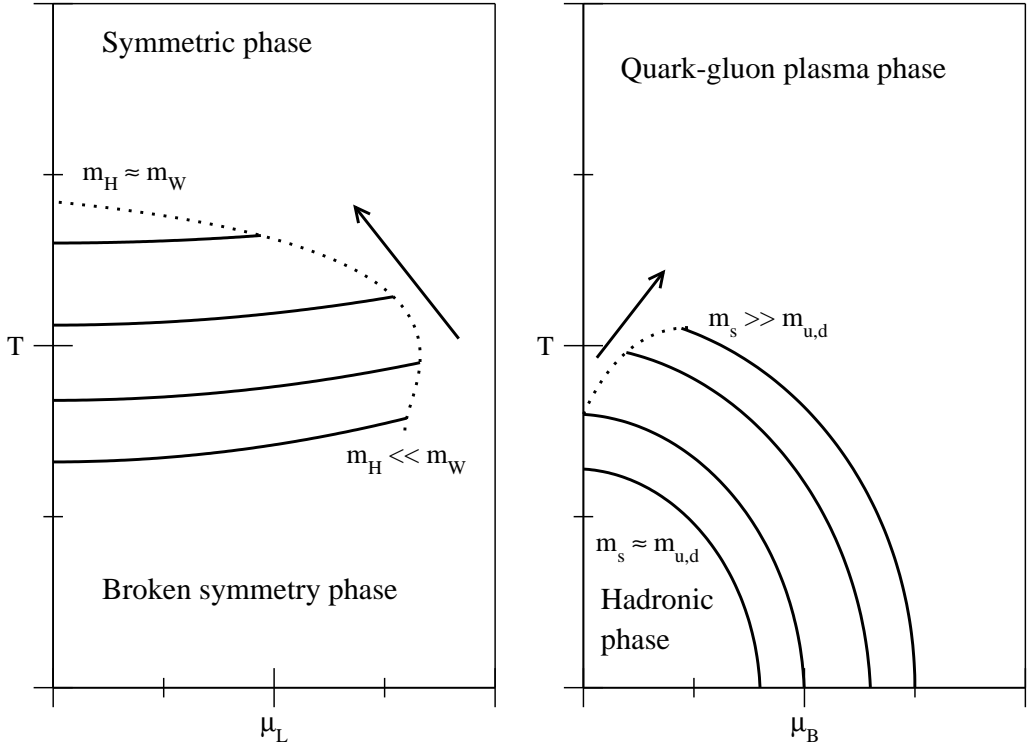}
\caption{Schematic plots of the electroweak (on the left) and QCD (right) phase diagrams in terms of temperature and the relevant chemical potentials. The solid lines correspond to the critical lines for a number of different Higgs/strange quark masses and the dotted lines indicate the location of the endpoint of the first order phase transition line as the masses are varied. The arrows indicate the order in which masses increase along the dotted lines. The low-T/high-$\mu$ phases of QCD (and also of electroweak theory), such as color superconducting phases, are not included in the picture.}
\label{fig:comppt}
\end{center}
\end{figure}

Consider first the cases when the masses parameterizing the theories are small. Then in both theories there is a first order phase transition, see Fig.~\ref{fig:comppt}. However, as the chemical potentials are increased, the critical temperature of the electroweak phase transition increases, while the critical temperature of the QCD phase transition decreases. Thus the responses of the systems on introducing the chemical potentials are opposite.

\begin{figure}[!tb]
\begin{center}
\includegraphics[width=0.87\textwidth]{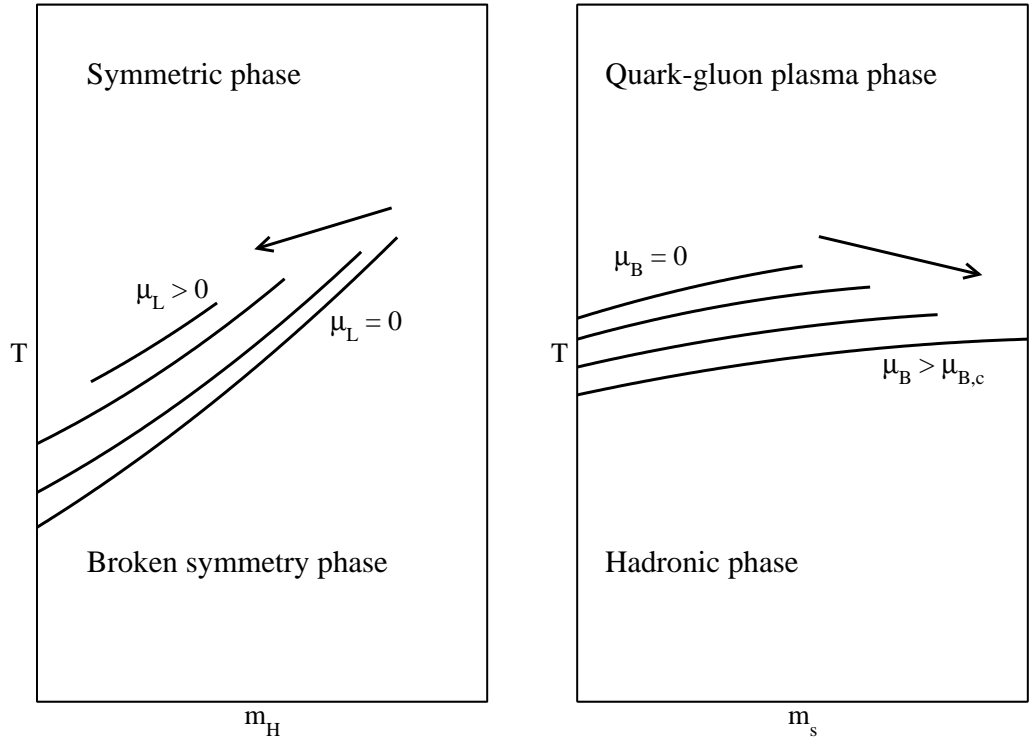}
\caption{Schematic plots of the electroweak (on the left) and QCD (right) phase diagrams in terms of temperature and the masses parameterizing the theories for a number of chemical potentials $\mu_L$ and $\mu_B$. The arrows indicate the order in which the chemical potentials increase.}
\label{fig:comppt2}
\end{center}
\end{figure}

Some interesting thermodynamics can be deduced from this. The Clausius-Clapeyron relation states that
\begin{eqnarray}
\frac{\mathrm{d}T}{\mathrm{d}\mu} & = & -\frac{n_1-n_2}{s_1-s_2},
\end{eqnarray}
where all the quantities are measured along the phase transition line and $n_i$ and $s_i$ are the particle number and entropy densities in each phase, respectively. Since the entropy is maximized in the high temperature phase, we can deduce from the relation above that in an electroweak system, since $\mathrm{d}T/\mathrm{d}\mu > 0$, the lepton number density in the low temperature phase (the broken symmetry phase) is higher than in the high temperature phase. In QCD the opposite is true, baryon density of the high temperature phase is higher.

Another feature of the phase transitions that we can see from Fig.~\ref{fig:comppt} is the response to increasing the masses that parametrize the theories. This is also depicted in Fig.~\ref{fig:comppt2}. There is no first order electroweak phase transition if the Higgs mass is big enough, regardless of the value of the leptonic chemical potential. Also, if the leptonic chemical potential is larger than some critical value ($\mu_{L,c} \gtrsim 50$~GeV), $\mu_L > \mu_{L,c}$, there is no first order transition for any value of the Higgs mass. On the other hand, in QCD, if the baryonic chemical potential is large enough, $\mu_B > \mu_{B,c}$, there is always a first order phase transition even in the limit $m_s \rightarrow \infty$. Note, though, that if $\mu_B$ is very high and $T$ is exponentially small, $T\sim g_s^{-5}\mathrm{exp}(-1/g_s)$, the low temperature phase of the system is not the normal hadronic phase but something more exotic (color superconductivity). Similar phenomena may be present in an electroweak system as well. In the present considerations, we assume $\mu_B$ ($\mu_L$) to be small enough so that the low temperature phase is the hadronic phase (normal broken symmetry phase). 

\chapter{Conclusions}

In this thesis we have studied some aspects of electroweak thermodynamics, namely the pressure of electroweak matter and the structure of the electroweak phase diagram. Due to lack of any experimental input about the properties of such matter, the motivation has been to understand the theory better. The electroweak theory, as a part of the standard model, describes nature to high accuracy and thus it is important that we understand the theory completely.

In this thesis, it was found that the pressure of matter composed of elementary particles at very high temperatures ($T \gtrsim 100$~GeV) deviates significantly from the ideal gas pressure that is commonly used, for example, in cosmological computations. However, a great part of the deviation is due to the strong interactions and the contribution from the purely electroweak degrees of freedom is seen to be close to the ideal gas estimate. From a theoretical point of view, the perturbative expansion of the electroweak pressure exhibits, however, some novel features that were not encountered in previous computations of pressure of gauge field theories. 

Concerning the phase diagram of the electroweak theory, it was found that introducing finite chemical potentials for the conserved particle numbers leads to a weaker transition (when the parameters of the theory, \emph{i.e.}, Higgs mass, are such that a first order phase transition is possible) and that for large enough chemical potentials there is no phase transition at all, regardless of the value of the Higgs mass, just a crossover. To achieve this result, a generalization of dimensional reduction methods to finite chemical potentials was needed in the context of electroweak interactions. The resulting theories can be used to study the properties of electroweak matter at high temperatures and finite chemical potentials, in general.

There is still much to learn about electroweak thermodynamics. The pressure of the standard model in the broken symmetry phase with all the particles taken into account with their proper masses has not been evaluated to high precision, although it is of phenomenological interest in cosmology. The structure of the electroweak phase diagram at large chemical potentials and (exponentially) low temperatures is fairly unknown as well. Such considerations have been carried out within QCD and they suggest that the phase structure might be very exotic. It is therefore to be expected that research within electroweak thermodynamics remains active in the future as well. 

The author would like to thank M.~Veps\"al\"ainen for collaboration, K.~Kajantie for his advices and M.~Laine, T.~Lappi, Y.~Schr\"oder 
and A.~Vuorinen for discussions. This work has been supported by the Graduate School in Particle and Nuclear Physics (GRASPANP), the
Helsinki Institute of Physics project ``Ultrarelativistic Heavy Ion Collisions'' and the Academy of Finland, contract no. 77744. 

\appendix

\chapter{Pressure of the electroweak theory}
\label{app:presEW}

In this appendix we will write the expansion of the pressure for the full standard model. The detailed derivation of the result can be found in \cite{Gynther:2005dj} and the case when the temperature is near the point where the electroweak crossover takes place was studied in \cite{Gynther:2005av}.

To be more precise, by the full standard model we mean the theory specified by the Euclidean action
\begin{eqnarray}
\mathcal{L} & = & \frac{1}{4}G_{\mu\nu}^a G_{\mu\nu}^a + \frac{1}{4}F_{\mu\nu} F_{\mu\nu} + \frac{1}{4}W_{\mu\nu}^a W_{\mu\nu}^a  
+ D_\mu\Phi^\dagger D_\mu\Phi - \nu^2\Phi^\dagger\Phi + \lambda(\Phi^\dagger\Phi)^2 \\
& & + \bar{l}_L\fslash{D}l_L + \bar{e}_R\fslash{D}e_R 
+ \bar{q}_L\fslash{D}q_L + \bar{u}_R\fslash{D}u_R + \bar{d}_R\fslash{D}d_R + ig_Y\left(\bar{q}_L\tau^2\Phi^\ast t_R - \bar{t}_R(\Phi^\ast)^\dagger\tau^2q_L\right), \nonumber
\end{eqnarray}
where we use the notation familiar from Sec.~\ref{sec:ewatft} and we will be keeping all the group theory factors for SU($N$), defined by
\begin{equation}
\begin{array}{rclrcl}
\TF\delta^{ab} & = & \mathrm{Tr}\; T^a T^b, & \CF \delta_{ij} & = & \left[T^a T^b \right]_{ij} \\
\CA \delta^{ab} & = & f^{ace}f^{bce}, & \dA & = & \delta^{aa},\; \dF \; = \; \delta_{ii},
\end{array}
\end{equation} 
explicitly in the expressions instead of evaluating them for SU($2$). Also, number of fermion families will be denoted by $n_\mathrm{F}$ and the number of fundamental scalars by $n_\mathrm{S}$.  Mixing between electroweak and QCD degrees of freedom is taken into account in evaluating the pressure, but the pure QCD contribution to the pressure, which can be obtained from \cite{Kajantie:2002wa,Braaten:1996jr,Arnold:1995eb,*Arnold:1994ps}, is left out in the following computations, to be added to the total pressure in the end. The total pressure is then given by
\begin{eqnarray}
p(T) & = & p_\mathrm{E}(T) + p_\mathrm{M}(T) + p_\mathrm{QCD}(T) + T^4\mathcal{O}(g^6),
\end{eqnarray}
where the components $p_\mathrm{E}(T)$ and $p_\mathrm{M}(T)$ are given below. Note that the gluon Debye mass $m_{\mathrm{D,gluon}}^2$, entering $p_\mathrm{QCD}(T)$, receives electroweak corrections,
\begin{eqnarray}
m_{\mathrm{D,gluon}}^2 & = & m_{\mathrm{D,gluon}}^2|_\mathrm{QCD} - g_s^2T^2 T_\mathrm{F}^{(\mathrm{QCD})}\left(\CF\dF n_\mathrm{F} g^2 + \frac{11}{18}n_\mathrm{F} g'^2 + \dF g_Y^2\right),
\end{eqnarray}
where $m_\mathrm{D,gluon}^2|_\mathrm{QCD}$ is the pure QCD contribution.  

\section{Contribution from the scale $2\pi T$}

The scale $2\pi T$ contributes to the pressure via $p_\mathrm{E}$ and the parameters of the effective theory. The effective theory is defined by
\begin{eqnarray}
S_\mathrm{E} &=& \int\!\dd^3\mathbf{x}\,\left\{ \frac{1}{4}G_{ij}^a G_{ij}^a +\frac{1}{4}F_{ij}F_{ij} +(D_i\Phi)^\dagger(D_i\Phi)
	+m_3^2\Phi^\dagger\Phi +\lambda_3(\Phi^\dagger\Phi)^2 \right. \nonumber \\
	&&+\half(D_i A_0^a)^2 +\half\mD^2 A_0^a A_0^a +\frac{1}{4}\lambda_A (A_0^a A_0^a)^2 +\half(\partial_i B_0)^2
	+\half\mD'^2 B_0 B_0 \nonumber \\
	&& \left. +h_3\Phi^\dagger\Phi A_0^a A_0^a +h_3'\Phi^\dagger\Phi B_0 B_0
	-\half g_3 g_3' B_0\Phi^\dagger A_0^a \tau^a\Phi \right\}, \label{eq:sesm}
\end{eqnarray}
where the couplings and masses are given by
\begin{equation}
\begin{array}{rclrcl}
	g_3^2 &=& g^2 T, & g_3'^2 &=& g'^2 T, \\
	\lambda_3 &=& \lambda T, & \lambda_A &=& \mathcal{O}(g^4), \\
	h_3 &=& {\displaystyle \frac{1}{4}g^2 T}, & h_3' &=& {\displaystyle \frac{1}{4}g'^2 T},
\end{array}
\end{equation}
\begin{eqnarray}
	\mD^2 &=& T²\left[ g²\left(\beta_{E1} +\beta_{E2}\epsilon +\mathcal{O}(\epsilon^2)\right) 
	+\frac{g^4}{(4\pi)^2}\left(\beta_{E3}+\mathcal{O}(\epsilon)\right) +\mathcal{O}(g^6) \right. \nonumber \\
	&& \left. +\frac{g^2}{(4\pi)^2}\left(\beta_{E\lambda}\lambda +\beta_{Es}g_s^2 +\beta_{EY}g_Y^2
		+\beta_{E'}g'^2 +\beta_{E\nu}\frac{-\nu^2}{T^2} \right) \right], \\ 
	\mD'^2 &=& T²\left[ g'²\left(\beta'_{E1} +\beta'_{E2}\epsilon +\mathcal{O}(\epsilon^2)\right) 
	+\frac{g'^4}{(4\pi)^2}\left(\beta'_{E3}+\mathcal{O}(\epsilon)\right) +\mathcal{O}(g'^6)\right. \nonumber \\
	&& \left. +\frac{g'^2}{(4\pi)^2}\left(\beta'_{E\lambda}\lambda +\beta'_{Es}g_s^2 +\beta'_{EY}g_Y^2
		+\beta'_{E}g^2 +\beta'_{E\nu}\frac{-\nu^2}{T^2} \right) \right],
\end{eqnarray}
\begin{eqnarray}
	m_3^2(\Lambda) &=& -\nu^2\left[1 +\frac{g^2}{(4\pi)^2}\beta_{\nu A} +\frac{g'^2}{(4\pi)^2}\beta_{\nu B} +\frac{\lambda}{(4\pi)^2}\beta_{\nu\lambda} +\frac{g_Y^2}{(4\pi)^2}\beta_{\nu Y} \right] \nonumber \\
	&&{}+T^2\left[ g^2(\beta_{A1}+\beta_{A2}\epsilon) +g'^2(\beta_{B1}+\beta_{B2}\epsilon)
	+\lambda(\beta_{\lambda 1}+\beta_{\lambda 2}\epsilon) +g_Y^2(\beta_{Y1}+\beta_{Y2}\epsilon) \right. \nonumber \\
	&&{}+\frac{g^4}{(4\pi)^2}\beta_{AA} +\frac{g'^4}{(4\pi)^2}\beta_{BB} +\frac{g^2 g'^2}{(4\pi)^2}\beta_{AB}
	+\frac{\lambda g^2}{(4\pi)^2}\beta_{A\lambda} +\frac{\lambda g'^2}{(4\pi)^2}\beta_{B\lambda}
	+\frac{\lambda^2}{(4\pi)^2}\beta_{\lambda\lambda} \nonumber \\
	&&\left. {}+\frac{g^2 g_Y^2}{(4\pi)^2}\beta_{AY} +\frac{g'^2 g_Y^2}{(4\pi)^2}\beta_{BY} 
	+\frac{g_s^2 g_Y^2}{(4\pi)^2}\beta_{sY} +\frac{\lambda g_Y^2}{(4\pi)^2}\beta_{\lambda Y}
	+\frac{g_Y^4}{(4\pi)^2}\beta_{YY} \right]
\end{eqnarray}
and the mass counterterm for $m_3^2$ is
\begin{equation}
	\delta m_3^2 = \frac{T^2}{(4\pi)^2\epsilon}\left( -\frac{81}{64}g^4 +\frac{7}{64}g'^4 +\frac{15}{32}g^2 g'^2
	-\frac{9}{4}\lambda g^2 -\frac{3}{4}\lambda g'^2 +3\lambda^2 \right).
\end{equation}
The pressure $p_\mathrm{E}$ is parametrized as
\begin{eqnarray}
\pE(T) & = & T^4\Big[\alpha_{E1} + g^2\alpha_{EA} + g'^2\alpha_{EB}
      + \lambda\alpha_{E\lambda} + g_Y^2\alpha_{EY} \nonumber \\ 
& + & \frac{1}{(4\pi)^2}\Big(g^4\alpha_{EAA} + g'^4\alpha_{EBB} +
      (gg')^2\alpha_{EAB} + \lambda^2\alpha_{E\lambda\lambda} + \lambda g^2 \alpha_{EA\lambda}
           + \lambda g'^2 \alpha_{EB\lambda} \nonumber \\
& + & \; g_Y^4\alpha_{EYY} + (gg_Y)^2\alpha_{EAY} + (g'g_Y)^2\alpha_{EBY}
           + \lambda g_Y^2\alpha_{EY\lambda} \nonumber \\
& + & \; (gg_s)^2\alpha_{EAs} + (g'g_s)^2\alpha_{EBs} + (g_Yg_s)^2\alpha_{EYs} \Big) \Big] \nonumber \\
& + & \nu^2T^2\Big[\alpha_{E\nu} + \frac{1}{(4\pi)^2}\big(g^2\alpha_{EA\nu} + g'^2\alpha_{EB\nu} + \lambda\alpha_{E\lambda\nu}
    + g_Y^2\alpha_{EY\nu}\big)\Big] \nonumber \\
& + & \frac{\nu^4}{(4\pi)^2}\alpha_{E\nu\nu} + T^4\cdot{\cal O}(g^6).
\end{eqnarray}
All the parameters above are given below:
\begin{eqnarray}
	\alpha_{E1} & = & \frac{\pi^2}{45}\left\{1+\dA+\dF\nS+\frac{7}{8}\Big[1+\dF+(2+\dF)\Nc\Big]\NF\right\} \\
	\alpha_{EA} & = & -\frac{1}{144}\left[\CA\dA + \frac{5}{2}\CF\dF\nS + \frac{5}{4}\CF\dF(1+\Nc)\NF\right] \\
	\alpha_{EB} & = & -\frac{5}{576}\left\{\frac{1}{2}\dF\nS + \left[1 + \frac{1}{4}\dF + \left(\frac{5}{9}+\frac{1}{36}\dF\right)\Nc\right]\NF\right\} \\
	\alpha_{E\lambda} & = & -\frac{\dF(\dF+1)}{144}\nS \\
	\alpha_{EY} & = & -\frac{5}{288}\Nc
\end{eqnarray}
\begin{eqnarray}
	\alpha_{EAA} & = & \frac{1}{12}\left\{\CA^2\dA\left(\frac{1}{\epsilon} + \frac{97}{18}\ln\frac{\Lambda}{4\pi T} + \frac{29}{15} + \frac{1}{3}\gamma + \frac{55}{9}\frac{\zeta'(-1)}{\zeta(-1)} - \frac{19}{18}\frac{\zeta'(-3)}{\zeta(-3)}\right) \right. \nonumber \\
& & \hspace{0.5cm} + \left[\CA\CF\dF\left(\frac{1}{2\epsilon} + \frac{169}{72}\ln\frac{\Lambda}{4\pi T} + \frac{1121}{1440}-\frac{157}{120}\ln 2 + \frac{1}{3}\gamma + \frac{73}{36}\frac{\zeta'(-1)}{\zeta(-1)}-\frac{1}{72}\frac{\zeta'(-3)}{\zeta(-3)}\right) \right. \nonumber \\
& & \hspace{1.5cm} \left. + \CF^2\dF\left(\frac{35}{32}-\ln 2\right)\right]\left(1+\Nc\right)\NF \nonumber \\
& & \hspace{0.5cm} + \CF\TF\dF\left(\frac{5}{36}\ln\frac{\Lambda}{4\pi T} + \frac{1}{144} - \frac{11}{3}\ln 2 + \frac{1}{12}\gamma + \frac{1}{9}\frac{\zeta'(-1)}{\zeta(-1)} - \frac{1}{18}\frac{\zeta'(-3)}{\zeta(-3)}\right)\left(1+\Nc\right)^2\NF^2 \nonumber \\
& & \hspace{0.5cm} + \CF\TF\dF\left(\frac{25}{72}\frac{\Lambda}{4\pi T} - \frac{83}{16} - \frac{49}{12}\ln 2 + \frac{1}{3}\gamma + \frac{1}{36}\frac{\zeta'(-1)}{\zeta(-1)} - \frac{1}{72}\frac{\zeta'(-3)}{\zeta(-3)}\right)\left(1+\Nc\right)\NF\nS \nonumber \\
& & \hspace{0.5cm} + \left[\CA\CF\dF\left(\frac{1}{\epsilon} + \frac{317}{72}\ln\frac{\Lambda}{4\pi T} + \frac{337}{720} + \frac{2}{3}\gamma + \frac{125}{36}\frac{\zeta'(-1)}{\zeta(-1)} + \frac{19}{72}\frac{\zeta'(-3)}{\zeta(-3)}\right) \right. \nonumber \\
& & \hspace{1.5cm} + \CF^2\dF\left(\frac{3}{2\epsilon} + \frac{19}{2}\ln\frac{\Lambda}{4\pi T} + \frac{881}{120} + \frac{3}{4}\gamma + \frac{23}{2}\frac{\zeta'(-1)}{\zeta(-1)} - \frac{11}{4}\frac{\zeta'(-3)}{\zeta(-3)}\right) \nonumber \\
& & \hspace{1.5cm} \left. \left. + \CF\TF\dF\left(\frac{23}{36}\ln\frac{\Lambda}{4\pi T} - \frac{283}{360} + \frac{1}{3}\gamma + \frac{11}{18}\frac{\zeta'(-1)}{\zeta(-1)} - \frac{11}{36}\frac{\zeta'(-3)}{\zeta(-3)}\right)\right]\nS\right\} \\
	\alpha_{EBB} & = & \frac{1}{128}\left\{\left[\dF\left(\frac{1}{\epsilon} + \frac{19}{3}\ln\frac{\Lambda}{4\pi T} + \frac{881}{180} + \frac{1}{2}\gamma + \frac{23}{3}\frac{\zeta'(-1)}{\zeta(-1)} - \frac{11}{6}\frac{\zeta'(-3)}{\zeta(-3)} \right) \right. \right. \nonumber \\
& & \hspace{0.5cm} \left. + \dF^2\left(\frac{23}{54}\ln\frac{\Lambda}{4\pi T} - \frac{283}{540} + \frac{2}{9}\gamma + \frac{11}{27}\frac{\zeta'(-1)}{\zeta(-1)} - \frac{11}{54}\frac{\zeta'(-3)}{\zeta(-3)}\right)\right]\nS \nonumber \\
& & \hspace{0.5cm} + \dF\left[1+\frac{5}{9}\Nc + \frac{\dF}{4}\left(1+\frac{\Nc}{9}\right)\right] \nonumber \\
& & \hspace{1.2cm} \times \left[\frac{25}{27}\ln\frac{\Lambda}{4\pi T} - \frac{83}{60} - \frac{147}{135}\ln 2 + \frac{8}{9}\gamma + \frac{2}{27}\frac{\zeta'(-1)}{\zeta(-1)} - \frac{1}{27}\frac{\zeta'(-3)}{\zeta(-3)}\right]\NF\nS \nonumber \\
& & \hspace{0.5cm} + \left[1+\frac{17}{81}\Nc+\frac{\dF}{16}\left(1+\frac{\Nc}{81}\right)\right]\left(\frac{35}{3}-\frac{32}{3}\ln 2\right)\NF \nonumber \\
& & \hspace{0.5cm} + \left[\left(1+\frac{5}{9}\Nc\right)^2+\frac{\dF}{2}\left(1+\frac{2}{3}\Nc+\frac{5}{81}\Nc^2\right) + \frac{\dF^2}{16}\left(1+\frac{\Nc}{9}\right)^2\right] \nonumber \\
& & \left. \hspace{1.2cm} \times \left(\frac{40}{27}\ln\frac{\Lambda}{4\pi T} + \frac{2}{27}-\frac{176}{45}\ln 2 + \frac{8}{9}\gamma + \frac{32}{27}\frac{\zeta'(-1)}{\zeta(-1)}-\frac{16}{27}\frac{\zeta'(-3)}{\zeta(-3)}\right)\NF^2\right\}
\end{eqnarray}
\begin{eqnarray}
	\alpha_{EAB} & = & \frac{1}{16}\left[\CF\dF\left(\frac{1}{\epsilon} + \frac{19}{3}\ln\frac{\Lambda}{4\pi T} + \frac{881}{180} + \frac{1}{2}\gamma + \frac{23}{3}\frac{\zeta'(-1)}{\zeta(-1)} - \frac{11}{6}\frac{\zeta'(-3)}{\zeta(-3)}\right)\nS \right. \nonumber \\
& & \left. \hspace{0.6cm} 
                         + \CF\dF\left(1+\frac{1}{9}\Nc\right)\left(\frac{35}{48} - \frac{2}{3}\ln 2\right)\NF\right] \\
	\alpha_{E\lambda\lambda} & = & \frac{\dF(\dF+1)}{9}\nS\left[\ln\frac{\Lambda}{4\pi T} + \frac{31}{40} + \frac{1}{4}\gamma
                                       + \frac{3}{2}\frac{\zeta'(-1)}{\zeta(-1)} - \frac{3}{4}\frac{\zeta'(-3)}{\zeta(-3)}
                                       + \frac{1}{4}\dF\left(\ln\frac{\Lambda}{4\pi T} + \gamma\right)\right] \nonumber \\
& & \\
	\alpha_{EA\lambda} & = & \frac{\dF(\dF+1)}{36}\CF\left(\frac{3}{\epsilon} + 15\ln\frac{\Lambda}{4\pi T} + 11 + 3\gamma
                                 + 12\frac{\zeta'(-1)}{\zeta(-1)}\right)\nS \\
	\alpha_{EB\lambda} & = & \frac{\dF(\dF+1)}{144}\nS\left(\frac{3}{\epsilon} + 15\ln\frac{\Lambda}{4\pi T} + 11 + 3\gamma
                             + 12\frac{\zeta'(-1)}{\zeta(-1)} \right) \\
	\alpha_{EYY} & = & -\frac{1}{32}\Nc\left[\ln\frac{\Lambda}{4\pi T} - \frac{239}{120}- \frac{11}{5}\ln 2
                           +2\frac{\zeta'(-1)}{\zeta(-1)} - \frac{\zeta'(-3)}{\zeta(-3)} \right.\nonumber \\
& &  \left. \hspace{1.1cm} - \Nc\left(\frac{10}{9}\ln\frac{\Lambda}{4\pi T} + \frac{53}{90} - \frac{106}{45}\ln 2 + \frac{4}{9}\gamma
            +\frac{4}{3}\frac{\zeta'(-1)}{\zeta(-1)} - \frac{2}{3}\frac{\zeta'(-3)}{\zeta(-3)} \right)\right]\\
	\alpha_{EAY} & = & \frac{1}{16}\Nc\left(\frac{1}{\epsilon} + \frac{19}{4}\ln\frac{\Lambda}{4\pi T} + \frac{619}{120} - \frac{13}{4}\ln 2 + \gamma + \frac{7}{2}\frac{\zeta'(-1)}{\zeta(-1)} + \frac{1}{4}\frac{\zeta'(-3)}{\zeta(-3)}\right)\\
	\alpha_{EBY} & = & \frac{1}{48}\Nc\left(\frac{1}{\epsilon} + \frac{131}{36}\ln\frac{\Lambda}{4\pi T} + \frac{6563}{1080}
                         -\frac{41}{20}\ln 2 + \gamma + \frac{23}{18}\frac{\zeta'(-1)}{\zeta(-1)} + \frac{49}{36}\frac{\zeta'(-3)}{\zeta(-3)} \right)\\
	\alpha_{EY\lambda} & = & \frac{1}{6}\Nc\left(\ln\frac{\Lambda}{4\pi T} - \ln 2 + \gamma\right) \\
	\alpha_{EAs} & = & \frac{\CF\dF}{12}\left(\Nc^2-1\right)\NF\left(\frac{35}{32} - \ln 2\right) \\
	\alpha_{EBs} & = & \frac{1}{12}\left(\Nc^2-1\right)\NF\left[\frac{175}{288}-\frac{5}{9}\ln 2 + \frac{\dF}{36}\left(\frac{35}{32}-\ln 2\right)\right] \\
	\alpha_{EYs} & = & -\frac{15}{144}\left(\Nc^2-1\right)\left(\ln\frac{\Lambda}{4\pi T}-\frac{62}{75}-\frac{27}{25}\ln 2 
                         +2\frac{\zeta'(-1)}{\zeta(-1)} - \frac{\zeta'(-3)}{\zeta(-3)}\right) \\
	\alpha_{E\nu} & = & \frac{\dF}{12}\nS \\
	\alpha_{EA\nu} & = & -\frac{\CF\dF}{2}\left(\frac{1}{\epsilon} + 3\ln\frac{\Lambda}{4\pi T} + \frac{5}{3} + \gamma + 2\frac{\zeta'(-1)}{\zeta(-1)}\right)\nS \\
	\alpha_{EB\nu} & = & -\frac{\dF}{8}\left(\frac{1}{\epsilon} + 3\ln\frac{\Lambda}{4\pi T} + \frac{5}{3} +\gamma + 2\frac{\zeta'(-1)}{\zeta(-1)} \right) \nS\\
	\alpha_{E\lambda\nu} & = & -\frac{\dF(\dF+1)}{3}\nS\left(\ln\frac{\Lambda}{4\pi T} + \gamma\right) \\
	\alpha_{EY\nu} & = & -\frac{1}{3}\Nc\left(\ln\frac{\Lambda}{4\pi T} -\ln 2 +\gamma \right)\\
	\alpha_{E\nu\nu} & = & \dF\nS\left(\ln\frac{\nu}{4\pi T} -\frac{3}{4} +\gamma \right)
\end{eqnarray}

\begin{eqnarray}
	\beta_{E1} &=& \frac{1}{3}\left[ \CA +\nF(N_c+1)\TF +\nS\TF \right] \\
	\beta'_{E1} &=& \frac{1}{3}\left[ \left(\frac{11}{36}N_c+\frac{3}{4}\right)\NF +\frac{\dF}{4}\nS \right] \\
	\beta_{E2} &=& \frac{2}{3}\left[ \CA\left( \frac{\zeta'(-1)}{\zeta(-1)}+\ln\frac{\Lambda}{4\pi T} \right) +\TF\nF(N_c+1) \left(\half-\ln 2+\frac{\zeta'(-1)}{\zeta(-1)}+\ln\frac{\Lambda}{4\pi T} \right) \right. \nonumber \\ &&{}\left. +\TF\nS\left(\half +\frac{\zeta'(-1)}{\zeta(-1)}+\ln\frac{\Lambda}{4\pi T}\right) \right] \\
	\beta'_{E2} &=& \frac{2}{3}\left[ \left(\frac{11}{36}N_c+\frac{3}{4}\right)\NF \left(\half-\ln 2+\frac{\zeta'(-1)}{\zeta(-1)}+\ln\frac{\Lambda}{4\pi T} \right) \right. \nonumber \\
                    & & \left. +\frac{\dF}{4}\nS\left(\half +\frac{\zeta'(-1)}{\zeta(-1)}+\ln\frac{\Lambda}{4\pi T}\right) \right] \\
	\beta_{E3} &=& \CA^2\left(\frac{5}{9}+\frac{22}{9}\gamma+\frac{22}{9}\ln\frac{\Lambda}{4\pi T}\right) +\CA\TF \nF(N_c+1)\left(1-\frac{16}{9}\ln 2+\frac{14}{9}\gamma+\frac{14}{9}\ln\frac{\Lambda}{4\pi T}\right) \nonumber \\
	&&{}+ \TF^2\left(\nF(N_c+1)\right)^2 \left(\frac{4}{9} -\frac{16}{9}\ln 2 -\frac{8}{9}\gamma-\frac{8}{9}\ln\frac{\Lambda}{4\pi T}\right) -2\CF\TF \nF(N_c+1) \nonumber \\
	&&{}+\TF^2 \nF(N_c+1)\nS \left(\frac{2}{9} -\frac{16}{9}\ln 2 -\frac{10}{9}\gamma -\frac{10}{9}\ln\frac{\Lambda}{4\pi T}\right) \nonumber \\
&&	+\CA\TF\nS\left(\frac{1}{3}+\frac{20}{9}\gamma+\frac{20}{9}\ln\frac{\Lambda}{4\pi T}\right)
	+ \TF^2\nS^2\left(-\frac{2}{9}-\frac{2}{9}\gamma-\frac{2}{9}\ln\frac{\Lambda}{4\pi T}\right) \nonumber \\
	&&{} +\CF\TF\nS \\
	\beta'_{E3} &=& \left(\frac{11}{36}N_c+\frac{3}{4}\right)^2\NF^2\left(\frac{4}{9} -\frac{16}{9}\ln 2 -\frac{8}{9}\gamma-\frac{8}{9}\ln\frac{\Lambda}{4\pi T}\right)
	+\frac{\dF^2}{16}\nS^2\left(-\frac{2}{9}-\frac{2}{9}\gamma-\frac{2}{9}\ln\frac{\Lambda}{4\pi T} \right) \nonumber \\
	&&{}+\left(\frac{11}{36}N_c+\frac{3}{4}\right)\frac{\dF}{4}\NF\nS\left(\frac{2}{9} -\frac{16}{9}\ln 2 -\frac{10}{9}\gamma -\frac{10}{9}\ln\frac{\Lambda}{4\pi T}\right) \nonumber \\
	&&{}-2\left(\frac{137}{1296}N_c+\frac{9}{16}\right)\NF +\frac{\dF}{16} \nS
\end{eqnarray}
\parbox{0.38\textwidth}{
\begin{eqnarray*}
	\beta_{E\lambda} &=& \frac{2}{3}\TF(\dF+1)\nS \\
	\beta_{Es} &=& -2C_\mathrm{3F}\TF N_c \nF \\
	\beta_{EY} &=& -\frac{1}{6}N_c\TF \\
	\beta_{E'} &=& -2\TF\left( \frac{N_c}{36}+\frac{1}{4}\right)\nF +\TF\frac{1}{4}\nS \\
	\beta_{E\nu} &=& 4\TF\nS
\end{eqnarray*}}
\parbox{0.61\textwidth}{
\begin{eqnarray}
	\beta'_{E\lambda} &=& \frac{2}{3}\frac{\dF}{4}(\dF+1)\nS \\
	\beta'_{Es} &=& -2C_\mathrm{3F}\frac{11}{36}N_c\NF \\
	\beta'_{EY} &=& -\frac{11\dF}{72}N_c \\
	\beta'_E &=& -2\CF\dF\left( \frac{N_c}{36}+\frac{1}{4}\right)\nF +\CF\frac{\dF}{4}\nS \\
	\beta'_{E\nu} &=& 4 \frac{\dF}{4}\nS
\end{eqnarray}}

\parbox{0.39\textwidth}{
\begin{eqnarray*}
	\beta_{\nu A}&=& 3\CF\left( 2\gamma+2\ln\frac{\Lambda}{4\pi T}\right) \\
	\beta_{\nu\lambda}&=& -2(\dF+1)\left( 2\gamma+2\ln\frac{\Lambda}{4\pi T}\right) \\
	\beta_{A1} &=& \frac{1}{4}\CF \\
	\beta_{B1} &=& \frac{1}{4} \left(\half\right)^2 \\
	\beta_{\lambda 1} &=& \frac{1}{6}(\dF+1) \\
	\beta_{Y1} &=& \frac{1}{12}N_c
\end{eqnarray*}}
\parbox{0.60\textwidth}{
\begin{eqnarray}
	\beta_{\nu B}&=& 3\frac{1}{4}\left( 2\gamma+2\ln\frac{\Lambda}{4\pi T}\right) \\
	\beta_{\nu Y}&=& -N_c\left( 4\ln 2 +2\gamma+2\ln\frac{\Lambda}{4\pi T}\right) \\
	\beta_{A2} &=& \CF\half \left(\frac{2}{3}+\frac{\zeta'(-1)}{\zeta(-1)}+\ln\frac{\Lambda}{4\pi T}\right)  \\
	\beta_{B2} &=& \left(\half\right)^2 \half \left(\frac{2}{3}+\frac{\zeta'(-1)}{\zeta(-1)} +\ln\frac{\Lambda}{4\pi T}\right)\\
	\beta_{\lambda 2} &=& \frac{\dF+1}{3} \left(1+\frac{\zeta'(-1)}{\zeta(-1)} +\ln\frac{\Lambda}{4\pi T}\right) \\
	\beta_{Y2} &=& \frac{N_c}{6} \left(1-\ln 2 +\frac{\zeta'(-1)}{\zeta(-1)} +\ln\frac{\Lambda}{4\pi T}\right)
\end{eqnarray}}

\begin{eqnarray}
	\beta_{AA} &=& \left( -\frac{11}{9} -\frac{5}{2}\frac{\zeta'(-1)}{\zeta(-1)}-\frac{2}{3}\gamma -\frac{19}{6}\ln\frac{\Lambda}{4\pi T}\right)\CA\CF
	+\left( 1+\frac{3}{2}\frac{\zeta'(-1)}{\zeta(-1)}+\frac{3}{2}\gamma +3\ln\frac{\Lambda}{4\pi T}\right)\CF^2 \nonumber \\
	&+&\left( \frac{1}{9}+\frac{2}{3}\ln 2 -\frac{2}{3}\gamma -\frac{2}{3}\ln\frac{\Lambda}{4\pi T}\right)\nF(N_c+1)\CF\TF
	+\left(1+\frac{\zeta'(-1)}{\zeta(-1)}+\ln\frac{\Lambda}{4\pi T}\right)\frac{\CF(\dF+1)}{4} \nonumber \\
	&+&\left( -\frac{2}{9}-\half\frac{\zeta'(-1)}{\zeta(-1)}-\frac{2}{3}\gamma -\frac{7}{6}\ln\frac{\Lambda}{4\pi T}\right)\CF\TF\nS \\
	\beta_{BB} &=& \frac{1}{16}\left( 1+\frac{3}{2}\frac{\zeta'(-1)}{\zeta(-1)}+\frac{3}{2}\gamma +3\ln\frac{\Lambda}{4\pi T}\right)
	+\left( -\frac{2}{9}-\half\frac{\zeta'(-1)}{\zeta(-1)}-\frac{2}{3}\gamma -\frac{7}{6}\ln\frac{\Lambda}{4\pi T}\right)\frac{\dF}{16}\nS \nonumber \\
        &+&\left( \frac{1}{9}+\frac{2}{3}\ln 2 -\frac{2}{3}\gamma -\frac{2}{3}\ln\frac{\Lambda}{4\pi T}\right)\frac{1}{4}\NF\left(\frac{11}{36}N_c+\frac{3}{4}\right) \nonumber \\ 
	&+&\left( \frac{1}{4} +\frac{1}{4}\frac{\zeta'(-1)}{\zeta(-1)} +\frac{1}{4}\ln\frac{\Lambda}{4\pi T}\right)\frac{\dF+1}{4} \\
	\beta_{AB} &=& \left( 2+3\frac{\zeta'(-1)}{\zeta(-1)}+3\gamma +6\ln\frac{\Lambda}{4\pi T}\right)\CF\frac{1}{4}
	+\left( \frac{1}{4} +\frac{1}{4}\frac{\zeta'(-1)}{\zeta(-1)} +\frac{1}{4}\ln\frac{\Lambda}{4\pi T}\right)\frac{\dF+1}{2} \\
	\beta_{A\lambda} &=& \left( -\frac{5}{3} -2\frac{\zeta'(-1)}{\zeta(-1)} -2\ln\frac{\Lambda}{4\pi T}\right)\CF(\dF+1) \\
	\beta_{B\lambda} &=& \left( -\frac{5}{3} -2\frac{\zeta'(-1)}{\zeta(-1)} -2\ln\frac{\Lambda}{4\pi T}\right)\frac{1}{4}(\dF+1)
\end{eqnarray}
\begin{eqnarray}
	\beta_{\lambda \lambda} &=& \left( 4 +4\frac{\zeta'(-1)}{\zeta(-1)} +4\ln\frac{\Lambda}{4\pi T}\right)(\dF+1) \nonumber \\
	& & {}+\left( -\frac{2}{3} -\frac{2}{3}\gamma -\frac{2}{3}\frac{\zeta'(-1)}{\zeta(-1)} -\frac{4}{3}\ln\frac{\Lambda}{4\pi T}\right)(\dF+1)^2 \\
	\beta_{AY} &=& \left( -\frac{1}{12} -\frac{1}{6}\ln2 +\half\gamma +\half\ln\frac{\Lambda}{4\pi T}\right)\CF N_c \\
	\beta_{BY} &=& \left( -\frac{11}{36} -\frac{55}{54}\ln 2 +\frac{17}{18}\gamma +\frac{17}{18}\ln\frac{\Lambda}{4\pi T} \right) \frac{1}{4}N_c \\
	\beta_{sY} &=& \left( -\half +\frac{8}{3}\ln 2 +\gamma +\ln\frac{\Lambda}{4\pi T}\right)C_\mathrm{3F} N_c \\
	\beta_{\lambda Y} &=& \left( -\frac{1}{3}\ln 2 -\frac{2}{3}\gamma -\frac{2}{3}\ln\frac{\Lambda}{4\pi T}\right) (\dF+1)N_c \\
	\beta_{YY} &=& \frac{3}{4}\gamma +\frac{3}{4} \ln\frac{\Lambda}{4\pi T}
\end{eqnarray}

\section{Contribution from the scale $gT$}

The contribution from the soft scale $gT$ to the pressure is given by $p_\mathrm{M}$
\begin{eqnarray}
	\frac{\pM(T)}{T} &=& \frac{1}{4\pi} \dF \nS \left(m_3^2\right)^{3/2} \left[ \frac{2}{3} +\epsilon\left(\frac{16}{9}+\frac{4}{3}\ln\frac{\mu_3}{2m_3}\right) \right]
	+\frac{1}{4\pi}\left( \frac{1}{3}\dA \mD^3 +\frac{1}{3} \mD'^3 \right) \nonumber \\
	&+&\frac{1}{(4\pi)^2}\left[-\dF (\dF+1)\nS \lambda_3 m_3^2 -\dF\dA\nS h_3 m_3 \mD -\dF\nS h_3' m_3 \mD' \right. \nonumber \\
	&-&\hspace{-0.2cm} \left. \left(\CF g_3^2 +\frac{1}{4}g_3'^2\right)\nS\dF m_3^2 \left( \frac{1}{2\epsilon} +\frac{3}{2} +2\ln \frac{\mu_3}{2m_3}\right) -\CA\dA g_3^2\mD^2 \left( \frac{1}{4\epsilon} +\frac{3}{4} 
		+ \ln \frac{\mu_3}{2\mD}\right) \right] \nonumber \\
	&+& \frac{1}{(4\pi)^3}\left[ g_3^4 m_3 B_{AAf} +g_3'^4 m_3 B_{BBf} +g_3^2 g_3'^2 m_3 B_{ABf} +g_3^4 \mD B_{AAa} +g_3^2\lambda_3 m_3 B_{A\lambda f} \right. \nonumber \\
	&+& g_3'^2 \lambda_3 m_3 B_{B\lambda f} +\lambda_3^2 m_3 B_{\lambda \lambda f} +h_3^2 m_3 B_{hhf} +h_3^2 \mD B_{hha} +h_3'^2 m_3 B'_{hhf} +h_3'^2 \mD' B'_{hhb}\nonumber \\
	&+& g_3^2 g_3'^2 m_3 2b(m_3) +g_3^2 g_3'^2\mD b(\mD) +g_3^2 g_3'^2 \mD' b(\mD')+\frac{\dF}{4m_3}(\dA h_3\mD +h_3'\mD')^2 \nonumber \\
	&+& \dF^2 m_3^2 \left( \frac{\dA h_3^2}{2\mD} +\frac{h_3'^2}{2\mD'}\right) 
	+g_3^4\CA\CF\dF\frac{1}{3}\left( \frac{m_3^2}{\mD}\ln\frac{\mD+m_3}{m_3} +\frac{\mD^2}{m_3}\ln\frac{\mD+m_3}{\mD} \right)\nonumber \\
	&+& \dF(\dF+1)\lambda_3(\dA h_3\mD +h_3'\mD') +g_3^2 h_3\mD B_{Aha} +g_3'^2 h_3'\mD' B'_{Bhb} +g_3^2 h_3'\mD' B'_{Ahb} \nonumber \\
	&+&\left. g_3'^2 h_3\mD B_{Bha} +g_3^2 h_3 m_3 B_{Ahf} \right],
\end{eqnarray}
for which the parameters are given by
\begin{eqnarray} 
	B_{AAf} &=& \CF^2\dF\nS \left( -\frac{3}{4\epsilon} -\frac{35}{4} -\frac{\pi^2}{3} +6\ln 2 -\frac{9}{2}\ln\frac{\mu_3}{2m_3} \right) 
	-\CF\TF\dF \left( \frac{1}{4\epsilon} +\frac{4}{3} -\frac{4}{3}\ln 2 +\frac{3}{2}\ln\frac{\mu_3}{2m_3} \right) \nonumber \\
	&& {}+\CA\CF\dF \left( \frac{3}{4\epsilon} +\frac{19}{24} -3\ln 2 +5\ln\frac{\mu_3}{2m_3} -\half\ln\frac{\mu_3}{2(m_3+\mD)} \right) \\
	B_{BBf} &=& \frac{\dF}{16}\nS\left( -\frac{3}{4\epsilon} -\frac{35}{4} -\frac{\pi^2}{3} +6\ln 2 	-\frac{9}{2}\ln\frac{\mu_3}{2m_3} \right)
	 \hspace{-0.1cm}-\frac{\dF^2}{16}\nS \left( \frac{1}{4\epsilon} +\frac{4}{3} -\frac{4}{3}\ln 2 +\frac{3}{2}\ln\frac{\mu_3}{2m_3} \right) \\
	B_{ABf} &=& \CF\dF\nS\left( -\frac{3}{8\epsilon} -\frac{35}{8} -\frac{\pi^2}{6} +3\ln 2 -\frac{9}{4}\ln\frac{\mu_3}{2m_3} \right) \\
	B_{AAa} &=& \CA\CF\dF\nS \left( -\frac{1}{8\epsilon}-\frac{23}{24} -\frac{1}{4}\ln\frac{\mu_3}{2\mD} -\half\ln\frac{\mu_3}{2(m_3+\mD)} \right) \nonumber \\
	&& {}+\CA^2 \dA \left( -\frac{89}{24} +\frac{11}{6}\ln 2 -\frac{\pi^2}{6} \right) \\
	B_{A\lambda f} &=& \CF\dF\nS \left( -4 +8\ln 2  \right)
	+\CF\dF(\dF+1)\nS\left( \frac{1}{\epsilon} +3 +6\ln\frac{\mu_3}{2m_3} \right) \\
	B_{B\lambda f} &=& \frac{1}{4}\dF(\dF+1)\nS\left(\frac{1}{\epsilon} -1 +8\ln 2 +6\ln\frac{\mu_3}{2m_3} \right) \\
	B_{\lambda \lambda f} &=& \dF (\dF+1)\nS\left( -\frac{1}{\epsilon} -8 +4\ln 2 -6\ln\frac{\mu_3}{2m_3} \right) 
	+\half\dF(\dF+1)^2 \\
	B_{Aha} &=& \CF\dF\dA\nS\left( \frac{1}{2\epsilon} +\frac{3}{2} +2\ln\frac{\mu_3}{2m_3} +\ln\frac{\mu_3}{2\mD} \right) \\
	B'_{Bhb} &=& \frac{1}{4}\dF\nS \left( \frac{1}{2\epsilon} +\frac{3}{2} +2\ln\frac{\mu_3}{2m_3} +\ln\frac{\mu_3}{2\mD'} \right) \\
	B'_{Ahb} &=& \CF\dF\nS \left( \frac{1}{2\epsilon} +\frac{3}{2} +2\ln\frac{\mu_3}{2m_3} +\ln\frac{\mu_3}{2\mD'} \right) \\
	B_{Bha} &=& \frac{1}{4}\dF\dA\nS \left( \frac{1}{2\epsilon} +\frac{3}{2} +2\ln\frac{\mu_3}{2m_3} +\ln\frac{\mu_3}{2\mD} \right) \\
	B_{Ahf} &=& \CA\dA\dF\nS \left( \frac{1}{2\epsilon} +\frac{3}{2} +2\ln\frac{\mu_3}{2\mD} +\ln\frac{\mu_3}{2m_3} \right) \\
	B_{hhf} &=& \dF\dA\nS\left( -\frac{1}{2\epsilon} -4 -2\ln\frac{\mu_3}{2(m_3+\mD)} -\ln\frac{\mu_3}{2m_3} \right) \\
	B_{hha} &=& \dF\dA\nS\left( -\frac{1}{2\epsilon} -4 -2\ln\frac{\mu_3}{2(m_3+\mD)} -\ln\frac{\mu_3}{2\mD} \right) \\
	B'_{hhf} &=& \dF\nS\left( -\frac{1}{2\epsilon} -4 -2\ln\frac{\mu_3}{2(m_3+\mD')} -\ln\frac{\mu_3}{2m_3} \right)  \\
	B'_{hhb} &=& \dF\nS\left( -\frac{1}{2\epsilon} -4 -2\ln\frac{\mu_3}{2(m_3+\mD')} -\ln\frac{\mu_3}{2\mD'} \right) \\
	b(x) &=& \CF\dF\nS\left( -\frac{1}{8\epsilon} -1 -\half\ln\frac{\mu_3}{2m_3+\mD+\mD'} -\frac{1}{4}\ln\frac{\mu_3}{2x} \right)
\end{eqnarray}

\section{Pressure near the electroweak crossover}

When evaluating the pressure close to the temperatures corresponding to the electroweak crossover, we again need to recalculate $p_\mathrm{M}(T)$. The computation was carried out in \cite{Gynther:2005av} and we cite the results here. The expression for the pressure will be given by
\begin{eqnarray}
p(T) & = & p_\mathrm{E}(T) + p_{\mathrm{M}1}(T) + p_{\mathrm{M}2}(T) + p_\mathrm{QCD}(T) + T^4\mathcal{O}\left(g^{5.5}\right),
\end{eqnarray}
where $p_\mathrm{E}(T)$ and $p_\mathrm{QCD}(T)$ remain the same as in the high temperature calculation. We get $p_{\mathrm{M}1}(T)$ from $S_\mathrm{E}$ in Eq.~(\ref{eq:sesm}) by integrating over the adjoint scalars $A_0^a$ and $B_0$. The result is
\begin{eqnarray}
\frac{p_\mathrm{M1}(T)}{T} &=&  \frac{1}{4\pi}\left( \frac{1}{3}\dA \mD^3 +\frac{1}{3} \mD'^3 \right)
        +\frac{1}{(4\pi)^2} \CA\dA g_3^2\mD^2 \left( -\frac{1}{4\epsilon} -\frac{3}{4} - \ln \frac{\mu_3}{2\mD}\right)
        \nonumber \\
        &-& \frac{1}{(4\pi)^3\epsilon}\left[ \frac{1}{4}\CA\CF g_3^4\mD +\dA h_3^2\mD +h_3'^2\mD'
        +\frac{1}{4}\CF g_3^2 g_3'^2(\mD+\mD') \right]\frac{\dF}{2}  \nonumber \\
        &+&\frac{1}{(4\pi)^3}\left\{ g_3^4\mD \left[ \CA^2\dA\left( -\frac{89}{24} +\frac{11}{6}\ln 2 -\frac{\pi^2}{6} \right)
        +\CA\CF\dF\left( -\half -\frac{3}{4}\ln\frac{\mu_3}{2\mD} \right) \right] \right. \nonumber \\
        &+& g_3^2 g_3'^2\CF\dF \frac{1}{4}\left[ (\mD+\mD')\left( -4 -2\ln\frac{\mu_3}{\mD+\mD'} \right)
        -\mD\ln\frac{\mu_3}{2\mD} - \mD'\ln\frac{\mu_3}{2\mD'} \right]  \nonumber \\
        &+& \left. h_3^2\mD\dA\dF\left( -4 -3\ln\frac{\mu_3}{2\mD} \right) +h_3'^2\mD'\dF\left( -4 -3\ln\frac{\mu_3}{2\mD'} \right) \right\}. \label{eq:m1result}
\end{eqnarray}

Theory describing just the fundamental scalar and the gauge fields is defined by
\begin{eqnarray}
S_\mathrm{E}' & = & \int\!\dd^3\mathbf{x}\, \left\{ \frac{1}{4}G^a_{ij}G^a_{ij} + \frac{1}{4}F_{ij}F_{ij} + (D_i\Phi)^\dagger(D_i\Phi)
                      + \widetilde{m}_3^2\Phi^\dagger\Phi + \widetilde{\lambda}_3(\Phi^\dagger\Phi)^2\right\},
\end{eqnarray}
where the couplings are the same as the corresponding couplings in $S_\mathrm{E}$,
$(\widetilde{g}_3^2,\,\tilde{g}_3'^2,\,\widetilde{\lambda}_3)=(g_3^2,\,g_3'^2,\,\lambda_3)$. Mass of the fundamental scalar receives corrections,
\begin{eqnarray}
\widetilde{m}_3^2 & = & m_3^2 - \frac{1}{4\pi}\left(\dA h_3 \mD + \frac{1}{4}g_3'^2 \mD'\right) \nonumber \\
                  & &   -\frac{1}{2\pi}\left[\dA h_3 \mD\left(1+\ln\frac{\mu_3}{2\mD}\right)
                          +\frac{1}{4}g_3'^2\mD'\left(1+\ln\frac{\mu_3}{2\mD'}\right)\right]\epsilon + \mathcal{O}(g^4). \label{eq:m3mass}
\end{eqnarray}
With parameters defined as such, we get for $p_{\mathrm{M}2}(T)$:
\begin{eqnarray}
\frac{p_{\mathrm{M}2}(T)}{T} & = & \frac{\dF}{6\pi}\widetilde{m}_3^3 \nonumber \\
&-& \frac{\widetilde{m}_3^2}{(4\pi)^2}\left[\dF(\dF+1)\widetilde{\lambda}_3
          + \frac{1}{2}\dF\left( \CF\widetilde{g}_3^2 +\frac{1}{4}\tilde{g}_3'^2\right) \left( \frac{1}{\epsilon} +3 +4\ln\frac{\widetilde{\mu}_3}{2\widetilde{m}_3}\right)\right].\label{eq:m2result}
\end{eqnarray}

\bibliographystyle{h-physrev4}
\cleardoublepage
\addcontentsline{toc}{chapter}{Bibliography}
\bibliography{articles}

\end{document}